\documentclass[conference,compsoc]{IEEEtran}

\usepackage{hyperref}
\usepackage{mathpartir, mathrsfs, amssymb, stmaryrd, amsthm, stackrel}
\usepackage{bm} 
\usepackage{csquotes} 
\usepackage{lipsum} 
\usepackage{cancel} 
\usepackage{xspace}
\usepackage{todonotes}
\usepackage{mathtools}
\usepackage{mleftright}
\usepackage{multirow, ragged2e, tabularx}
\usepackage{pifont}
\usepackage{listings, color}
\usepackage{url}
\usepackage{subcaption}
\usepackage{multicol}
\usepackage{stackengine}
\usepackage{graphicx}
\usepackage[inline]{enumitem}
\graphicspath{ {101-figures/letters} }
\usepackage[noend]{algpseudocode}

\usepackage{footmisc}

\usepackage{tikz,array}
\usetikzlibrary{calc}

\ifluatex\usepackage{fontspec}\fi
\ifxetex\usepackage{fontspec}\fi

\makeatletter
\def\maybeloadfont[#1]#2{
  \@ifpackageloaded{fontspec}{
    \expandafter\newfontfamily\csname#1\endcsname{#1}[
      Path = ./100-assets/ ,
      UprightFont = #2 ,
      BoldFont = #2 ,
      ItalicFont = #2
    ]
  }{}
}

\maybeloadfont[mcursive]{Cursif-BLKG.ttf}

\maybeloadfont[mlogicname]{Ruthie-6dax.ttf}

\makeatother

\newtheorem{innerdef}{Definition}
\newenvironment{mdef}{\begin{innerdef}}{\end{innerdef}}

\newtheorem{innerdef*}{Definition}
\newenvironment{mdef*}{\begin{innerdef*}}{\end{innerdef*}}

\newtheorem{innernotation}{Notations}
\newenvironment{mnotation}{\begin{innernotation}}{\end{innernotation}}

\newtheorem*{innernotations*}{Notations}
\newenvironment{mnotation*}{\begin{innernotations*}}{\end{innernotations*}}

\newtheorem{innerremark}{Remark}
\newenvironment{mrmk}{\begin{innerremark}}{\end{innerremark}}

\newtheorem*{innerremark*}{Remark}
\newenvironment{mrmk*}{\begin{innerremark*}}{\end{innerremark*}}

\newtheorem{innerexample}{Example}
\newenvironment{mex}{\begin{innerexample}}{\end{innerexample}}

\newtheorem*{innerexample*}{Example}
\newenvironment{mex*}{\begin{innerexample*}}{\end{innerexample*}}

\newtheorem{innerlem}{Lemma}
\newenvironment{mlemma}{\begin{innerlem}}{\end{innerlem}}

\newtheorem{innercor}{Corollary}
\newenvironment{mcorollary}{\begin{innercor}}{\end{innercor}}

\newtheorem{innerth}{Theorem}
\newenvironment{mth}[1][]{\begin{innerth}[#1]}{\end{innerth}}

\newtheorem*{innerth*}{Theorem}
\newenvironment{mth*}[1][]{\begin{innerth*}[#1]}{\end{innerth*}}

\newtheorem{innerprop}{Property}
\newenvironment{mprop}{\begin{innerprop}}{\end{innerprop}}

\newtheorem*{innerprop*}{Property}
\newenvironment{mprop*}{\begin{innerprop*}}{\end{innerprop*}}

\newenvironment{mproof}[1][]{\begin{IEEEproof}}{\end{IEEEproof}}

\usepackage{environ}
\NewEnviron{mtheoremenv}[3][]{
    \ifstrempty{#3}{
        \begin{mth*}[#2]
            \BODY
        \end{mth*}
    }{
        \begin{mth}[#2]\label{th:#3}
            \BODY
        \end{mth}
    }
}

\newenvironment{mdefenv}[3][]{%
    \begin{mdef}[#2]\ifstrempty{#3}{}{\label{def:#3}}%
}{
    \end{mdef}
}

\newenvironment{mlemmeenv}[3][]{%
    \begin{mlemma}[#2]\ifstrempty{#3}{}{\label{lemme:#3}}%
}{
    \end{mlemma}
}

\NewEnviron{mpropenv}[3][]{
    \ifstrempty{#3}{
        \begin{mprop*}[#2]
            \BODY
        \end{mprop*}
    }{
        \begin{mprop}[#2]\label{prop:#3}
            \BODY
        \end{mprop}
    }
}

\NewEnviron{mrmqenv}[3][]{
    \ifstrempty{#3}{
        \begin{mrmk*}[#2]
            \BODY
        \end{mrmk*}
    }{
        \begin{mrmk}[#2]\label{rmq:#3}
            \BODY
        \end{mrmk}
    }
}

\NewEnviron{mexplenv}[3][]{
    \ifstrempty{#3}{
        \begin{mex*}[#2]
            \BODY
        \end{mex*}
    }{
        \begin{mex}[#2]\label{expl:#3}
            \BODY
        \end{mex}
    }
}

\newenvironment{mdemoenv}[2][]{%
    \ifstrempty{#2}{%
        \begin{IEEEproof}
    }{%
        \begin{IEEEproof}[Proof of \autoref{#2}]\label{demo:#2}
    }%
}{\end{IEEEproof}}

\NewEnviron{mnotationenv}[3][]{
    \ifstrempty{#3}{
        \begin{mnotation*}[#2]
            \BODY
        \end{mnotation*}
    }{
        \begin{mnotation}[#2]\label{notation:#3}
            \BODY
        \end{mnotation}
    }
} 

\newif\ifcomments

\newif\ifdraft

\newcommand{\mtodo}[2][]{\ifcomments \todo[#1]{#2} \fi}

\newcommand{\notetoself}[1]{\ifcomments

{\footnotesize #1}

\fi%
}
\newcommand{\gsep}{\,\big|\,}

\newcommand{\msquirrel}{\textsc{Squirrel}\xspace}
\newcommand{\mvampire}{\textsc{Vampire}\xspace}
\newcommand{\tool}{\textsc{CryptoVampire}\xspace} 

\newcommand{\AAA}{\ensuremath{\mathcal{A}}}
\renewcommand{\implies}{\Rightarrow}

\newcommand{\customType}[2][]{\ensuremath{\mathsf{#2}}\xspace}
\newcommand{\msgt}[1][]{\customType[#1]{Message}}
\newcommand{\condt}[1][]{\customType[#1]{Condition}}
\newcommand{\tpt}[1][]{\customType[#1]{Time}}

\newcommand{\boolt}[1][]{\customType[#1]{Bool}}
\newcommand{\bitstrt}[1][]{\customType[#1]{Bitstring}}
\newcommand{\noncet}[1][]{\randomt[#1]}
\newcommand{\randomt}[1][]{\customType[#1]{Nonce}}

\newcommand{\mat}{\makeatletter @ \makeatother}
\newcommand{\mfindst}[4]{\ensuremath{\textsc{Find } #1 \textsc{ \allowbreak Such \allowbreak that } #2 \textsc{ \allowbreak Then } #3 \textsc{ \allowbreak Else } #4}}
\newcommand{\mifthenelse}[3]{\ensuremath{\textsc{If } #1 \allowbreak\textsc{ Then } #2 \allowbreak\textsc{ Else } #3}}
\newcommand{\mifthenelsenb}[3]{\ensuremath{\textsc{If } #1 \textsc{ Then } #2 \textsc{ else } #3}}

\definecolor{sugar}{HTML}{666699}
\definecolor{constructors}{HTML}{b800e6}

\DeclareMathOperator{\mpred}{{\color{constructors} pred}}
\DeclareMathOperator{\minput}{{\color{constructors} input}}
\DeclareMathOperator{\mmsg}{{\color{sugar} msg}}
\DeclareMathOperator{\mcond}{{\color{sugar} cond}}
\DeclareMathOperator{\mhappens}{{\color{constructors} happens}}

\DeclareMathOperator{\minit}{\ensuremath{\mathsf{init}}}
\DeclareMathOperator{\mundef}{\ensuremath{\mathsf{undef}}}
\DeclareMathOperator{\matt}{\ensuremath{\mathsf{att}}}

\DeclareMathOperator{\mhash}{\ensuremath{\mathcal{H}}}
\DeclareMathOperator{\mverify}{\ensuremath{\mathsf{verify}}}

\DeclareMathOperator{\mmok}{\ensuremath{\mathsf{ok}}}
\DeclareMathOperator{\mmko}{\ensuremath{\mathsf{ko}}}
\DeclareMathOperator{\mfail}{\ensuremath{\mathsf{fail}}}

\DeclareMathOperator{\mmin}{\ensuremath{\mathsf{in}}}

\DeclareMathOperator{\mfv}{fv}

\newcommand{\mepred}[2][\mathbb{T}]{\overline{\mathsf{pred}}_{ #1 }\mleft( #2 \mright)}

\newcommand{\barWithSpacing}[1]{\mathbin{\bar{#1}}}
\newcommand{\barWithoutSpacing}[1]{\bar{#1}}

\newcommand{\bvee}{\barWithSpacing{\vee}}
\newcommand{\bwedge}{\barWithSpacing{\wedge}}
\newcommand{\bimplies}{\barWithSpacing{\Rightarrow}}

\newcommand{\beq}{\barWithSpacing{=}}

\newcommand{\bneg}{\barWithoutSpacing{\neg}}
\newcommand{\bexists}{\barWithoutSpacing{\exists}}
\newcommand{\bforall}{\barWithoutSpacing{\forall}}
\newcommand{\bbot}{\mathtt{false}}
\newcommand{\btop}{\mathtt{true}}

\newcommand{\meval}[1]{\mleft| #1 \mright|}

\newcommand{\mevaltimeE}[2][\mathbb{E}]{\left( #2 \right)^{#1}}

\newcommand{\mevalidxT}[2][\mathbb{T}]{\underline{ #2 }_{#1}}
\newcommand{\mevaltimeT}[2][\mathbb{T}]{\mevaltimeE[#1]{#2}}

\newcommand{\mevalT}[2][\mathbb{T}]{\left[ #2 \right]^{#1}}
\newcommand{\mevalTsq}[2][\mathbb{T}]{\left[ #2 \right]^{#1}_{\mathrm{sq}}}

\newcommand{\mevalbc}[1]{\ensuremath{\mleft\llbracket #1 \mright\rrbracket_\text{bc}^{\mathbb{M}}}}
\newcommand{\mevalsq}[2][\mathbb{T}]{\ensuremath{\mevalbc{\mevalTsq[#1]{#2}}}}

\newcommand{\bcint}[3]{\ensuremath{\mevalbc{#1}\mleft( 1^{#2}, #3 \mright)}} 

\newcommand{\smtlib}{\msmt}
\newcommand{\eufcma}{\textsc{Euf-Cma}\xspace}
\newcommand{\mac}{\textsc{MAC}\xspace}

\newcommand{\msmt}{\textsc{Smt-Lib}\xspace}

\newcommand{\msubstit}[3]{{{#1}{\mleft\{ #2 \mapsto #3 \mright\}}}}

\usepackage{xparse, nameref}
\makeatletter

\DeclareDocumentCommand \inferrule { s O {} m m o }{%
  \IfBooleanTF{#1}%
  {%
    \mpr@inferstar[#2]{#3}{#4}%
  }{%
    \mpr@inferrule[#2]{#3}{#4}%
  }%
  \IfValueT{#5}%
  {%
    \textsc{#5}%
    \my@name@inferrule{\textsc{#5}}%
  }%
}
\NewDocumentCommand \my@name@inferrule { m }{%
  \def\@currentlabelname{\ensuremath{#1}}%
}
\makeatother

\newcommand{\bsqsubseteq}{\mathrel{\sqsubseteq^\circ}}

\newcommand{\mnonce}[1]{\bar{#1}}

\newcommand{\marity}[1]{\left\| #1 \right\|}

\DeclareMathOperator{\mnegl}{negl}
\newcommand{\mprob}[1][\rho]{\ensuremath{{\mathbb{P}\mathrm{rob}}_{#1}}\xspace}

\newcolumntype{M}[1]{>{\centering\arraybackslash}m{#1}}

\newcommand{\mzed}{\textsc{z3}\xspace}
\newcommand{\mcvc}{\textsc{cvc5}\xspace}

\newcommand{\cmark}{\ding{51}}%
\newcommand{\xmark}{\ding{55}}%

\usetikzlibrary{shapes, arrows.meta}

\makeatletter
\newcommand{\mhypertarget}[2]{%
  \hypertarget{#1}{#2}%
    \protected@write\@mainaux{}{%
        \string\expandafter\string\gdef
          \string\csname\string\detokenize{#1}\string\endcsname{#2}%
    }%
  }
\newcommand{\myhyperlink}[1]{%
\hyperlink{#1}{\csname #1\endcsname}%
}
\makeatother

\newcommand\mrestriction[2]{{
  \left.\kern-\nulldelimiterspace 
  #1 
  \vphantom{\big|} 
  \right|_{#2} 
  }}

\newcommand{\appref}[1]{\hyperref[#1]{Appendix~\ref*{#1}}}

\newcommand{\SymbLogic}{Symbolic Logic\xspace}
\newcommand{\EvalLogic}{Evaluated Logic\xspace}
\newcommand{\BCLogic}{BC Logic\xspace}

\makeatletter
\@ifpackageloaded{fontspec}{%
  \newcommand{\genf}{\scalebox{0.7}{\mcursive f}}
}{
  \newcommand{\genf}{\mathfrak{f}}
}
\makeatother

\newcommand{\ib}{\bm{\mathrm{i}}}

\newcommand{\ub}{\bm{u}}
\newcommand{\vb}{\bm{v}}

\newcommand{\noncefont}[1]{\mathsf{#1}}
\newcommand{\n}{\ensuremath{\noncefont{n}}\xspace}
\newcommand{\m}{\ensuremath{\noncefont{m}}\xspace}
\newcommand{\mk}{\ensuremath{\noncefont{k}}\xspace}

\newcommand{\cellfont}[1]{\mathfrak{#1}}
\newcommand{\cc}{\cellfont{c}\xspace}

\newcommand{\stepfont}[1]{\mathsf{#1}}
\newcommand{\stpa}{\stepfont{a}\xspace}
\newcommand{\stpb}{\stepfont{b}\xspace}

\newcommand{\veci}{\vec{\imath}}
\newcommand{\vecj}{\vec{\jmath}}
\newcommand{\vecu}{\vec{u}}
\newcommand{\vecv}{\vec{v}}
\newcommand{\vecib}{\vec{\bm{\mathrm{\i}}}}
\newcommand{\vecjb}{\vec{\bm{\mathrm{\j}}}}
\newcommand{\vect}{\vec\tau}
\newcommand{\vectb}{\vec{\bm{\tau}}}
\newcommand{\vecub}{\vec{\ub}}
\newcommand{\vecvb}{\vec{\vb}}

\newcommand{\vecscalled}[1]{\vec{\scalebox{1}[0.9]{$#1$}}}
\newcommand{\vecI}{\vecscalled{I}}

\newcommand{\alphab}{\bm{\alpha}}
\newcommand{\vecalpha}{\vec\alpha}
\newcommand{\vecalphab}{\vec{\alphab}}

\newcommand{\betab}{\bm{\beta}}
\newcommand{\vecbeta}{\vec{\beta}}
\newcommand{\vecbetab}{\vec{\betab}}

\makeatletter
\@ifpackageloaded{fontspec}{%
  \newcommand{\logicn@me}{\mathpalette\logicn@meinner\relax}%
  \newcommand{\logicn@meinner}[2]{%
    \scalebox{1.2}{$#1\text{\mlogicname \ln@me}$}%
  }%
  \newcommand{\logicnameone}[1]{\def\ln@me{#1}\logicn@me}
  \newcommand{\logicname}[2]{\logicnameone{#1}}
}{
  \newcommand{\logicnameone}[1]{\ensuremath{\mathfrak{#1}}}
  \newcommand{\logicname}[3][-0.3ex]{%
    \ifdraft%
        \ensuremath{\mathscr{#2}}%
    \else%
      \ifmmode%
        \mathchoice{\raisebox{#1}{\includegraphics[height=2.2ex]{#3}}}
          {\raisebox{#1}{\includegraphics*[height=2.2ex]{#3}}}
          {\raisebox{#1}{\includegraphics*[height=1.5ex]{#3}}}
          {\raisebox{#1}{\includegraphics*[height=1ex]{#3}}}
      \else%
        \raisebox{#1}{\includegraphics*[height=2ex]{#3}}%
      \fi%
    \fi%
  }
}
\makeatother

\newcommand{\msk}[1][]{\mathrm{sk}_{#1}}
\newcommand{\mSk}{\mathcal{S}\mathrm{k}}

\newcommand{\mindices}{\ensuremath{\mathcal{I}}\xspace}
\newcommand{\mnonces}{\ensuremath{\mathcal{N}}\xspace}
\newcommand{\mfunctions}{\ensuremath{\mathcal{F}}\xspace}
\newcommand{\msteps}{\ensuremath{\mathcal{S}}\xspace}
\newcommand{\mcells}{\ensuremath{\mathcal{C}}\xspace}
\newcommand{\mskolems}{\ensuremath{\mSk}\xspace}

\newcommand{\mnoncesbc}{\ensuremath{\mathcal{N}_\mathrm{bc}}\xspace}
\newcommand{\mfunctionsbc}{\ensuremath{\mathcal{F}_\mathrm{bc}}\xspace}
\newcommand{\mattsbc}{\ensuremath{\mathcal{G}_\mathrm{bc}}\xspace}

\newcommand{\BCname}{\logicname{BC}{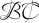}\xspace}
\newcommand{\mbcl}[1][\mnoncesbc, \mfunctionsbc, \mattsbc]%
  {\ensuremath{\BCname\mleft( #1 \mright)}\xspace}

\newcommand{\msymblcontent}{\mnonces, \mfunctions, \mindices, \mcells, \msteps}

\newcommand{\Symbname}{\logicname{S}{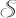}\xspace}
\newcommand{\msymbl}[1][\msymblcontent]%
  {\ensuremath{\Symbname\mleft( #1 \mright)}\xspace}

\newcommand{\Evalname}{\logicname{E}{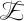}\xspace}
\newcommand{\mevall}[1][\msymblcontent]%
  {\ensuremath{\Evalname\mleft( #1 \mright)}\xspace}

\newcommand{\Evalnamesk}{\logicname{E}{e-letter}_\mathrm{sk}\xspace}
\newcommand{\mevallskinner}[2]%
  {\ensuremath{\Evalnamesk\mleft( #1, #2 \mright)}\xspace}
\newcommand{\mevallsk}[1][\mSk]%
  {\mevallskinner{\mevall[\msymblcontent]}{#1}}

\DeclareMathOperator{\domterm}{\logicname[-0.25ex]{D}{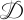}\xspace}

\newcommand{\modelclass}{\logicname[-0.3ex]{C}{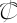}\xspace}

\newcommand{\proves}{\vdash}
\newcommand{\validp}[1][\mathbb{P}]{\vDash_{#1}}
\newcommand{\valid}{\vDash}
\newcommand{\validbc}{\vDash_\mathrm{bc}}

\newcommand{\domE}{\ensuremath{\mathbb{S}}}
\newcommand{\domI}{\ensuremath{\mathbb{I}}}
\newcommand{\domSE}[1][\msteps]{\ensuremath{#1_{\domI}}}
\newcommand{\precE}[1][\domE]{<_{#1}}
\newcommand{\preceqE}[1][\domE]{\leq_{#1}}

\newcommand{\sigmI}{\ensuremath{\sigma_\domI}}

\newcommand{\Ez}{\ensuremath{\mathcal{E}_0}\xspace}

\DeclareMathOperator{\dirac}{\delta}

\newcommand{\mhelpmacro}[3][\mathbb{T}]{\mathsf{#2}^{#1}_{#3}}
\newcommand{\mminput}[2][\mathbb{T}]{\mhelpmacro[#1]{input}{#2}}
\newcommand{\mmexec}[2][\mathbb{T}]{\mhelpmacro[#1]{exec}{#2}}
\newcommand{\mmframe}[2][\mathbb{T}]{\mhelpmacro[#1]{frame}{#2}}

\newcommand{\modelc}{\ensuremath{\mathcal{L}}\xspace} 
\newcommand{\modelsk}{\ensuremath{\mathsf{M}_{\msk}}\xspace} 
\newcommand{\modelec}{\ensuremath{\mathfrak{K}}\xspace} 
\newcommand{\compmodel}{\ensuremath{\mathbb{M}}\xspace} 

\newcommand{\submodel}{\trianglelefteq}
\newcommand{\symdiff}{\mathbin{\Delta}}

\newcommand{\mstbc}{\mathsf{st}_{\BCname}}
\newcommand{\mmst}[1][\mathcal{P}]{\mathsf{st}_{#1}}

\newcommand{\mstsq}{\textsf{st}_\mathcal{P}^{\square}}
\newcommand{\mmsttmp}[2]{\textsf{st}_\mathcal{P}^{#1\, #2}}

\newcommand{\mmmap}[1]{\mathscr{T}\left( \underline{#1} \right)}

\begin{document}


    \newif\ifreport
    \reporttrue

    \newif\ifannon
%
\title{\tool:  Automated Reasoning for the Complete Symbolic Attacker Cryptographic  Model}

%
\newcommand{\tuwienaffiliation}{%
  \institution{TU Wien}%
  \city{Vienna}%
  \country{Austria}%
}








    \ifannon\else
        \author{%
    \IEEEauthorblockN{Simon~Jeanteur\IEEEauthorrefmark{1}, %
    Laura~Kov\'acs\IEEEauthorrefmark{1}, %
    Matteo~Maffei\IEEEauthorrefmark{1}\IEEEauthorrefmark{2} and %
    Michael~Rawson\IEEEauthorrefmark{1}}
    \IEEEauthorblockA{%
        \IEEEauthorrefmark{1}TU Wien, \{firstname\}.\{lastname\}@tuwien.ac.at
    }
    \IEEEauthorblockA{\IEEEauthorrefmark{2}Christian Doppler Laboratory Blockchain Technologies for the Internet of Things}
}
    \fi

    \maketitle
    
    \ifdraft
        \todo[inline]{If you see this, you are in \enquote{draft} mode. (ex: $\Symbname$)}
    \fi

    \mtodo[inline]{If you see this, then comments are set to \texttt{true}}

    \begin{abstract}
        Cryptographic protocols are hard to design and prove correct, as witnessed by the ever-growing list of attacks even on protocol standards. \emph{Symbolic models of cryptography} enable automated formal security proofs of such protocols against an idealized cryptographic model, which abstracts away from the algebraic properties of cryptographic schemes and thus misses attacks. \emph{Computational models of cryptography} yield rigorous guarantees but support at present only interactive proofs and/or restricted classes of protocols (e.g., stateless ones). A promising approach is given by the \emph{computationally complete symbolic attacker (CCSA)} model, formalized in the \BCLogic, which aims at bridging and getting the best of the two worlds, obtaining cryptographic guarantees by symbolic protocol analysis. The \BCLogic is supported by a recently developed interactive theorem prover, namely \msquirrel, which enables machine-checked interactive security proofs, as opposed to automated ones, thus requiring expert knowledge both in the cryptographic space as well as on the reasoning side.

In this paper, we introduce the \tool cryptographic protocol verifier, which for the first time fully automates proofs of trace properties in the \BCLogic. The key technical contribution is a first-order formalization of protocol properties with tailored handling of subterm relations. As such, we overcome the burden of interactive proving in higher-order logic and automatically establish soundness of cryptographic protocols using {only} first-order reasoning. Our first-order encoding of cryptographic protocols is challenging for various reasons. On the theoretical side, we restrict full first-order logic with cryptographic axioms to ensure that, by losing the expressivity of the higher-order BC Logic, we do not lose soundness of cryptographic protocols in our first-order encoding. On the practical side, \tool integrates dedicated proof techniques using first-order saturation algorithms and heuristics, which all together enable leveraging the state-of-the-art \mvampire first-order automated theorem prover as the underlying proving engine of \tool. Our experimental results showcase the effectiveness of \tool as a standalone verifier as well as in terms of automation support for \msquirrel.
    \end{abstract}


    \mleftright

    \ifreport
        This is the technical report.
    \fi
    \section{Introduction}%
    \label{sec:introduction}
    
\notetoself{
}

\emph{Cryptographic protocols} are the software interfaces used by the components of our digital world to communicate securely with one another.
Unfortunately, designing such protocols is notoriously difficult and error-prone.
From the classic attack on the Needham-Schroeder protocol~\cite{loweAttackNeedhamSchroederPublickey1995}, over to the ubiquitous but repeatedly broken TLS protocol~\cite{alfardanLuckyThirteenBreaking2013}, up to new and subtle blockchain protocols~\cite{malavoltaAnonymousMultiHopLocks2019}, the list of attacks on popular standards is ever-growing.

Formal methods have proved to be a very successful tool to guarantee properties of protocols and recently accompanied the design of protocol standards like TLS~1.3~\cite{bhargavanProvingTLSHandshake2014}, WireGuard~\cite{lippMechanisedCryptographicProof2019}, or 5G-AKA~\cite{koutsos5GAKAAuthenticationProtocol2019a}.

\subsection{Related Work}
Security properties for cryptographic protocols are typically formalized in terms of trace properties~\cite{loweHierarchyAuthenticationSpecifications1997} or observational equivalence relations~\cite{DBLP:conf/csfw/ClarksonS08}.
The former express guarantees on a single execution trace  (possibly reflecting multiple protocol sessions) and cover secrecy of random data, authentication, and more general constraints on the order of security-relevant events.
The latter express the inability of an attacker to tell the difference between two protocol configurations, such as the inability to understand which among two low-entropy secrets is used.

Formal verification of cryptographic protocols is further split into techniques working in the \emph{symbolic} or \emph{computational} model.
Symbolic techniques typically abstract away from the algebraic properties of cryptography by reasoning over the so-called \enquote{Dolev-Yao} attacker~\cite{dolev-yao}, which has infinite computational resources at its disposal but is restricted to building only symbolic terms based on those already in their knowledge.
This results in easier proof techniques, which enabled the design of several successful automated protocol verifiers like \textsc{ProVerif}~\cite{proverif} and \textsc{Tamarin}~\cite{tamarin}.
However, such a symbolic setting does not provide any computational guarantees: for instance, symbolically-secure protocols may, in fact, be attacked by leveraging weaknesses in the underlying cryptographic realization~\cite{bhargavanTranscriptCollisionAttacks2016}.

The computational cryptographic model is instead based on computational and probabilistic notions, which faithfully describe real-world implementations.
In a computational setting, the attacker is a probabilistic polynomial-time Turing machine, and cryptographic proofs show that if the attacker is able to break the security of a protocol, then it can also with high probability break commonly accepted mathematical assumptions.
Such a proof technique over the computational protocol model gives strong security guarantees against real-world cryptographic implementations, but it is much harder to formalize and automate.

Nonetheless, a few approaches automating computational proofs emerged over the years, such as \textsc{EasyCrypt}~\cite{easycrypt}, a cryptographic proof checker that simplifies reasoning about probabilistic computations in an adversarial setting and is typically used to prove the security of cryptographic schemes, and \textsc{CryptoVerif}~\cite{cryptoverif}, a protocol verifier that automates the type of game-based proofs conducted by cryptographers and scales to cryptographic protocols, although it is restricted to stateless ones.

Recently, various approaches tried to merge the symbolic and computational protocol models to obtain the best of the two worlds, i.e., the ease of proofs and expressiveness of the former with the strong cryptographic guarantees of the latter.
In particular, Bana and Comon-Lundh designed the \emph{Computationally Complete Symbolic Attacker} (CCSA) model and its accompanying \BCLogic~\cite{long_bana_towards_2012,bana_computationally_2014,banaComputationallyCompleteSymbolic2013}, which later evolved into a fragment of higher-order logic~\cite{BaeldeKoutsosLallemand2023}.
\BCLogic supports both trace properties and observational equivalence and aims at opening the door to formal, computationally sound verification, although proofs in \BCLogic had long to be done by hand.

Recently, a breakthrough in the mechanization of proofs in the \BCLogic has materialized in the interactive \msquirrel proof assistant~\cite{squirrel}, which supports both trace properties and observational equivalence.
\msquirrel pushed further developments of the CCSA, such as the support for stateful protocols~\cite{baelde_cracking_2022}, post-quantum security~\cite{cremersLogicInteractiveProver2022}, and proof composition~\cite{composition_framework_2020}.
\msquirrel is an interactive theorem prover that is effective in yielding machine-checked proofs but does not provide full automation and requires expertise in both cryptography and logic.

The key technical challenge towards proof automation in the \BCLogic is the expressiveness of the underlying higher-order logical fragment, which blends together a very relaxed equality theory with strict symbolic reasoning in the form of a complex subterm relation.
Such a logical combination extends the full first-order theory of term algebras (including algebraic data types, such as lists) with the subterm predicate, which is inherently undecidable~\cite{venkataraman1987decidability}.
While proving security properties in first-order logic instead of \BCLogic would come with the benefits of semi-decidability and (potential) full automation, the use of subterm relations (and other higher-order constructs) from \BCLogic imposes an open challenge to automated theorem proving~\cite{z3,vampire,cvc5} as such relations are not first-order axiomatizable.

\subsection{Our contributions} We introduce \tool, \emph{the first  automated verification tool for the \BCLogic} 
(\autoref{sec:evaluated}). Our work so far focuses on trace properties, leaving observational equivalence relations for future work. The design of \tool blends together four core  contributions:

\begin{enumerate}[
    label=\textbf{(\Alph{enumi})},
    ref=(\Alph{enumi}),
    wide 
]
    \item We provide an encoding of the \BCLogic into \emph{first-order logic}.
    Our first-order encoding of \BCLogic resolves the challenge of effective subterm reasoning (\autoref{sec:form:challenges}).
    In particular, given that subterm relations are not finitely axiomatizable in first-order logic, we provide a tailored handling of subterm relations to ensure that subterm reasoning within \tool can be directly supported within saturation-based first-order theorem proving~\cite{vampire}.

    \item When moving from the higher-order setting of \BCLogic to our first-order formalization, we ensure that our loss in logical expressivity when compared to the \BCLogic does not impact the soundness of our proofs (\autoref{sec:evaluated}).
    Namely, we show that if a protocol is proven secure in our first-order logic, then it is secure in \BCLogic too (\autoref{prop:validity}).

    \item We enhance the performance of \tool by introducing \emph{dedicated reasoning procedures} within saturation.
    \ifreport
        In particular, we control proof search by specialized term orderings, preprocessing techniques, and heuristics
    \else
        In particular, we control proof search by preprocessing techniques and heuristics
    \fi
    (\autoref{sec:optimisations}).

    \item We conduct an \emph{experimental evaluation} of \tool on all trace-based queries from the \msquirrel library (\autoref{sec:experiments}): \tool can verify all these protocols in a few milliseconds.
    We further demonstrate how \tool exhibits better performance than \textsc{CryptoVerif} on these protocols.
    Finally, we demonstrate the usefulness of \tool as automation support for \msquirrel by proving a set of lemmas used in \msquirrel proofs, some of which are used in the proofs of observational equivalence properties.
    Overall, our experimental results demonstrate that \tool is not only effective as a standalone cryptographic protocol verifier but can also be used to partially automate \msquirrel proofs.
\end{enumerate}

\ifreport
\else
    All proofs and the accompanying technical material are available online~\cite{CRYPTOVAMPIREAutomatedReasoningTR}.
\fi

    \section{Overview}%
    \label{sec:overview}

This section gives a high-level overview of \tool, while the following sections will go into the  more formal details. The tool takes as input \begin{enumerate*}[label=\textbf{(\roman{enumi})}, ref=(\roman{enumi})] \item a protocol specification composed of function declarations, capturing the cryptographic messages used in the protocol, \item constraints on those functions, expressing their cryptographic semantics, and \item a protocol description in the form of partially preordered \emph{steps}.\end{enumerate*}
The given protocol description is translated into a first-order logic (FOL) formula over term algebras with tailored term evaluations, and the final proof is off-loaded to the \mvampire first-order theorem prover~\cite{vampire}.
In other words, \tool transforms the problem of protocol verification into validity checking of term algebra properties, integrating first-order axiomatic reasoning completely within saturation theorem proving.
As such, \tool provides fully-automated reasoning to prove the security of CCSA models, leveraging and extending subterm reasoning in term algebras within \mvampire.

\begin{mex}[Basic Hash]\label{basic-hash-example}
    Let us illustrate our approach through the following simple protocol inspired by the RFID protocols described in~\cite{brusoFormalVerificationPrivacy2010}:
    \begin{equation}
        T_{j} \rightarrow R:\quad \left< \n, \mhash\left( \n, \mk_j \right)\right>
    \end{equation}

    Here, a tag $T_j$ outputs a \emph{fresh} nonce $\n$ both in plain text and hashed ($\mhash$) with the key $\mk_j$ shared with a reader $R$. For simplicity, we model a single reader $R$ operating over multiple tags $T$ (parameterized by $j$).

    Intuitively, this protocol guarantees a property called non-injective agreement~\cite{loweHierarchyAuthenticationSpecifications1997}, i.e., when $R$ authenticates tag $T_{j}$, then $T_{j}$ started an authentication session with $R$.
\end{mex}

We will now illustrate the main steps of the analysis of \autoref{basic-hash-example} in \tool.

\subsection{First-Order Formalization of the Protocol\label{sec:motivating:FOTheory}}

Following the CCSA model, cryptography is represented \emph{symbolically}, i.e., using a term algebra, which is built on a set of honest functions $\mfunctions$ and a set of nonce names $\mnonces$.

\subsubsection{Honest functions} 
The functions in $\mfunctions$ are \emph{uninterpreted functions} in \tool and they represent deterministic polynomial time Turing machines.

\begin{mex}[Functions]
    To encode \autoref{basic-hash-example}, we will use the following functions in $\mathcal{F}$:  
    $\left<\_, \_ \right>,\allowbreak \pi_1(\_),\allowbreak \pi_2(\_)$ respectively denoting a pair constructor $\left<\_,\_\right>$ and projections $\pi_i$;
    $\mmok,\allowbreak \mmko$ for success and failure of tag authentication;  
    $\mhash\left( \_, \_ \right),\allowbreak \mverify\left( \_, \_, \_ \right) $ for hash computation and verification;
    and $\mifthenelse{\_\allowbreak}{\_\allowbreak}{\_\allowbreak}$ for conditionals.
\end{mex}

\subsubsection{Honest randomness} 
The only source of honest randomness in \tool are \emph{nonces}.
They represent pairwise independent random bitstrings of length $\eta$, the security parameter.
In the logic, nonces are constants, which are used to express certain cryptographic properties (e.g., no collision).
Nonces are indexed in order to support freshness across an unbounded number of sessions and participants.

\begin{mex}[Nonces]
    In \autoref{basic-hash-example}, $\mnonces = \{\n\left[ \_, \_ \right], \mk\left[ \_ \right] \}$.
    Furthermore, $\n\left[ i, j \right]$ is indexed by the session $i$ and the identifier $j$ of the participating tag, and $\mk\left[ j \right]$ is only indexed by the tag $j$.
    We use nonces to represent keys as they are a source of honest randomness.
\end{mex}

\subsubsection{Cryptographic properties}\label{par:overview:crypto-properties}
The semantics of constants and functions are expressed through FOL rules, which are typically used to express cryptographic properties.

\begin{mex}[Cryptographic Properties]
    Intuitively, in the first-order formalization of \autoref{basic-hash-example}, we will encode the semantics of projection operators as $\meval{\pi_i\left( \left< x_1, x_2 \right> \right)} = \meval{x_i}$, for $i=1, 2$.
    We write $\meval{t}$ to denote the evaluation of $t$ (e.g., applying a projection operator).
    We also formalize message authentication code through the verification condition $\meval{\mverify( \sigma, m, k )} \Leftrightarrow ( \meval{\sigma} = \meval{\mhash(m, k)})$.
    We capture existential unforgeability via formula (\eufcma), which we axiomatize in FOL as shown in \autoref{prop:euf-cma}.  %
\end{mex}

\begin{mprop}[\eufcma]\label{prop:euf-cma}
    If $\mhash(\_, \_)$ and $\mverify(\_, \_, \_)$ form a \mac scheme that is existentially unforgeable under chosen message attacks, 
    then, for all nonce $\mk$, the protocol $\mathcal{P}$ satisfies the following:
    \begin{multline}\label{eq:eufcma1}
        \meval{ \mverify\left(\sigma, m, \mk \right)} \Rightarrow \\
            \mleft(\begin{multlined} \mathsf{k}\sqsubseteq_{ \mverify\left(\_,\_,\bullet\right),\mhash\left(\_,\bullet\right)} m, \sigma \\
        \vee \exists u. \mleft( \mhash\left(u, \mk\right) \sqsubseteq m, \sigma \wedge \meval{u} = \meval{m} \mright)\end{multlined} \mright)
    \end{multline}
\end{mprop}

This means that, assuming $\mhash(\_, \_)$ and $\mverify(\_, \_, \_)$ are \eufcma-secure, for any execution of $\mathcal{P}$, any message $m$, \mac $\sigma$, and key\footnote{Remember that keys are represented as nonces} $\mk$, if $(m, \sigma)$ form a valid \mac pair with key $\mk$, then the key \emph{symbolically appears} in a non-key position (i.e., in one of the first two arguments of $\mverify$ or in the first argument of $\mhash$) in $m$, $\sigma$, or $\mathcal{P}$ (left disjunction operand $\sqsubseteq$), or we can find another message $u$ that evaluates to the same bitstring as $m$ and whose \mac with key $\mk$ symbolically appears in either $m$ or $\sigma$ (right disjunction operand $\sqsubseteq_{ \mverify\left(\_,\_,\bullet\right),\mhash\left(\_,\bullet\right)}$).
In other words, if verification succeeds, then either the key has been misused or the message has indeed been signed before.
The cryptographic soundness of \eufcma is proven in \ifreport \appref{sec:crypto-axioms} \else \cite{CRYPTOVAMPIREAutomatedReasoningTR} \fi.
Note that, in the CCSA model, the attacker can take any action, unless specific rules restrict its capabilities (e.g., \eufcma and nonce guessing).

\subsubsection{Protocol steps} Within \tool, a protocol is a set of atomic \emph{steps}, and each execution of this protocol is a valid sequence of steps. A step takes an \emph{input} and computes an \emph{output} (or \emph{message}) and assignments to some \emph{memory cells}. A step may be guarded by a \emph{condition} (also computed from the input), which specifies under which assumptions the step is executed.

\begin{mex}[Steps]\label{ex:basic-hash:steps}
    We can express \autoref{basic-hash-example} in terms of the following steps:

    \underline{$\mathsf{T}\left[i, j\right]$}: ($i^\text{th}$ execution of the $j^\text{th}$ tag)
    \begin{itemize}
        \item[] \textbf{Condition}: $\btop$
        \item[] \textbf{Message}: $\left<\n\left[ i, j \right], \mhash\left( \n\left[ i, j \right], \mk\left[ j \right] \right) \right>$
    \end{itemize}

    \underline{$\mathsf{Rs}\left[i, j\right]$}: (Successful authentication of the $j^\text{th}$ tag on the $i^\text{th}$ execution)
    \begin{itemize}
        \item[] \textbf{Condition}: $\mverify\left( \pi_2\left( \mmin \right),  \pi_1\left( \mmin \right), \mk\left[ j \right]\right)$
        \item[] \textbf{Message}: $\mmok$
    \end{itemize}
    where $\mmin$ stands for $\minput\left( \mathsf{Rs}\left[i, j\right] \right)$.

    \underline{$\mathsf{Rf}\left[i\right]$}: (No authentication on the $i^\text{th}$ execution)
    \begin{itemize}
        \item[] \textbf{Condition}: $\bneg\bexists j.\mverify\left( \pi_2\left( \mmin \right),  \pi_1\left( \mmin \right), \mk\left[ j \right]\right)$
        \item[] \textbf{Message}: $\mmko$
    \end{itemize}
    where $\mmin$ stands for $\minput\left( \mathsf{Rf}\left[i\right] \right)$.
\end{mex}

Protocol steps are executed in an order captured by the relation $<$. In addition, we introduce the mutually exclusive relation $\diamond$ to relate steps, out of which at most one can be executed in the same protocol run.

\begin{mex}[Mutual Execution]\label{ex:basic-hash:ordering}
    In \autoref{basic-hash-example}, the reader can only execute one branch. Thus, in each session $i$, the authentication either succeeds or fails, and at most one tag can be authenticated, which is formalized as follows:
    \begin{equation}\label{eq:ex:basic-hash:ordering}
        \text{for all $i$, $j\neq k$, } \left( \mathsf{Rf}\left[i\right] \diamond \mathsf{Rs}\left[i, j\right] \right) \text{ and } \left( \mathsf{Rs}\left[i, j\right] \diamond \mathsf{Rs}\left[i, k\right] \right)
    \end{equation}
    There are no other ordering constraints for this protocol.
\end{mex}

\subsubsection{Protocol query} Finally, we need to provide a query, which can be any valid first-order formula. \tool tries to prove that this query holds for any possible execution of the protocol in any model that verifies the constraints defined in \autoref{par:overview:crypto-properties}.

\begin{mex}[Query]
    For \autoref{basic-hash-example}, we are concerned with non-injective agreement~\cite{loweHierarchyAuthenticationSpecifications1997}, which is formalized as:
    \begin{multline}\label{eq:basic-hash-query}
        \forall i,j. \mhappens\left( \mathsf{Rs}\left[i, j\right] \right) \wedge \meval{\mcond\left(  \mathsf{Rs}\left[i, j\right]\right)} \Rightarrow \\
            \exists k.
            \left( \begin{multlined}
                \mathsf{T}\left[k, j\right] < \mathsf{Rs}\left[i, j\right]\\
                \wedge  \meval{\pi_1\left( \minput\left( \mathsf{Rs}\left[i, j\right] \right) \right)} = \meval{\pi_1\left( \mmsg\left(  \mathsf{T}\left[k, j\right] \right) \right)}\\
                \wedge  \meval{\pi_2\left( \minput\left( \mathsf{Rs}\left[i, j\right] \right) \right)} = \meval{\pi_2\left( \mmsg\left(  \mathsf{T}\left[k, j\right] \right) \right)}
            \end{multlined} \right)
    \end{multline}
\end{mex}

Query~\eqref{eq:basic-hash-query} states that, if $\mathsf{Rs}\left[i, j\right]$ is selected for execution and its condition holds (i.e., authentication succeeds), then $\mathsf{T}\left[k, j\right]$ was executed before and had an output that matched $\mathsf{Rs}\left[i, j\right]$'s input (i.e., a matching authentication request was issued before).

\subsection{Automated Verification of  Protocol Queries}\label{sec:motivating:Verify}

While one might expect that a first-order protocol formalization directly yields  an automated verification procedure by encoding the model in a first-order theorem prover, such an encoding presents a number of technical challenges encompassing both definitions and resolution algorithms, as  we review in this section.


\subsubsection{Subterm reasoning}\label{sec:overview:mot-symbolic}
\autoref{prop:euf-cma} illustrates how reasoning works within the CCSA model.
We take a computational property (here $\meval{ \mverify\left(\sigma, m, \mk \right)}$) and turn it into a symbolic analysis. This symbolic analysis often relies on variations over a \emph{subterm relation} ($\sqsubseteq$ and $\sqsubseteq_{ \mverify\left(\_,\_,\bullet\right),\mhash\left(\_,\bullet\right)}$ in \autoref{eq:eufcma1}).

To automate such rules, our first-order formalization extends the first-order 
theory of finite term algebras~\cite{kovacsComingTermsQuantified2017} with dedicated subterm reasoning. As it is not finitely axiomatizable~\cite{venkataraman1987decidability},
we took inspiration from \cite{squirrel} for some overapproximations of $\sqsubseteq$ that we then axiomatized in FOL.
Moreover, we modified existing subterm reasoning in \mvampire~\cite{kovacsComingTermsQuantified2017} to allow for the flexibility required by our overapproximations, such as supporting not only $\sqsubseteq$ but also $\sqsubseteq_{ \mverify{\_}{\_}{\bullet},\mhash{\_}{\bullet}}$ (cf.  \autoref{sec:optimisations}).

Yet, the subterm relation used for \autoref{prop:euf-cma} interacts  poorly with the expected equational theory of $\meval{\_}=\meval{\_}$ (i.e., equality over evaluated terms, later denoted as $\beq$), as highlighted in~\cite{cremersSubtermbasedProofTechniques2022} and shown next.

\begin{mex}\label{ex:subterm-contradiction}
    Assume $\left<\_,\_\right>$, $\pi_1(\_)$ are respectively the tuple constructor and its first projection, and $m_1$ and $m_2$ are two \emph{distinct} constants such that $m_1 \not\sqsubseteq m_2$.
    We have $m_1 \sqsubseteq \pi_1\left(\left< m_2, m_1 \right>\right)$, but also $m_2\beq \pi_1\left(\left< m_2, m_1 \right>\right)$. Thus, reasoning modulo $\beq$ (i.e., equality over evaluated terms), yields $m_1 \sqsubseteq m_2$, a contradiction.
\end{mex}
To overcome the reasoning difficulties arising from combining subterm relations and the equality $\beq$, we split the reasoning on terms between
\begin{itemize}
    \item their \emph{symbolic forms} (described in \autoref{sec:symbolic}), on which we can  \emph{apply the subterm relation},
    \item and their \emph{evaluated form} (described in \autoref{sec:evaluated}), denoted by $\meval{\_}$, on which we \emph{reason modulo $\beq$ and do not apply subterm reasoning}.
\end{itemize}
As a result, the contradiction presented in \autoref{ex:subterm-contradiction} becomes impossible since the superposition step (substituting $a$ with $b$ when $a\beq b$) is no longer allowed.

\subsubsection{Soundness challenges}
It has long been known that perfect notions of security are non-realistic for real-world applications~\cite{shannonCommunicationTheorySecrecy1949}.
Therefore, cryptographic properties are often probabilistic. The \BCLogic is no exception, and its semantics are grounded in probability. Consequently, the \BCLogic lies beyond classical FOL~\cite{long_bana_towards_2012}. This is a critical blow, as most automated provers address classical FOL~\cite{vampire, z3, cvc5}.

In \autoref{sec:evaluated} we propose a classical first-order encoding that overapproximates the \BCLogic, then in \autoref{sec:soundness} we introduce a method to regain the probabilistic cryptographic semantics while retaining the ability to reason with FOL.

\subsubsection{Further optimizations}
We propose reasoning heuristics (\autoref{sec:preprocessing}) within \tool, which allow us to eliminate most of the symbolic reasoning, leaving only what can be encoded in standard FOL (i.e., modulo $\beq$) while remaining consistent before off-loading to the theorem prover.
For that, we look for terms that might be relevant concerning the cryptographic axioms and preprocess them.

\begin{mex}[Preprocessing]\label{ex:example-preprocessing}
In \autoref{basic-hash-example} we note that the term $ \mverify( \pi_2( \minput(\mathsf{Rs}\left[ i, j \right]) ),\allowbreak  \pi_1(  \minput(\mathsf{Rs}\left[ i, j \right])), \mk\left[ j \right])$
appears in the protocol specification (including the assertions and the query). Since this term has the form of the premise of axiom~\eqref{eq:eufcma1} of \autoref{prop:euf-cma}, we preprocess it in \tool with built-in decision procedures before passing it to the underlying first-order theorem prover.

The result of the preprocessing for \autoref{prop:euf-cma} is
\begin{multline}\label{eq:example-preprocessing}
    \forall i, j. \meval{\mverify\left(  \begin{multlined}
        \pi_2\left( \minput\left(\mathsf{Rs}\left[ i, j \right]\right) \right),\\
        \pi_1\left(  \minput\left(\mathsf{Rs}\left[ i, j \right]\right)\right)
    \end{multlined}, \mk\left[ j \right] \right)}\Rightarrow \\
    \exists i', j'. \left( \begin{multlined}
        \mathsf{T}\left[ i', j' \right] < \mathsf{Rs}\left[ i, j \right]
        \wedge j' = j\\
        \wedge \meval{\mathsf{n}\left[ i', j' \right]}  = \meval{\pi_1\left( \minput\left( \mathsf{Rs}\left[ i, j \right] \right) \right)}
    \end{multlined} \right)
\end{multline}

In~\eqref{eq:example-preprocessing}, $\mathsf{T}\left[ i', j' \right] < \mathsf{Rs}\left[ i, j \right]$ and $j' = j$ appear as a consequence of the symbolic analysis. The variable $u$ from~\eqref{eq:eufcma1} gets inlined.
\end{mex}

\begin{mrmk}[Eliminating subterm reasoning]
Note that the need for subterm reasoning is entirely eliminated from~\eqref{eq:example-preprocessing}.
When the preprocessing is sufficiently comprehensive to establish the proof of the query without the help of the general axioms, we can factor out a significant portion of the subterm reasoning to a point that we can again reason modulo $\beq$ while avoiding the unsoundness problems shown in \autoref{ex:subterm-contradiction}.
This heuristic, further described in \autoref{sec:final-optimization}, leads thus to a significant performance improvement, as shown in \autoref{sec:experiments}, at the cost of completeness.
\end{mrmk}

    \section{Preliminaries}
    \label{sec:preliminaries}

This section delves into \tool's reasoning capabilities, which are grounded in an extension of the \BCLogic (\BCname) closely aligned with the one adopted in \cite{squirrel, baelde_cracking_2022}. We call this extension the \enquote{\SymbLogic{}} (\Symbname).

\autoref{sec:contributions} later motivates and introduces our main contribution in the form of the \enquote{\EvalLogic{}} (\Evalname) that \tool uses to interact with \Symbname.
We show how we can leverage classical FOL methods to produce results in its specific semantics, and how these semantics closely match the \BCLogic's and \msquirrel's (resp. \autoref{prop:validity} and \autoref{th:sq-is-cv}).

\begin{mnotation*}
    We will use throughout this paper vector notations $\vec y$ as a shorthand for $y_1,\dots,y_n$ when $n$ is obvious from the context. We use $\marity{\_}$ to denote the size of a (finite) set or the length of a vector.

    We write $\mfv\left( \phi \right)$ for the free variables of $\phi$.

    We also write the \emph{inference rule} $\infer{A_1\ldots A_n}{B}$, with $n\geq 0$, to mean $A_1\wedge\ldots\wedge A_n\implies B$ is valid.
\end{mnotation*}

\subsection{The \BCLogic}\label{sec:bc-logic}
\tool supports protocols and queries expressed in an extension of the \BCLogic~\cite{long_bana_towards_2012}.
In this subsection, we give a quick introduction to the latter's syntax and semantics.
We assume the following sets:
\begin{enumerate}
    \item \mnoncesbc, a finite set of \emph{nonce} names
    \item \mfunctionsbc, a finite set of \emph{honest} functions
    \item \mattsbc, a finite set of \emph{attacker} functions 
\end{enumerate}

Terms in a \BCLogic \mbcl are terms built from the following grammar:
\begin{equation}\label{eq:bc-base-grammar}
    t \coloneqq x \gsep \n \gsep f\left( \vec{t} \right) \gsep g\left( \vec{t} \right)
\end{equation}
with $\n\in\mnoncesbc$, $f\in \mfunctionsbc$ and $g\in\mattsbc$; $x$ is a variable.

\begin{mrmk}[Connectives]\label{rmk:bc-connectives}
    Note \mbcl does not include the usual boolean connectives. To emulate them, we assume that \mfunctionsbc contains at least $\_\beq\_$, $\mifthenelse{\_}{\_}{\_}$, $\btop$ and $\bbot$.
    We can then build the rest of the boolean connectives on top of these functions.
    To distinguish them from regular first-order boolean operators, we will overline them (e.g., we write $\bwedge$ instead of $\wedge$).
\end{mrmk}

A computational model \compmodel is an assignment of the elements of \mnoncesbc, \mfunctionsbc, and \mattsbc to polynomial-time Turing machines, which determine an interpretation $\mevalbc{t}$ of each BC term $t$.
Specifically, this is defined as a polynomial-time Turing machine taking the security parameter $\eta$ in unary and a pair $\rho = \left( \rho_h, \rho_\AAA \right)$ of random tapes, one for honest agents and one for the adversary, respectively, constructed as follows:

\begin{enumerate}
    \item $\bcint{\n}{\eta}{\left( \rho_h, \rho_\AAA \right)}$, where $\n\in \mnoncesbc$, is a slice of length $\eta$ of $\rho_h$, pairwise distinct for all elements of $\mnoncesbc$.
    \item for $f\in \mfunctionsbc$ we have
    \begin{multline*}
        \bcint{f\left( t_1, \dots, t_n \right)}{\eta}{\rho}=\\\left\llbracket f \right\rrbracket \left( 1^\eta, \bcint{t_1}{\eta}{\rho}, \dots, \bcint{t_n}{\eta}{\rho} \right)
    \end{multline*}
    where $\left\llbracket f \right\rrbracket$ is a polynomial-time Turing machine with no direct access to $\rho$.
    \item for $g \in \mattsbc$ we have
    \begin{multline*}
        \bcint{g\left( t_1, \dots, t_n \right)}{\eta}{\rho}=\\\left\llbracket g \right\rrbracket \left( 1^\eta, \rho_\AAA, \bcint{t_1}{\eta}{\rho}, \dots, \bcint{t_n}{\eta}{\rho} \right)
    \end{multline*}
    where $\rho=\left( \rho_h, \rho_\AAA \right)$ and $\left\llbracket g \right\rrbracket$ is a polynomial-time Turing machine with direct access to $\rho_\AAA$ but not $\rho_h$. Thus terms of this form model attacker computations.
\end{enumerate}

\begin{mprop}[Subterm]\label{prop:subterm-main-prop}
    For all models $\mathbb{M}$ and all $\eta$, $t'$ appears in $t$ iff, for all $\rho$, the Turing machine $\mevalbc{t}$ applied to $\left(1^\eta, \rho\right)$ calls $\mevalbc{t'}$ on $\left(1^\eta, \rho\right)$.
\end{mprop}

The \BCname is generally used with a somewhat peculiar notion of satisfiability: a formula holds when it evaluates to~$1$ with overwhelming probabilities.
Formally:

\begin{mdef}[Cryptographic Satisfiability]\label{def:bc-model-validity}
    We say that \compmodel \emph{satisfies} a BC~formula $t$ and write $\compmodel \validp t$ when
    \begin{equation}
        \mprob\left( \bcint{t}{\eta}{\rho} \neq 1\right)=\mnegl\left( \eta \right)
    \end{equation}
    Where $\mnegl\left( \eta \right)$ is a function $h$ \emph{negligible} in $\eta$, that is $h\left( \eta \right)=o\left( \eta^{-c} \right)$ for all $c\in \mathbb{N}$.
\end{mdef}

We say that a formula $t$ is \emph{cryptographically valid} for a set $\modelclass_{\mathrm{bc}}$ of models and write $\modelclass_{\mathrm{bc}}\validp t$ when it holds for all computational models $\mathbb{M}$ of $\modelclass_{\mathrm{bc}}$.

The key idea of the CCSA is to reason within the biggest class $\modelclass_{\mathrm{bc}}$ of models that respect some cryptographic assumptions about \mfunctionsbc and \mattsbc (e.g., $\eufcma$) that we show are consistent with a computational attacker.


\subsection{Modeling Protocols -- The \SymbLogic}\label{sec:unbound-protocol}\label{sec:symbolic}

\begin{figure}
    \begin{align*}
        I &\coloneqq i \gsep \ib \\
        T &\coloneqq \tau \gsep \mpred{(T)} \gsep \stpa\left[ \vecI \right]\\
        A &\coloneqq \mhappens{\left( T \right)} \gsep T < T' \gsep T = T' \gsep I = I'\\
        t &\coloneqq x \gsep \n\left[ \vecI \right] \gsep f\left[ \vecI \right]\left( \vec{t} \right)  \gsep \cc\left[ \vecI \right]!\left( T \right) \gsep A \gsep \minput{\left( T \right)}  \\
            & \qquad\gsep
                \mfindst{\veci}{t}{t'}{t''}\\
            & \qquad\gsep
                \mfindst{\vect}{t}{t'}{t''}
    \end{align*}
    \caption{Grammar of \msymbl}\label{fig:symbolic-logic}
\end{figure}

Thus far, we have shown how to reason about symbolic computation. This is enough to describe a single protocol execution~\cite{long_bana_towards_2012}: at any point, we can model the current message being sent as a BC term where the input is the list of all the messages sent before (we call it the \emph{frame}) applied to an attacker function. However, our goal is to reason about all possible protocol executions. Hence, we illustrate now how to model protocols and the resulting extension of the base logic. We do so by closely following the formalism adopted in \msquirrel~\cite{squirrel}, which makes \tool interoperable with it (\autoref{th:sq-is-cv}).

We extend the syntax of \eqref{eq:bc-base-grammar} with the key ingredients to model protocols, as formalized in \autoref{fig:symbolic-logic}.

\paragraph{\textbf{Indices}} We use indices to extend our reasoning to unbounded numbers of objects. Formally, indices are members $\ib$ of a countable set \mindices.

\paragraph{\textbf{Steps}} Timepoints (or \emph{steps}) are ranged over by $T$ and represent atomic input/output operations in a protocol. We assume a finite set $\msteps$ of step names, containing at least the initialization step $\minit \in \msteps$. Steps are referenced by their name $\stpa\in \msteps$ and some indices, or relatively according to their order of execution, with $\mpred(T)$ denoting the step executed before $T$.

\paragraph{\textbf{Predicates over protocol executions}} The set of logical predicates characterizing a specific protocol execution, ranged over by $A$, includes  $\mhappens(T)$, which expresses whether or not the current execution includes $T$;  $T<T'$, which specifies if $T$ is executed before $T'$,  and $T=T$, which captures equality between steps.

\paragraph{\textbf{Input}} We abstract away the attacker functions with an $\minput(T)$ constructor. This represents which input the attacker gave to step $T$. It is an attacker function applied to the knowledge the attacker has gained until $T$'s execution.

\paragraph{\textbf{Memory cells}} $\cc\left[ \vecI \right]!(T)$ with $\cc\in\mcells$ represents what $T$ has stored into the memory $\cc\left[ \vecI \right]$. $\mcells$ is assumed to be finite.

\paragraph{\textbf{Lookups}} The $\mfindst{\_\allowbreak}{\_\allowbreak}{\_\allowbreak}{\_}$ are lookups constructions.
They incidentally also let us build quantifier-like objects.
\begin{align}
    \bexists \alpha.\ t&\coloneqq \begin{multlined}
        \mfindst{\alpha}{t\\}{\btop}{\bbot}
    \end{multlined}\label{eq:def-exitsts}\\
    \bforall\alpha.\ t&\coloneqq \bneg \bexists \alpha.\ \bneg t\label{eq:def-forall}
\end{align}

Finally, $i$, $\tau$ and $x$ are variables. The resulting logic is \msymbl.

\begin{mdef}[Step]\label{def:step}
    A \emph{step} $\stpa\left[ \veci \right]$ is an atomic input/output honest operation in a protocol. It is a triple $(c\left[ \veci \right], m\left[ \veci \right], u_{\stpa\left[ \veci \right]})$ composed of:
    \begin{enumerate}
        \item a term $c\left[ \veci \right]$ (referred to by $\mcond(\stpa\left[ \veci \right])$), describing the condition under which the step may be executed.
        \item a term $m\left[ \veci \right]$ (referred to by $\mmsg(\stpa\left[ \veci \right])$), describing the output of the step.
        \item a function $u_{\stpa\left[ \veci \right]}$ from $\left\{ \cc\left[\vecj\right]\middle| \cc\in\mcells \right\}$ to the terms such that $\mfv\left( u_{\stpa\left[ \veci \right]}\left(\cc\left[\vecj\right] \right)\right)\subseteq \veci\cup\vecj$. This function is independent from the indices, i.e., for all permutation $\sigma$, $u_{\stpa\left[ \sigma\left(\veci\right) \right]}\left(\cc\left[\sigma\left(\vecj\right)\right] \right)=\sigma\left(u_{\stpa\left[ \veci \right]}\left(\cc\left[\vecj\right] \right)\right)$.
    \end{enumerate}
    None of these elements may use ground indices.
\end{mdef}

Given a set of step names $\msteps$, we write $\msteps_\mindices$ for the set of instantiated steps: $\left\{ \stpa\left[ \vecib \right]\middle| \stpa\left[ \_ \right]\in\msteps, \vecib\in\mindices^{\marity{\vecib}} \right\}$. 
We are now ready to define protocols as follows.
\begin{mdef}[Protocol]\label{def:protocol}
    A \emph{protocol} $\mathcal{P}$ is a tuple $\left( \msteps, \prec \right)$ respectively composed of a set of \emph{steps} names and a preorder over $\msteps_\mindices$ such that:
    \begin{enumerate}
        \item $\minit$ is the smallest element of $\msteps_\mindices$ according to $\prec$.
        \item $\prec$ is independent from the indices. That is for all indices $\vecib$, $\vecjb$ and permutation $\sigma$ of \mindices, $\stpa\left[ \vecib \right]\prec \stpb\left[ \vecjb \right]$ iff. $\stpa\left[ \sigma\left( \vecib  \right) \right]\prec \stpb\left[ \sigma\left( \vecjb \right) \right]$.
        \item  Terms of the from $T$ in a step $\stpa\left[ \vecib \right]$ may only be of the form $\mpred^{n}(\stpb\left[ \vecjb \right])$ where $\stpb\left[ \vecjb \right] \preceq \stpa\left[ \vecib \right]$ and $n\geq 0$ (i.e., steps may only reference previous steps).
        \item\label{item:cell-termination} Memory cell assignments must terminate (i.e., no cyclic call graphs are allowed).
    \end{enumerate}
\end{mdef}

\begin{mex*}[\ref{ex:basic-hash:steps}]
    We recall here \autoref{ex:basic-hash:steps} which describes the steps of \autoref{basic-hash-example}:

    \underline{$\mathsf{T}\left[i, j\right]$}: ($i^\text{th}$ execution of the $j^\text{th}$ tag)
    \begin{itemize}
        \item[] \textbf{Condition}: $\btop$
        \item[] \textbf{Message}: $\left<\n\left[ i, j \right], \mhash\left( \n\left[ i, j \right], \mk\left[ j \right] \right) \right>$
    \end{itemize}

    \underline{$\mathsf{Rs}\left[i, j\right]$}: (Successful authentication of the $j^\text{th}$ tag on the $i^\text{th}$ execution)
    \begin{itemize}
        \item[] \textbf{Condition}: $\mverify\left( \pi_2\left( \mmin \right),  \pi_1\left( \mmin \right), \mk\left[ j \right]\right)$
        \item[] \textbf{Message}: $\mmok$
    \end{itemize}
    where $\mmin$ stands for $\minput\left( \mathsf{Rs}\left[i, j\right] \right)$.

    \underline{$\mathsf{Rf}\left[i\right]$}: (No authentication on the $i^\text{th}$ execution)
    \begin{itemize}
        \item[] \textbf{Condition}: $\bneg\bexists j.\mverify\left( \pi_2\left( \mmin \right),  \pi_1\left( \mmin \right), \mk\left[ j \right]\right)$
        \item[] \textbf{Message}: $\mmko$
    \end{itemize}
    where $\mmin$ stands for $\minput\left( \mathsf{Rf}\left[i\right] \right)$.
\end{mex*}

\begin{mex*}[\ref{ex:basic-hash:ordering}]
    We remind \autoref{ex:basic-hash:ordering} describing the ordering of \autoref{basic-hash-example} as mutual exclusion.
    \begin{equation*}
        \text{for all $i$, $j\neq k$, } \left( \mathsf{Rf}\left[i\right] \diamond \mathsf{Rs}\left[i, j\right] \right) \text{ and } \left( \mathsf{Rs}\left[i, j\right] \diamond \mathsf{Rs}\left[i, k\right] \right)
        \tag{\ref{eq:ex:basic-hash:ordering}}
    \end{equation*}
\end{mex*}

\begin{mrmk}{Mutual Exclusion}\label{rmk:mutual-exclusion}
    We model the mutual exclusion predicate $\stpa\left[ \vecib \right]\diamond\stpb\left[ \vecjb \right]$ from \autoref{ex:basic-hash:ordering} as $\stpa\left[ \vecib \right]\prec \stpb\left[ \vecjb \right]$ and $\stpb\left[ \vecjb \right] \prec \stpa\left[ \vecib \right]$ (remember, $\preceq$ is just a preorder, antisymmetry is not required).
\end{mrmk}

\begin{mex}
    \autoref{ex:basic-hash:steps} and \autoref{ex:basic-hash:ordering} (reminded above) fully describe \autoref{basic-hash-example}.
\end{mex}

Each execution of a protocol yields a trace. Intuitively, a trace is simply a chosen sequence of instantiated steps that does not contradict the protocol. Formally, we also consider a finite set of indices required to execute that trace:

\begin{mdef}[Trace]\label{def:trace}
    A \emph{trace} $\mathbb{T}$ of a protocol is a tuple $\left( \domI, \domE, \sigmI, \preceqE \right)$ such that
    \begin{enumerate}
        \item $\domI\subset \mindices$ is finite;
        \item $\domE\subseteq\domSE\coloneqq\left\{ \stpa\left[ \vecib \right] \middle| \stpa\in S, \vecib\in \domI^{\marity{\vecib}} \right\}$ contains $\minit$;
        \item $\preceqE$ is a total ordering over $\domE$ compatible with $\preceq$, i.e., if $\stpa\left[ \vecib \right] \preceq \stpb\left[ \vecjb \right]$ then $\stpa\left[ \vecib \right] \preceqE \stpb\left[ \vecjb \right]$;
        \item $\sigmI:\mindices \to \domI$ is the identity over $\domI$.
    \end{enumerate}
\end{mdef}

\begin{mrmk}
    Note that \autoref{rmk:mutual-exclusion} encodes the expected behavior of mutual exclusion: the compatibility of $\preceqE$ ensures that $\stpa\left[ \vecib \right]$ and $\stpb\left[ \vecjb \right]$ cannot be both in $\domE$.
    Furthermore, $\domE$ can be selected independently of the conditions of the steps. Instead, the failure of a step is modeled by returning $\mfail$ to all subsequently scheduled steps.
\end{mrmk}

\begin{mex}\label{ex:basic-hash:exec}
    Reusing \autoref{basic-hash-example}, let us assume we have two tags and two rounds. We can split the protocol into the steps $\minit$, $\mathsf{T}\left[ i, j \right]$, $\mathsf{Rs}\left[ i, j \right]$ and $\mathsf{Rf}\left[ i \right]$ like in \autoref{ex:basic-hash:steps}. Taking $\mindices=\mathbb{N}$, consider the following:
    \begin{gather}
        \mathcal{S}_1=\left\{ \minit <_1 \mathsf{T}\left[ 1, 2 \right] <_1 \mathsf{Rs}\left[ 1,2 \right] \right\}\label{eq:ex:execution:1}\\
        \mathcal{S}_2=\left\{ \minit <_2 \mathsf{T}\left[ 1, 1 \right] <_2 \mathsf{Rs}\left[ 1, 2 \right] <_2 \mathsf{T}\left[ 2, 2 \right] \right\}\label{eq:ex:execution:2}\\
        \mathcal{S}_3=\left\{ \minit <_3 \mathsf{Rf}\left[ 1 \right] <_3 \mathsf{T}\left[ 1, 1 \right] <_3 \mathsf{Rs}\left[ 1,1 \right] \right\}\label{eq:ex:execution:3}
    \end{gather}

    \eqref{eq:ex:execution:1} describes a valid trace with $\mathbb{I}=\left\{ 1, 2 \right\}$, and so is \eqref{eq:ex:execution:2} despite step $\mathsf{Rs}\left[ 1,2 \right]$ failing. However, \eqref{eq:ex:execution:3} is not a valid trace as it contradicts $\prec$ despite the condition of all the steps seemingly holding.
\end{mex}

\begin{mdefenv}{Unfolding}{unfolding}
    Given a trace $\mathbb{T}$, we can instantiate and unfold any term $t$ into a \BCLogic term. We note $\mevalT{t}$ the result of this transformation. $\mevalT{t}$ follows closely \cite{squirrel}'s corresponding transformation and we formalize it in our setting in \appref{sec:full-semantics}.
\end{mdefenv}
This unfolding turns any term into its BC equivalent for a given trace following the intuition given at the beginning of this section. For instance, $\minput\left( T \right)$ becomes $\matt$ applied to everything sent over the network before $T$.

\begin{mex}\label{ex:unfolding} Continuing \autoref{basic-hash-example}, let us unfold $\minput\left( \mathsf{Rs}\left[ 1,2 \right] \right)$ according to the trace $\mathbb{T}_{\eqref{eq:ex:execution:1}}$ described in \autoref{eq:ex:execution:1} from \autoref{ex:basic-hash:exec}:
    \begin{multline}\label{eq:ex:unfolding}
        \mevalT[\mathbb{T}_{\eqref{eq:ex:execution:1}}]{\minput\left( \mathsf{Rs}\left[ 1,2 \right] \right)} =\\
        \matt\left(\begin{multlined}[6.5cm]
            \ll \mevalT[\mathbb{T}_{\eqref{eq:ex:execution:1}}]{\mcond\left( \mathsf{T}\left[ 1, 2 \right] \right)} \bwedge \mevalT[\mathbb{T}_{\eqref{eq:ex:execution:1}}]{\mcond\left( \minit \right)};\\
                \ll \mifthenelse{\mevalT[\mathbb{T}_{\eqref{eq:ex:execution:1}}]{\mcond\left( \mathsf{T}\left[ 1, 2 \right] \right)} \bwedge \mevalT[\mathbb{T}_{\eqref{eq:ex:execution:1}}]{\mcond\left( \minit \right)}\\}{\mevalT[\mathbb{T}_{\eqref{eq:ex:execution:1}}]{\mmsg\left( \mathsf{T}\left[ 1, 2 \right] \right)}}{\mfail};\\\mevalT[\mathbb{T}_{\eqref{eq:ex:execution:1}}]{\mmsg\left( \minit \right)}
            \gg\gg
        \end{multlined}  \right)
    \end{multline}
\end{mex}
This minimal example already shows that the attacker has access to the messages of all the previous steps and their condition.
We also see that failing a condition blocks the execution of all subsequent steps as they return $\mfail$. Finally, \autoref{ex:unfolding} highlights the highly recursive nature of $\mevalT{\_}$.



    \section{\tool  Formalization: A 
    First-Order Theory of Protocol Queries}%
    \label{sec:contributions}
    We now describe how to encode the \SymbLogic from \autoref{sec:symbolic} in FOL in order to automate security proofs, and how to recover BC-like semantics from such a proof.
This is particularly challenging, since several axioms of the \SymbLogic, as well as the original \BCLogic, rely on beyond what can be finitely axiomatized within the logic itself  (\autoref{sec:encoding-challenges}).
Moreover, the semantics of the \BCLogic do not match that of classical FOLs, as assumed by first-order theorem provers (\autoref{sec:soundness-challenges}).
In order to avoid higher-order and non-classical reasoning, which would degrade performance, we introduce a tailored encoding in FOL (Sections~\ref{sec:quantifed-terms} and~\ref{sec:subterm}) as well as properties (\autoref{sec:soundness}) to ensure the soundness of the result.

\subsection{The Challenges of Protocol Queries}\label{sec:form:challenges}
\subsubsection{Encoding challenges}\label{sec:encoding-challenges}

Many cryptographic properties are defined in a way that is not finitely axiomatizable in the \BCLogic, nor in our current \SymbLogic.
This motivates the need for an \EvalLogic.
In this subsection, we will look at the challenges that this new logic is meant to overcome.
We illustrate them with the \eufcma property (cf. \autoref{prop:euf-cma}) that we state in its pure \BCLogic form in \autoref{ex:challenge}.

\begin{mprop}[\eufcma]\label{ex:challenge}
    Let $\modelclass{}_{\eufcma}$ be a class of models where $\mhash\left( \_, \_ \right)$ and $\mverify\left( \_, \_, \_ \right)$ form a \eufcma-secure \mac scheme. In the \BCLogic (\autoref{sec:bc-logic}), we have for every signature $\sigma$, message $m$ and key $\mk$:
    \begin{equation}\label{eq:euf-cma-bc}
        \modelclass{}_{\eufcma}\validbc \mverify\left(\sigma, m, \mk \right) \bimplies
            \bar{\bigvee}_{\mhash\left( u, \mk \right)\in \mstbc\left( m, \sigma \right) } u \beq m
    \end{equation}
    where $\mk$ only appears in the position of the key in $m$ and $\sigma$. $\mstbc(\vec t)$ is the set of subterms appearing in $\vec t$.

\end{mprop}

\autoref{ex:challenge} highlights two challenges:

\begin{itemize}
    \item \autoref{eq:euf-cma-bc} is an axiom scheme. We cannot express it within the \SymbLogic with finitely many axioms.~\ref{sec:quantification-challenges}
    \item \autoref{ex:challenge} also makes heavy use of subterm reasoning. Not only is this not finitely axiomatizable~\cite{kovacsComingTermsQuantified2017}, even an incomplete finite axiomatization is beyond the \SymbLogic due to unfortunate interactions with the equational theory.~\ref{sec:subterm-challenges}
\end{itemize}

\begin{enumerate}[
    label=\textbf{(\Alph{enumi})},
    ref=(\Alph{enumi}),
    wide 
]
\item\textbf{Quantification over arbitrary terms}.\label{sec:quantification-challenges}
\autoref{eq:euf-cma-bc} represents a set of axioms ranging over every term $m$, $\sigma$, and nonces $\mk$. Unfortunately, there are infinitely many terms. This can be solved with quantification: universal quantification over $m$, $\sigma$, and $\mk$ and existential quantification over $u$ to avoid having a disjunction that depends on $m$, $\sigma$, and $\mk$.

Yet, quantification over arbitrary terms is beyond the \SymbLogic. Indeed, we can quantify over indices and timepoints only because we can unfold quantifiers into finitely nested $\mifthenelse{\_\allowbreak}{\_\allowbreak}{\_\allowbreak}$ statements for any given trace (see \appref{sec:full-semantics}).
Using the same trick with arbitrary terms would yield infinitely large terms.
Even somehow expanding the \SymbLogic to include such terms would contradict their interpretations into the \BCLogic. Indeed, the interpretation $\mevalbc{t}$ of a term $t$ \emph{must} be a \emph{polynomial} time Turing machine.

We devise the \EvalLogic to release ourselves from this last constraint. Thus, we include terms whose interpretation is beyond the \BCLogic. We then show that this new interpretation not only matches BC's on common terms (\autoref{prop:validity}) but expands it to any other terms (\autoref{th:main}).

\item\textbf{Subterm Relation $\sqsubseteq$ and Equalities $\beq$}.\label{sec:subterm-challenges}\label{sec:subterm-and-equalities}
In the \BCLogic, honest and malicious computations are first described symbolically. This symbolic description informs on whether terms were directly computed or not. For instance, if a BC~term $t'$ does \emph{not} appear in another BC~term $t$ (we write $t'\not\sqsubseteq t$), then we know that the Turing machine $\mevalbc{t}$ does \emph{not} call $\mevalbc{t'}$, as stated in \autoref{prop:subterm-main-prop}. This allows us to closely capture the semantics of cryptographic games, and is already exposed in \autoref{ex:challenge}.

\begin{mex}[\eufcma]\label{eufcma:subterm}
    Within the \eufcma axiom of \autoref{ex:challenge}, the subterm property side condition on $\mathsf{k}$ guards against key misuses: $\mathsf{k}$ can only be used to hash and verify.
    Furthermore, the subterm property $\mhash\left( u, \mathsf{k} \right)\in \mstbc\left( m, \sigma \right)$ guards against illegal uses of the signing oracle in the \eufcma cryptographic game: the attacker may ask to sign anything with $\mathsf{k}$ except for any term that evaluates as $m$.
\end{mex}

Let us look in more detail at the challenges associated with this subterm relation. Remember that in \autoref{ex:subterm-contradiction} we saw that the interplay between equality and subterm reasoning can easily lead to unsoundness. \autoref{ex:subterm-unsound} shows how a naive encoding is unsound:

\begin{mex}\label{ex:subterm-unsound}
   Consider now the subterm axiom
    \begin{equation}\label{ex:unsound}
        x\sqsubset f(t_1,\dots,t_n)\Leftrightarrow \bigvee\nolimits_{i=1}^n x\sqsubseteq t_i.
    \end{equation}
    \autoref{ex:unsound} is \emph{unsound} modulo $\beq$ when tuples, projections, and empty are defined. It is indeed enough to implement \autoref{ex:subterm-contradiction} and thus show $x\sqsubset \varnothing$. However, using the $\Rightarrow$ implication of~\eqref{ex:subterm-contradiction}, instantiated with $\varnothing$, we also derive $x\not\sqsubset \varnothing$, yielding thus a source of unsoundness.
\end{mex}

Unlike~\cite{cremersSubtermbasedProofTechniques2022}, we cannot circumvent the issue by relying on the specifics of $\beq$ as it is a general equivalence relation, nor can we exploit specific solver decision procedures as we aim for a sound first-order encoding.
Instead, we go around the issue in our \EvalLogic by sandboxing the reasoning modulo $\beq$ with the evaluation predicate $\meval{\_}$.
We keep $=$ as the symbolic equality because we have $(x = y) \Rightarrow \meval{x \beq y}$. Since $\meval{\_ \beq \_}$ is an equivalence relation, we can let theorem provers use their powerful equality reasoning with $\beq$ by introducing the sugar:
\begin{equation}\label{eq:subterm:eq}
    \meval{t} = \meval{t'} \coloneqq \meval{t \beq t'}
\end{equation}

We recover some of the overhead introduced by $\meval{\_}$ by axiomatizing the commutativity of $\meval{\_}$ with boolean connectives (\autoref{prop:conn-commutes}). The concrete axioms are listed in \autoref{fig:sound-b-axioms} of \appref{sec:base-axioms}.
\end{enumerate}

All in all, we address the above challenges by embedding the \SymbLogic of \autoref{sec:unbound-protocol} into an \emph{\EvalLogic} where we can safely quantify over arbitrary terms and reason symbolically about subterms. \autoref{sec:evaluated} presents the syntax and semantics of this new logic, while \autoref{sec:subterm} dives deeper into our modeling of subterm analysis.

\subsubsection{Soundness challenges}\label{sec:soundness-challenges}

As already noted in~\cite{long_bana_towards_2012}, the \BCLogic's semantics does not match that of classical FOL. Indeed, the notion of satisfiability introduced in \autoref{def:bc-model-validity} quickly gives us the intuition that a term can be true, false, or something in between.

\begin{mex}[Counterexample to classical semantics]\label{ex:bc-is-not-classical}
    Let \compmodel be a computational model that interprets the function $1^\mathrm{st}\ \mathrm{bit}\left( x \right)$ as the first bit of $x$.
    Let $\n$ be a nonce; we get that $\compmodel\validp \left( 1^\mathrm{st}\ \mathrm{bit}\left( \n \right) \right)\bvee \bneg \left( 1^\mathrm{st}\ \mathrm{bit}\left( \n \right) \right)$ as it is always true.
    However, we have neither $\compmodel \validp \left( 1^\mathrm{st}\ \mathrm{bit}\left( \n \right) \right)$ nor $\compmodel \validp \bneg \left( 1^\mathrm{st}\ \mathrm{bit}\left( \n \right) \right)$ as they both alternate between $0$ and $1$ with probability $\frac{1}{2}$.
\end{mex}

The upcoming \EvalLogic is designed to fix the issues introduced in Examples~\ref{ex:subterm-unsound} and~\ref{ex:bc-is-not-classical}.

\subsection{\EvalLogic}\label{sec:evaluated}\label{sec:quantifed-terms}
\begin{figure}
    \begin{align*}
        t &\coloneqq \dots\\
        S &\coloneqq t = t' \gsep t \sqsubseteq t' \gsep \dots\\
        \varphi &\coloneqq \top \gsep \bot \gsep S \gsep \meval{t}  \gsep \varphi \vee \varphi' \gsep \neg\varphi \\
        &\qquad  \gsep \exists \vec\imath, \vec\tau, \vec x. \varphi \gsep \forall \vec\imath, \vec\tau, \vec x. \varphi
    \end{align*}
    where $t$ is a term from \msymbl as defined in \autoref{fig:symbolic-logic}. We extend with $\wedge$, $\Rightarrow$ and $\Leftrightarrow$ as expected, and write $\meval{t}=\meval{t'}$ as sugar for $\meval{t \beq t'}$.
    \caption{The \EvalLogic \mevall}\label{fig:evaluated-logic}
\end{figure}
Let us call $\Omega$ the set of random tape pairs. $\mprob\left( \_ \right)$ is a probability measure over $\Omega$.

\subsubsection{Syntax}
The \EvalLogic, denoted as \mevall, extends the \SymbLogic \msymbl from \autoref{sec:unbound-protocol} with predicates over terms. In a nutshell, the \EvalLogic is a standard FOL whose literals are predicates over terms from the \SymbLogic, or evaluation of those terms, denoted by the predicate $\meval{\_}$. The evaluation of a term forces its computational interpretation.

The syntax is described in \autoref{fig:evaluated-logic}:  $t\sqsubseteq t'$ expresses that $t$ is a subterm of $t'$, $\meval{t}$ captures the concrete evaluation of the symbolic term $t$ (e.g., symbolic cryptographic functions are evaluated into concrete bitstrings); $\varphi$ ranges over first-order formulas with existential and universal quantification over terms.

\subsubsection{Semantics}

\begin{figure}
    \input{101-figures/009-eval-model.tex}
    \caption{Cryptographic Model}\label{fig:cryptographic-model}
\end{figure}

\begin{mdef}[Cryptographic Model]\label{def:cryptographic-model}
    A \emph{cryptographic model} $\modelc$ is a tuple $\left( \mathbb{M}, \mathbb{T}, \mathcal{M} \right)$ where $\mathbb{M}$ is a computational model, $\mathbb{T}$ a trace, and $\mathcal{M}$ a symbolic model.

    $\modelc$ associates terms of \mevall to functions of $\mathbb{N}\times\Omega\to\left\{ 0, 1 \right\}$ according to \autoref{fig:cryptographic-model}.
    We write $\modelc\left( \varphi\right)$ the interpretation of $\varphi$ by $\modelc$ and $\modelc^\eta\left( \varphi \right)$ to express the random variable $[\rho\in\Omega \mapsto\modelc^\eta\left( \varphi \right)\left( \rho \right)]$.
\end{mdef}

\begin{mrmk}\label{rmk:no-proba-vs-sq}
    As compared to the semantics adopted in Squirrel~\cite{baelde_cracking_2022}, we avoid the probabilistic interpretation of symbolic terms. That is, the interpretation of $\meval{t}$ is the underlying Turing machine \emph{applied} to some $\eta$ and $\rho$, and not $\validp \left( t \right)^\mathbb{T}$ as seen in previous works \cite{baelde_cracking_2022}.
\end{mrmk}

\autoref{rmk:no-proba-vs-sq} solves \autoref{ex:bc-is-not-classical} sufficiently for our trace properties and allows us to move boolean connectives between logics, 
greatly improving the performance of theorem provers (\autoref{prop:conn-commutes}).
Achieving such a functionality for indistinguishability properties is more challenging~\cite{BaeldeKoutsosLallemand2023} and would require dedicated interpretations of the evaluation predicate.

\begin{mprop}\label{prop:conn-commutes}
    $\meval{\_}$ commutes with boolean connectives and quantifiers\label{foot:bool-connective-quant}.
\end{mprop}
\begin{mproof}
    Via \autoref{prop:base-axioms} from \appref{sec:base-axioms}.
\end{mproof}

\subsubsection{Relation to the \BCLogic and \msquirrel}\label{sec:relation-to-bc}

We now intuitively explain how we relate this semantic interpretation to the one of the \BCLogic (\autoref{def:bc-model-validity}).

\begin{mdef}[Cryptographic Satisfiability]\label{def:cryptographic-satisfiability}
    We adapt \autoref{def:bc-model-validity} to the new setting. Thus $\modelc$ \emph{cryptographically satisfies} $\varphi$ when
    \begin{equation}\label{eq:validity}
        \mprob\left(\modelc^\eta\left( \varphi \right)\neq 1 \right)=\mnegl\left( \eta \right)
    \end{equation}
    and we reuse the notation $\modelc\validp \varphi$ and also extend it to class $\modelclass$ of cryptographic models like so: $\modelclass\validp\varphi$.
\end{mdef}

We connect this notion of validity to the one of the \BCLogic (\autoref{def:bc-model-validity}) as described in \autoref{prop:validity}.

\begin{mprop}[Extension of BC]\label{prop:validity}
    Let $t$ be a symbolic term. Its evaluation in \mevall is \emph{valid} iff any unfolding in the \BCLogic is valid. Formally: 
    \begin{equation}
        \modelclass\valid \meval{t}\text{ iff for all trace $\mathbb{T}$ we have }\modelclass_\mathbb{M} \validbc \mevalT{t}
    \end{equation}
    where $\modelclass_\mathbb{M}$ are the computational models of $\modelclass$.
    
\end{mprop}

We can in fact find a transformation $\mmmap{\_}$ from a relevant subset of \msquirrel's term to \tool's such that their interpretation coincide.

Intuitively, $\mmmap{\_}$ commutes with every common symbol (e.g., functions, nonces, quantifiers,\dots). 
\msquirrel's macros are either unfolded (e.g., $\mathsf{cond}\mat T$ is unfolded into the corresponding condition) or replaced by the corresponding \tool term (e.g., $\mathsf{input}\mat T$ is replaced by $\minput\left( \mmmap{T} \right)$). The remaining \msquirrel terms have a direct counterpart in \tool. The only exception is the 
$\mathsf{frame}\mat T$ macro due to its recursive nature: this is, however,  typically not used directly and is instead only used to define  $\mathsf{input}\mat T$, which our transformation supports. While  our transformation does not aim at completeness, all examples from \autoref{sec:experiments} are supported.

\begin{mrmk}[Quantifiers]
    The translation only preserves the interpretation, not the unfolding. This especially matters with quantifiers: they unfold differently, but they evaluate to logically equivalent disjunctions/conjunctions in both tools (see \autoref{lemme:quantifiers} of \appref{sec:soundness-th} and \autoref{prop:conn-commutes}).
\end{mrmk}

\begin{mth}[Interoperability]\label{th:sq-is-cv}
    For all \msquirrel protocol $\mathcal{P}_{\mathrm{sq}}$, we can find a \tool protocol $\mathcal{P}$, such that for all \msquirrel trace $\mathbb{T}_{\mathrm{sq}}$ over $\mathcal{P}_{\mathrm{sq}}$, we can find a \tool trace over $\mathcal{P}$, such that for all computational models $\mathbb{M}$, security parameter $\eta$, pairs of random tapes $\rho$, and \msquirrel $t$ over which $\mmmap{\_}$ is defined, we have
    \begin{equation}\label{eq:sq-cv-th}
        \modelc^\eta\left(\meval{\mmmap{t}} \right)\left( \rho \right)=\mevalsq[\mathbb{T}_{\mathrm{sq}}]{t}\left( 1^\eta, \rho \right)
    \end{equation}
    where $\mevalTsq[\mathbb{T}_{\mathrm{sq}}]{\_}$ is the \msquirrel unfolding from~\cite{squirrel,baelde_cracking_2022}.
\end{mth}

For more details, we refer to 
    \ifreport \appref{sec:comp-w-squirrel}. \else the technical report~\cite{CRYPTOVAMPIREAutomatedReasoningTR}. \fi

\subsection{Linking Cryptographic Semantics and Classical First-Order Logic}\label{sec:soundness}
In this section, we first show that the \EvalLogic is not enough to faithfully encode the behavior of cryptographic protocols (\autoref{sec:crypto-axioms-pbls}), and we propose a method to sidestep the problem while retaining both the automation potential offered by classical FOL and the cryptographic semantics of \autoref{def:cryptographic-satisfiability} (\autoref{sec:main-th}).

\subsubsection{The Problem with cryptography-related axioms}\label{sec:crypto-axioms-pbls}

We first show that \autoref{def:cryptographic-satisfiability} does not satisfy many critical axioms while, at the same time, allowing for some cryptographically unsound formulas. Consider \autoref{th:no-guessing} (adjusted from~\cite{long_bana_towards_2012,bana_computationally_2014}, proven in \ifreport \appref{sec:crypto-axioms} \else \cite{CRYPTOVAMPIREAutomatedReasoningTR} \fi):

\begin{mth}[No Guessing]\label{th:no-guessing}
    It is not possible to guess honest nonces. Formally, for all nonce $\n\left[ \_ \right]\in \mathcal{N}$, ground indices $\vecib\in \mindices$, and message $m$, we have
    \begin{equation}\label{eq:no-guessing}
        \meval{\n\left[ \vecib \right]} = \meval{m} \Rightarrow \n\left[ \vecib \right]\sqsubseteq m
    \end{equation}
\end{mth}
\autoref{eq:no-guessing} describes that any message $m$ that evaluates to a nonce $\n\left[ \vecib \right]$ must \emph{symbolically} contain that nonce. This theorem is valid for all cryptographic models and is fundamental to many proofs as it encodes how randomness behaves in the logic. However, \eqref{eq:no-guessing} is still an axiom scheme. Unfortunately, turning it into a formula breaks down as shown in \autoref{ex:no-guessing-rejected}.

\begin{mex}[The \EvalLogic rejects \autoref{th:no-guessing}]\label{ex:no-guessing-rejected}
    Any cryptographic model that can express the natural numbers cannot satisfy \eqref{eq:ex:no-guessing-rejected}. (notice the $\forall$)
    \begin{equation}\label{eq:ex:no-guessing-rejected}
        \forall \veci, x.\ \meval{\n\left[ \veci \right]} = \meval{x} \Rightarrow \n\left[ \veci \right]\sqsubseteq x
    \end{equation}
\end{mex}
This is a consequence of the even worse \autoref{ex:no-no-guessing}.
\begin{mex}[The \EvalLogic allows for unsound formulas]\label{ex:no-no-guessing}
    Assuming \mfunctions contains $0$ and $S\left( \_ \right)$, any cryptographic model that interprets $0$ and $S\left( \_ \right)$ as respectively $0$ and the successor function satisfies \eqref{eq:ex:no-no-guessing}.
    \begin{equation}\label{eq:ex:no-no-guessing}
        \exists x.\ \meval{\n\left[ \vecib \right]} = \meval{x} \wedge \n\left[ \vecib \right]\not\sqsubseteq x
    \end{equation}
\end{mex}
\begin{mproof}
    Let $\modelc$ be such model. $\modelc^\eta\left( \n\left[ \vecib \right] \right)\left( \rho \right)$ is a natural number $N$. We then define $\underline{N}_S\coloneqq S^{N}\left(0\right)$, where $S^k$ represents $k$ compositions of $S$. We get that $\modelc^\eta\left( \n\left[ \vecib \right] \right)\left( \rho \right) = \modelc^\eta\left( \underline{N}_S\right)\left( \rho \right)$ and yet $\n\left[ \vecib \right]\not\sqsubseteq \underline{N}_S$. As this holds for any $\eta$ and $\rho$, we get \eqref{eq:ex:no-no-guessing}.
\end{mproof}

The $\underline{N}_S$ that appears in the proof of \autoref{ex:no-no-guessing} is a side effect of \autoref{rmk:no-proba-vs-sq}. We can observe this side effect even more clearly when putting \eqref{eq:ex:no-no-guessing} in Skolem normal form~\cite{skolem}:
\begin{equation}
    \meval{\n\left[ \vecib \right]} = \meval{\msk[\eqref{eq:ex:no-no-guessing}]} \wedge \n\left[ \vecib \right]\not\sqsubseteq \msk[\eqref{eq:ex:no-no-guessing}]
\end{equation}
The resulting Skolem $\msk[\eqref{eq:ex:no-no-guessing}]$ is beyond our control. For any $\eta$ and $\rho$, $\msk[\eqref{eq:ex:no-no-guessing}]$ may take the value of any constant that just so happens to evaluate like $\n\left[ \vecib \right]$ for this \emph{specific} $\eta$ and $\rho$. This makes it a \emph{non-polynomial} function that guesses $\n\left[ \vecib \right]$.

In the next section, we show how to reject formulas that may produce such unwanted behavior while allowing for well-behaved ones like \autoref{eq:ex:no-guessing-rejected}.

\subsubsection{Linking First-order and Cryptography Together}\label{sec:main-th}

\autoref{ex:no-no-guessing} is a result of $\msk[\eqref{eq:ex:no-no-guessing}]$ being able to take too many values.
Thus, we say that a formula has a \hyperref[def:formula-bounded-snf]{\emph{bounded Skolem normal form}} when we can find \emph{finite} sets of terms such that we can safely assume that the Skolems will only take values in those sets.
We formalize this notion in \autoref{def:formula-bounded-snf} of \appref{sec:soundness-th}.
In practice, we do not need to go back to the definition, and we use the following properties instead, yielding an almost syntactic description of the formulas verifying the property.

The base case is handled by \autoref{prop:bsnf-in-base-logic}, meaning that any quantifier-free formula, or with simple enough quantification, has a bounded Skolem normal form.

\begin{mprop}\label{prop:bsnf-in-base-logic}
    A formula whose Skolem normal form is already in \mevall (i.e., no new symbols are required) has a bounded Skolem normal form.
\end{mprop}

Then \autoref{prop:bsnf-and-or} and \autoref{prop:existential-quantifiers-and-acceptability} let us build new formulas with bounded Skolem normal forms from existing ones.
\autoref{def:bounding-formula} describes formulas that ensure terms cannot take too many values; consequently, they can prevent Skolem functions from displaying the behavior described in \autoref{sec:crypto-axioms-pbls}. They, therefore, allow existential quantification.
Notably, subterm relations produce such formulas.
\begin{mprop}\label{prop:bsnf-and-or}
    The notion of bounded Skolem normal form is stable by conjunction and disjunction.
\end{mprop}

\begin{mdefenv}{Bounding Formula}{bounding-formula}
    We say a formula $\varphi$ with $\mfv\left(\varphi\right)=x\uplus\vecu$ is \emph{bounding} on $x$ if for all $\vecvb$ we can find a set $D_{\varphi}^{\vecvb}$ such that for all $\eta$, $\rho$, $t$, and cryptographic models $\modelc$, we have 
    \begin{equation}
        \modelc^\eta\left( 
            \varphi\begin{Bmatrix} 
                x \mapsto t\\
                \vecu \mapsto \vecvb 
            \end{Bmatrix} 
            \implies \bigvee\nolimits_{\vspace*{-2pt}\substack{\phi\in D_{\varphi}^{\vecvb}\\\vecalpha = \mfv\left(\phi\right)}} \exists \vecalpha.\ \phi = t \right)\left( \rho \right)=1
    \end{equation}
\end{mdefenv}
\begin{mprop}\label{prop:existential-quantifiers-and-acceptability}
    If $\varphi$ is bounding on $x$ and has a \hyperref[def:formula-bounded-snf]{bounded Skolem normal form}, then $\exists x.\ \varphi$ has a bounded Skolem normal form.
\end{mprop}

Finally, formulas with bounded Skolem normal forms enjoy a property close to the modus Polens:
\begin{mth}[Cryptographic Validity]\label{th:main}
    Let $\varphi$ be a formula s.t. $\mfv\left(\varphi\right)=\vecu$, $t$ a term, and $\modelclass$ a set of cryptographic models such that
    \begin{enumerate}[ref=(\arabic{enumi})]
        \item\label{item:th:main:1} $\varphi$ has a bounded Skolem normal form;
        \item\label{item:th:main:2} for all $\vecvb$, $\modelclass\validp \msubstit{\varphi}{\vecu}{\vecvb}$;
        \item\label{item:th:main:3} $\left( \forall\vecu.\ \varphi \right)\implies \meval{t}$ is valid in \emph{FOL};
    \end{enumerate}
    then $\modelclass\validp \meval{t}$.
\end{mth}
\autoref{th:main} links the cryptographic semantics of axioms~\ref{item:th:main:2} with first-order reasoning~\ref{item:th:main:3} and extracts a cryptographic property so long as these axioms are well-behaved~\ref{item:th:main:1}, i.e., they have bounded Skolem normal forms.

We notice now that property~\eqref{eq:ex:no-guessing-rejected} of \autoref{ex:no-guessing-rejected}  verifies items \ref{item:th:main:1} and~\ref{item:th:main:2}. It is therefore usable in first-order proofs (and so is \autoref{th:no-guessing}).
Similarly, we can turn the axiom schema~\eqref{eq:eufcma1} of \autoref{prop:euf-cma} to a formula that verifies \ref{item:th:main:1} and~\ref{item:th:main:2}.
We prove the properties introduced in this section in \appref{sec:soundness-th}.

    \section{\tool Reasoning for Proving Protocol Queries} 
    \label{sec:optimisations}
    Based on the first-order formalization from \autoref{sec:contributions}, we now describe how to extend saturation-based first-order theorem proving~\cite{vampire} in order to achieve automation in a first-order setting.
The main problem to be solved is how to efficiently encode subterm reasoning in FOL.
\autoref{sec:subterm} reviews the higher-order definition of subterm, which is similar to the one adopted in Squirrel~\cite{squirrel}.
\autoref{sec:modifying-vampire} presents an efficient solving procedure grounded in first-order saturation that we embed in the \mvampire theorem prover.
\autoref{sec:preprocessing} introduces preprocessing and heuristic techniques that further improve performance and that we integrate in \tool.

\subsection{Subterm Relations in \tool}\label{sec:subterm}
As explained in the sections above, the logics presented in this paper ground their semantics in the \BCLogic, including subterm relations. In the \BCLogic, the notion is fairly straightforward: for instance, the strict subterms of $\left<x, y\right>$ are $x$ and $y$ (assuming $x$ and $y$ are constant symbols). However, this simplicity is lost after the unfolding $\mevalT{\_}$ of \autoref{def:unfolding}. We explore now the problems that arise and our solutions.

\begin{mex*}[\ref{ex:unfolding}]
    Recall \autoref{def:unfolding} and \autoref{ex:unfolding}: let us unfold $\minput\left( \mathsf{Rs}\left[ 1,2 \right] \right)$ according to the trace $\mathbb{T}_{\eqref{eq:ex:execution:1}}$ described in \autoref{eq:ex:execution:1} from \autoref{ex:basic-hash:exec}:
    \begin{multline*}
        \mevalT[\mathbb{T}_{\eqref{eq:ex:execution:1}}]{\minput\left( \mathsf{Rs}\left[ 1,2 \right] \right)} =\\
        \matt\left(\begin{multlined}[6.5cm]
            \ll \mevalT[\mathbb{T}_{\eqref{eq:ex:execution:1}}]{\mcond\left( \mathsf{T}\left[ 1, 2 \right] \right)} \bwedge \mevalT[\mathbb{T}_{\eqref{eq:ex:execution:1}}]{\mcond\left( \minit \right)};\\
                \ll \mifthenelse{\mevalT[\mathbb{T}_{\eqref{eq:ex:execution:1}}]{\mcond\left( \mathsf{T}\left[ 1, 2 \right] \right)} \bwedge \mevalT[\mathbb{T}_{\eqref{eq:ex:execution:1}}]{\mcond\left( \minit \right)}\\}{\mevalT[\mathbb{T}_{\eqref{eq:ex:execution:1}}]{\mmsg\left( \mathsf{T}\left[ 1, 2 \right] \right)}}{\mfail};\\\mevalT[\mathbb{T}_{\eqref{eq:ex:execution:1}}]{\mmsg\left( \minit \right)}
            \gg\gg
        \end{multlined}  \right)
        \tag{\ref{eq:ex:unfolding}}
    \end{multline*}
\end{mex*}

A naive approach to implementing the BC subterm in \tool would be to axiomatize the unfolding of $\mevalT{\_}$. However, \autoref{ex:unfolding} highlights multiple problems:
\begin{enumerate*}[label=\textbf{(\roman{enumi})}, ref=(\roman{enumi})]
    \item\label{item:sec:infinte-recursion:1} the unfolding is dependent on the trace that we would like to abstract away, and \item\label{item:sec:infinte-recursion:2} it is highly recursive.
\end{enumerate*}
Point~\ref{item:sec:infinte-recursion:2}, in particular, would require some inductive reasoning, but induction is not a first-order property and is extremely challenging to algorithmically address~\cite{bundyChapter13Automation2001}.

Instead, reusing insights from~\cite{squirrel}, we encode this inductive reasoning with subterms directly into our description of a subterm relation. That is, we design a $\sqsubseteq$ relation that overapproximates the BC notion of subterm. This is achieved through a set $\mmst\left(\_\right)$, which can be computed by \tool with the help of a theorem prover to solve problems \ref{item:sec:infinte-recursion:1}--\ref{item:sec:infinte-recursion:2}.

Concretely, $\mmst\left(t\right)$ lists all the terms $t'$ for which we can find a trace $\mathbb{T}$ such that $t'$ unfolds into a BC~subterm of the unfolding of $t$. To keep the number of plausible $t'$ in check, we add a guard on the trace that unfolds to $\bbot$ in traces where $t'$ is definitely not a subterm of $t$. Overall, we then define $\sqsubseteq$ as the following

\begin{mdef}[\tool's base subterm]\label{def:subterm}
    We define the \emph{base subterm} relation $\sqsubseteq$ on any term $t$ as such:
    \begin{equation}\label{eq:def:subterm}
        t \sqsubseteq t'\text{ iff } \bigvee\nolimits_{\substack{\left( c, u \right)\in \mmst\left( t' \right)\\ \text{with }\veci=\mfv(u)\cup\mfv(v)}}\exists \veci. \meval{c} \wedge u = t
    \end{equation}
    where $\mmst\left( \_ \right)$ is fully described in \appref{sec:subterm-appendix}.
\end{mdef}

We prove in \appref{sec:subterm-appendix} \ifreport\else of~\cite{CRYPTOVAMPIREAutomatedReasoningTR} \fi how $\sqsubseteq$ is an overapproximation of the BC notion of subterm. We exemplify how $\mmst\left(\_\right)$ is built in \autoref{ex:subterm-set}.

\begin{mex}[Subterm set]\label{ex:subterm-set}
    Let $ST\left(\_\right)$ be the syntactic subterms, that is, for instance:
    \begin{multline}
        ST\left( \mverify\left( \pi_2\left( \mmin \right), \pi_1\left( \mmin \right), \mk\left[ j \right]\right) \right)\coloneqq\\
        \left\{
            \mverify\left( \pi_2\left( \mmin \right), \pi_1\left( \mmin \right), \mk\left[ j \right]\right), \pi_2\left( \mmin \right), \pi_1\left( \mmin \right), \mk\left[ j \right], \mmin
        \right\}
    \end{multline}
    where $\mmin$ stands for $\minput\left( \mathsf{Rs}\left[i, j\right] \right)$ as in \autoref{ex:basic-hash:steps}.

    Then, with $\mathcal{P}_{\ref{basic-hash-example}}$ being the protocol of \autoref{basic-hash-example}, keeping in mind its steps description (\autoref{ex:basic-hash:steps}), we have
    \begin{multline}
        \mmst[\mathcal{P}_{\ref{basic-hash-example}}]\left(\minput\left(\tau\right)\right)\coloneqq\\
        \begin{aligned}
            &\bigcup\nolimits_{t\in ST\left(\mmsg(\mathsf{T}\left[i, j\right]\right)\cup ST\left(\mcond(\mathsf{T}\left[i, j\right]\right)} \left(\mathsf{T}\left[i, j\right] < \tau, t \right)\\
            \cup&\bigcup\nolimits_{t\in ST\left(\mmsg(\mathsf{Rs}\left[i, j\right]\right)\cup ST\left(\mcond(\mathsf{Rs}\left[i, j\right]\right)} \left(\mathsf{Rs}\left[i, j\right] < \tau, t \right)\\
            \cup&\bigcup\nolimits_{t\in ST\left(\mmsg(\mathsf{Rf}\left[i\right]\right)\cup ST\left(\mcond(\mathsf{Rf}\left[i\right]\right)} \left(\mathsf{Rf}\left[i\right] < \tau, t \right)
        \end{aligned}
    \end{multline}
\end{mex}

\begin{figure}
    \input{101-figures/004-main-subterm.tex}
    \caption{Axiomatization of the subterm relation}\label{fig:subterm-relation}
\end{figure}

With $\mmst\left(\_\right)$ we can take the knowledge of the protocol and the inductive reasoning out of the theorem prover and internalize it into \tool itself.
In practice, 
we describe $\sqsubseteq$ in FOL not as \autoref{eq:def:subterm} but through the axioms laid out in \autoref{fig:subterm-relation}. With this axiomatization, we dispense from \begin{enumerate*}
    \item[\ref{item:sec:infinte-recursion:1}] the trace dependence as well as \item[\ref{item:sec:infinte-recursion:2}] inductive reasoning
\end{enumerate*}.
We describe how to extend $\sqsubseteq$ into other subterm relations (e.g., $\sqsubseteq_{\mverify\left(\_,\_,\bullet\right),\mhash\left(\_,\bullet\right)}$) in\ifreport \appref{sec:custom-subterm}\else~\cite{CRYPTOVAMPIREAutomatedReasoningTR} \fi.

\subsection{Native Subterm Reasoning}\label{sec:modifying-vampire}
As the subterm relation is transitive, the number of logical consequences produced by saturation quickly becomes a burden for any saturation-based prover. While the $\mmst(\_)$ set can effectively be computed (cf. \cite{squirrel} and \appref{sec:subterm-appendix}), automated reasoning with such sets is non-trivial due to their complex interaction with the equational theory of $\beq$ and the required knowledge of the protocol. To overcome such limitations hindering the effectiveness of automation, we devise our own native subterm reasoning within saturation, extending existing approaches~\cite{kovacsComingTermsQuantified2017,cremersSubtermbasedProofTechniques2022}.

We recall that saturation-based provers iteratively apply a finite set of \emph{inference rules}, deriving new clauses\footnote{formulas are preprocessed and clausified} as logical consequences of existing formulas. Whenever the empty clause $\bot$ is inferred, saturation stops and reports unsatisfiability of the negated input formula (hence, the input is valid). For efficiency reasons, saturation implements two types of inferences: \begin{enumerate*}[label=\textbf{(\roman{enumi})}, ref=(\roman{enumi})] \item \emph{generating} rules that add new clauses to the search space, and \item \emph{simplifying} rules that remove so-called redundant clauses from the search space. \end{enumerate*} Importantly, removing redundant clauses does not destroy (refutational) completeness: if a formula is provable using redundant clauses, it is also provable without redundant clauses~\cite{vampire}. Clearly, simplifying rules are eagerly applied within any efficient saturation-based proving process. In what follows, within simplifying rules, we will denote the deleted (redundant) clause by drawing a line through it, for example: $\bcancel{A\vee C}$.

The crux of our approach for native subterm reasoning within \tool comes with two new simplifying inference rules. Given a list $\mathfrak{L}$ of symbols and a subterm relation $\sqsubseteq$, for all $f\in\mathfrak{L}$, we introduce the following simplifying rules \emph{in addition} to the superposition inference rules for FOL with equality~\cite{vampire}:

\begin{mathpar}
    \inferrule{\bcancel{x \sqsubseteq f\left( y_1, \dots, y_n \right)\vee C}}{x = f\left( y_1, \dots, y_n \right)\vee \bigvee\nolimits_{k=1}^n x \sqsubseteq y_k \vee C}[Sbtm]\label{rule:sbtm}

    \inferrule{\bcancel{x \not\sqsubseteq f\left( y_1, \dots, y_n \right)\vee C}}{x\neq f\left( y_1, \dots, y_n \right)\vee C \\ x\not\sqsubseteq y_k \vee C \\ \text{for all $k$}}[-Sbtm]\label{rule:-sbtm}
\end{mathpar}

The inference rules \nameref{rule:sbtm} and \nameref{rule:-sbtm} replace axiomatic reasoning with the axioms \eqref{eq:strict-subterm} and \eqref{eq:subterm-def-function} of \autoref{fig:subterm-relation}. We prevent these rules from contradicting the rest of the axioms of \autoref{fig:subterm-relation} by selectively leaving out symbols from $\mathfrak{L}$; this is notably the case for $\minput$ and memory cells.

The simplifying nature of \nameref{rule:sbtm} and \nameref{rule:-sbtm} ensures that $m$ in $x\sqsubseteq m$ is fully destructed (up to special cases) before the conclusions are added to the search space.

\subsection{Preprocessing}\label{sec:preprocessing}
\subsubsection{Instance preprocessing}\label{sec:preprocess-axioms-instances}
While the $\mmst\left( \_ \right)$ sets, defining subterm relations, are computable, it is not known \textit{a priori} over which terms $\mmst\left( \_ \right)$ should be applied.
Hence, the axiomatization discussed in \autoref{sec:subterm}.
Nevertheless, we may compute estimates about terms that may be involved in subterm reasoning, allowing us to predict/prioritize the application of subterm analysis, as follows.

Most cryptographic axioms have the pattern $\meval{\phi} \Rightarrow \varphi$, where $\phi$ is an easy-to-identify symbolic term (e.g., $\mverify\left(\sigma, m, \mk \right)$ for \eufcma in \autoref{prop:euf-cma}) and $\varphi$ depends on some subterm analysis based on instances of variables in $\phi$ (e.g., $\mhash\left(u, \mk\right) \sqsubseteq m$, among others, in \autoref{prop:euf-cma} too).
If we know $\phi$, then we can often pre-compute $\varphi$ to such an extent that it no longer contains any subterm search by inlining the subterm relation (\autoref{def:subterm}).

We therefore consider the following \tool heuristic: when we encounter a term that can unify with $\phi$, we assume that it is likely to be used within proof search, through the respective cryptographic axiom.
We then preprocess all occurrences of $\phi$ found throughout the protocol specification, including the step definitions, queries, and other user-defined assertions.

\begin{mex*}[\ref{ex:example-preprocessing}]
    \autoref{eq:example-preprocessing} from \autoref{ex:example-preprocessing} is the result of such heuristic applied to \autoref{prop:euf-cma} in \autoref{basic-hash-example}.
    \begin{multline}\tag{\ref{eq:example-preprocessing}}
        \forall i, j. \meval{\mverify\left(  \begin{multlined}
            \pi_2\left( \minput\left(\mathsf{Rs}\left[ i, j \right]\right) \right),\\
            \pi_1\left(  \minput\left(\mathsf{Rs}\left[ i, j \right]\right)\right)
        \end{multlined}, \mk\left[ j \right] \right)}\Rightarrow \\
        \exists i', j'. \left( \begin{multlined}
            \mathsf{T}\left[ i', j' \right] < \mathsf{Rs}\left[ i, j \right]
            \wedge j' = j\\
            \wedge \meval{\mathsf{n}\left[ i', j' \right]}  = \meval{\pi_1\left( \minput\left( \mathsf{Rs}\left[ i, j \right] \right) \right)}
        \end{multlined} \right)
    \end{multline}
\end{mex*}

\subsubsection{Removing subterm reasoning}\label{sec:final-optimization}
Finally, we propose an incomplete heuristic to completely factor out symbolic reasoning when applicable.

Generating terms that are not part of the input problem can, in general, lead to state  explosion, but it is nevertheless necessary during proof search; for example, such terms might be needed to instantiate another cryptographic axiom.
On the other hand, if an axiom instance does not get preprocessed in \autoref{sec:preprocess-axioms-instances}, then its instantiation will likely get significantly delayed during saturation\footnote{The machine is likely to run out of resources before that.}.
For that reason, we propose to continue the preprocessing under the assumption  that \autoref{sec:preprocess-axioms-instances} effectively preprocessed \emph{all} the relevant instances.
Under such assumption, it is possible to factor out all symbolic reasoning (as it was inlined during preprocessing, as in \autoref{ex:example-preprocessing}).
That is, when trying to prove $\valid \left(\forall \vecu.\Gamma\right)\implies \meval{t}$ for \autoref{th:main}'s~\ref{item:th:main:3}, we can find a new axiom $\Delta$ (the result of the preprocessing) such that $\valid \left(\forall \vecu.\Gamma\right)\implies \Delta$ and all symbols used in $\Delta$ are compatibles with $\beq$.
Then showing $\Delta \implies \meval{t}$ can be done modulo $\beq$, which dramatically improves the performance of the solvers.

Interestingly, this idea can be applied not only to protocol specifications but also some cryptographic theorems such as the no-guessing theorem. It can, in particular, be rephrased in a way to eliminate subterm reasoning.

\begin{mth}[No guessing]\label{th:no-guessing-2}
    For all nonces $\n\in\mnonces$, \eqref{eq:super-no-guessing} can be added to the set of axioms.
    \begin{equation}\label{eq:super-no-guessing}
        \forall \veci, m.\ \meval{\n\left[ \veci \right]} = \meval{m} \Rightarrow \n\left[ \veci \right]\bsqsubseteq m
    \end{equation}
    where $\bsqsubseteq$ is defined by \eqref{eq:sq-nonce-def} and is compatible with $\beq$.
    \begin{equation}\label{eq:sq-nonce-def}
        \m \bsqsubseteq \meval{t} \coloneqq \forall t'. \left(\meval{t}=\meval{t'} \Rightarrow \m \sqsubseteq t' \right).
    \end{equation}
\end{mth}
\begin{mproof}
    \ifreport
        \appref{sec:app:no-guessing}
    \else
        In the technical report~\cite{CRYPTOVAMPIREAutomatedReasoningTR}.
    \fi
\end{mproof}

While this heuristic is not complete, our experiments demonstrate that it is widely applicable and leads to significant performance gains (cf. \autoref{sec:experiments}).

    \section{Experiments}%
    \label{sec:experiments}
    \begin{figure*}
    \newcommand{\wl}{${}^\ddagger$}
\newcommand{\segflt}{${}^\dagger$}
\newcommand{\tw}{${}^*$}

\newlength{\mcolumn}
\settowidth{\mcolumn}{\mvampire}
\newcommand{\mproportiontable}{0.23}

\begin{center}
\begin{tabular}{|r|c||c|c|M{0.5\mcolumn}|M{0.5\mcolumn}||c|c|M{0.5\mcolumn}|M{0.5\mcolumn}||c|c|c|}
\hline
\multicolumn{2}{|c||}{\multirow{3}{*}{protocol}}
	& \multicolumn{4}{M{\mproportiontable\textwidth}||}%
            {no preprocessing}
	& \multicolumn{4}{M{\mproportiontable\textwidth}||}%
            {instance preprocessing}
	& \multicolumn{3}{M{\mproportiontable\textwidth}|}%
            {\autoref{sec:final-optimization}'s heuristic} \\
\cline{3-13}
\multicolumn{2}{|c||}{}
	& \multirow{2}{*}{\mzed} & \multirow{2}{*}{\mcvc} & \multicolumn{2}{c||}{\mvampire}
	& \multirow{2}{*}{\mzed} & \multirow{2}{*}{\mcvc} & \multicolumn{2}{c||}{\mvampire}
	& \multirow{2}{*}{\mzed} & \multirow{2}{*}{\mcvc} & \multirow{2}{*}{\mvampire} \\
\cline{5-6}\cline{9-10}
\multicolumn{2}{|c||}{}
	&  &  & $\circ$ & $\bullet$
	&  &  & $\circ$ & $\bullet$
	&  &  &  \\
	\hline\hline
\multirow{1}{*}{\texttt{basic-hash}} & $\square/\triangle$
	& \xmark & \cmark & \xmark & \cmark
	& \cmark\tw & \cmark & \cmark & \cmark
	& \cmark & \cmark & \cmark \\
	\hline\hline
\multirow{3}{*}{\texttt{hash-lock}} & $\square$
	& \xmark & \xmark & \xmark & \xmark
	& \xmark & \xmark & \xmark & \cmark
	& \cmark & \cmark & \cmark \\
\cline{2-13}
 & $\blacksquare$
 	& \xmark & \xmark & \xmark & \xmark
	& \xmark & \xmark & \xmark & \cmark
	& \cmark & \cmark & \cmark \\
\cline{2-13}
 & $\triangle$
 	& \xmark & \xmark & \xmark & \xmark
	& \xmark & \xmark & \xmark & \cmark\tw
	& \cmark\tw & \cmark & \cmark \\
	\hline\hline
\multirow{2}{*}{\texttt{mw}} & $\square$
	& \xmark & \xmark & \xmark & \xmark
	& \xmark & \xmark & \xmark & \cmark\wl\tw
	& \cmark & \cmark & \cmark \\
\cline{2-13}
 & $\triangle$
 	& \xmark & \xmark & \xmark & \cmark
	& \xmark & \xmark & \xmark & \cmark\tw
	& \cmark${}^{**}$ & \cmark & \cmark \\
	\hline\hline
\multirow{1}{*}{\texttt{lak-tag}} & $\square$
	& \xmark & \xmark & \xmark & \xmark
	& \xmark & \xmark & \xmark & \cmark\tw
	& \cmark & \cmark & \cmark \\
	\hline\hline
\multirow{1}{*}{\texttt{feldhofer}} & $\square$
	& \xmark & \xmark & \xmark & \xmark
	& \xmark & \cmark & \xmark & \cmark\tw
	& \cmark & \cmark & \cmark \\
	\hline\hline
\multicolumn{2}{|c||}{\texttt{euf-key-secrecy}}
	& \xmark & \cmark\wl & \xmark & \cmark\wl
	& \xmark & \cmark\wl & \cmark\wl & \cmark\wl
	& \cmark\wl & \cmark\wl & \cmark\wl \\
 	\hline\hline
\multirow{2}{*}{\texttt{ddh}} & $\square$
	& \xmark & \xmark & \xmark & \xmark
	& \xmark & \cmark\wl & \xmark & \xmark
	& \cmark\wl & \cmark\wl & \cmark\wl \\
 \cline{2-13}
 & $\triangle$
    & \xmark & \xmark & \xmark & \xmark
	& \xmark & \cmark\wl & \xmark & \xmark
	& \cmark\wl & \cmark\wl & \cmark\wl \\
\hline
\end{tabular}
\end{center}
{\footnotesize
\begin{multicols}{2}
	\begin{itemize}
		\item[$\circ$] vanilla \mvampire (without \autoref{sec:optimisations} adjustments)
		\item[$\bullet$] using the added decision procedures
        \item[$\square$] authentication
        \item[$\blacksquare$] injective authentication
        \item[$\triangle$] used as a lemma for unlinkability in \msquirrel
		\item[\wl] made use of lemmas (proven in \tool)
		\item[\tw] split into two queries, one for each direction of the double implication
        \item[${}^{**}$] the solver succeded on one side of the equivalence
	\end{itemize}
\end{multicols}
}

    \caption{\tool experiments}\label{fig:all-experiments}
\end{figure*}

We implemented \tool in extension of the \mvampire theorem prover~\cite{vampire}, by adjusting saturation, term ordering, subterm reasoning and preprocessing in \mvampire as described in \autoref{sec:optimisations}.

\subsubsection*{Experimental Setup} We evaluated \tool{} against examples and trace-based queries from~\cite{squirrel} that did not rely on the composition framework~\cite{composition_framework_2020}.
For competitive evaluations, we varied the proving backend using a selection of state-of-the-art solvers for FOL with theories, \mzed~\cite{z3}, \mcvc~\cite{cvc5} and \mvampire~\cite{vampire} (with and without our modifications), and used various levels of preprocessing.

\subsubsection*{Benchmarks} We test \tool on all 
trace-based properties from the \msquirrel's benchmarks~\cite{squirrel}, which consist of authentication properties in the \texttt{basic-\allowbreak{}hash}~\cite{brusoFormalVerificationPrivacy2010}, \texttt{hash-lock}~\cite{juelsDefiningStrongPrivacy2009}, \texttt{lak-\allowbreak{}tag}~\cite{hirschiMethodUnboundedVerification2019b}, \texttt{feldhofer}~\cite{feldhoferStrongAuthenticationRFID2004}, \texttt{mw}~\cite{molnarPrivacySecurityLibrary2004} and \texttt{ddh}~\cite{ISO9798-3} protocols (the $\square$ in \autoref{fig:all-experiments}).
We follow very closely \cite{squirrel}'s modelling of these protocols.

We note that \cite{squirrel} uses the examples we considered to further verify observational equivalence hyperproperties (e.g., unlikability or strong-secrecy in the case of these protocols) whose proofs rely on the initial trace properties.
\tool does not yet support such hyperproperties.
However, we show that \tool can verify some of the lemmas that are used to prove those properties in \msquirrel~\cite{squirrel} (marked with a $\triangle$ in \autoref{fig:all-experiments}).
Indeed, a proof of observational equivalence in the CCSA may require trace properties to prune out (nearly) impossible traces, or rewriting equal terms according to $\beq$.
Both kinds of properties are supported in \tool, which can therefore be used to facilitate proofs with \msquirrel.
Indeed, since both tools use the same underlying theory, results in \tool are immediately valid in \msquirrel (and vice versa) (see\ifreport \appref{sec:comp-w-squirrel}\else~\cite{CRYPTOVAMPIREAutomatedReasoningTR}\fi).

\subsubsection*{Experimental Analysis} Our experimental results are summarized in \autoref{fig:all-experiments}.

We show the effectiveness of our preprocessing by gradually increasing its impact in \autoref{fig:all-experiments}. We start with only preprocessing the definition of the subterm relations (see \autoref{sec:subterm}).
We continue with the instance preprocessing (see \autoref{sec:preprocess-axioms-instances}), which, combined with our saturation modifications (see \autoref{sec:modifying-vampire}), enables us to verify all considered protocols.
However, in this mode, \tool often needs guidance to split double implications into two queries (indicated via \cmark${}^*$ in \autoref{fig:all-experiments}), or use further lemmas to prove (\cmark${}^\ddagger$ in \autoref{fig:all-experiments}).

We also tested \tool using the incomplete simplification of the logic introduced in \autoref{sec:final-optimization}.
In this mode, we observe that \tool overall gains in capabilities despite its incomplete strategy/heuristics.
\autoref{fig:all-experiments} also highlights the limits of this heuristic.
Indeed, \texttt{euf-key-secrecy} and \texttt{ddh} required very simple auxiliary lemmas (hinting which axioms are to be instantiated with which messages) to trigger the right preprocessing.
In the case of \texttt{ddh}, this lemma is not even passed on to the theorem prover as it is only used for the preprocessing but not for the proof.

We conducted a performance evaluation based on the previous benchmarks, using a machine with 64 cores and 128GB of RAM and running \mvampire, \mzed, and \mcvc as a verification backend for \tool, measuring the time required by the fastest among them. \tool is capable of verifying the considered examples in a few dozen milliseconds.

In addition, \tool shows very competitive performance even against \textsc{CryptoVerif}~\cite{cryptoverif}, the state-of-the-art automated cryptographic protocol verifier supporting computational security proofs. The results of our comparative performance evaluation are summarized in \autoref{fig:gotta-to-fast}.

\begin{figure}



\newlength{\mlengthhead}
\settowidth{\mlengthhead}{\tool}
\newcommand{\tikzmark}[1]{\tikz[remember picture,overlay]\coordinate (#1);}
\newcommand{\myangle}{25}
\newcommand{\myextralength}{3mm}
\setlength{\mlengthhead}{\mlengthhead+\myextralength}
\setlength{\mlengthhead}{\mlengthhead+\myextralength}

\newcolumntype{T}[1]{@{\hspace{\tabcolsep}}M{1.1cm}@{\hspace{\tabcolsep}\tikzmark{#1}}}
\newcolumntype{W}[1]{@{\hspace{\tabcolsep}}r@{\hspace{\tabcolsep}\tikzmark{#1}}}

\begin{center}
\begin{tikzpicture}[remember picture]
    \node[inner xsep=-\pgflinewidth,inner ysep=-\pgflinewidth] at (0,0) (mytable){%
    \begin{tabular}{|W{w}|T{b}|T{c}|}
    \hline
    \texttt{basic-hash}    & $52.8\,\mathrm{ms}$ & $21.4\,\mathrm{ms}$\\\hline
    \texttt{hash-lock}     & $62.6\,\mathrm{ms}$ & $24.4\,\mathrm{ms}$\\\hline
    \texttt{lak-tag}       & $61.9\,\mathrm{ms}$ & $30.4\,\mathrm{ms}$\\\hline
    \texttt{feldhofer}     & $60.9\,\mathrm{ms}$ & $31.6\,\mathrm{ms}$\\\hline
    \texttt{mw}            & $71.9\,\mathrm{ms}$ & $38.3\,\mathrm{ms}$\\\hline
        \end{tabular}
    };

    \draw (mytable.north east) --++ (180-\myangle:\mlengthhead+\myextralength);
    \draw (mytable.north-|w) --++ (180-\myangle:\mlengthhead+\myextralength);
    \draw (mytable.north-|b) --++ (180-\myangle:\mlengthhead+\myextralength);
    \draw (mytable.north east) ++ (180-\myangle:\mlengthhead+\myextralength) --([shift={(180-\myangle:\mlengthhead+\myextralength)}]mytable.north-|w);

    \node[rotate=-\myangle,anchor=west, align=left] at ($(mytable.north-|w)!0.5!(mytable.north-|b)+(180-\myangle:\mlengthhead)$) {\textsc{CryptoVerif}};
    \node[rotate=-\myangle,anchor=west] at ($(mytable.north-|b)!0.5!(mytable.north east)+(180-\myangle:\mlengthhead)$) {\tool};
\end{tikzpicture}
\end{center}

    \caption{Runtimes of \tool and \textsc{CryptoVerif} for authentication queries}\label{fig:gotta-to-fast}
\end{figure}

    \section{Conclusion and Further Work}%
    \label{sec:further-works}
    We introduced \tool, the first fully automated cryptographic protocol verifier supporting the CCSA.
A core contribution is the first-order encoding of CCSA along with dedicated proof techniques. 
We further designed tailored reasoning procedures and heuristics to enable leveraging state-of-the-art first-order theorem proving in \tool, as demonstrated by our experimental results. 

As future work, we are interested in extending \tool to support compositional proof techniques~\cite{composition_framework_2020}. Furthermore, we plan to integrate axioms to reason about post-quantum security~\cite{cremersLogicInteractiveProver2022}. 
We also plan to integrate relational reasoning into \tool to support indistinguishability proofs~\cite{bana_computationally_2014}.

    \ifannon\else
    \subsubsection*{Acknowledgments}
This work was partially supported by the European Research Council (ERC) through the  Consolidator Grants ARTIST 101002685 and Browsec 771527;  the TU Wien Doctoral College SecInt;  the Austrian Science Fund (FWF) through the SpyCoDe SFB projects  \mbox{F8504} and \mbox{F8510-N}; the  
 Vienna Science and Technology Fund (WWTF) through [ForSmart Grant ID: 10.47379/ICT22007]; the Austrian Research Promotion Agency (FFG) through the  the COMET K1 SBA and ABC; and the  
 the Christian Doppler Research Association through the Christian Doppler Laboratory Blockchain Technologies for the Internet of Things (\mbox{CDL-BOT}).
    \fi


    \bibliographystyle{IEEEtran}
    \bibliography{IEEEabrv,990-bibliography.bib}

\begin{thebibliography}{10}
\providecommand{\url}[1]{#1}
\csname url@samestyle\endcsname
\providecommand{\newblock}{\relax}
\providecommand{\bibinfo}[2]{#2}
\providecommand{\BIBentrySTDinterwordspacing}{\spaceskip=0pt\relax}
\providecommand{\BIBentryALTinterwordstretchfactor}{4}
\providecommand{\BIBentryALTinterwordspacing}{\spaceskip=\fontdimen2\font plus
\BIBentryALTinterwordstretchfactor\fontdimen3\font minus
  \fontdimen4\font\relax}
\providecommand{\BIBforeignlanguage}[2]{{%
\expandafter\ifx\csname l@#1\endcsname\relax
\typeout{** WARNING: IEEEtran.bst: No hyphenation pattern has been}%
\typeout{** loaded for the language `#1'. Using the pattern for}%
\typeout{** the default language instead.}%
\else
\language=\csname l@#1\endcsname
\fi
#2}}
\providecommand{\BIBdecl}{\relax}
\BIBdecl

\bibitem{loweAttackNeedhamSchroederPublickey1995}
\BIBentryALTinterwordspacing
G.~Lowe, ``An attack on the {{Needham-Schroeder}} public-key authentication
  protocol,'' \emph{Information Processing Letters}, vol.~56, no.~3, pp.
  131--133, Nov. 1995. [Online]. Available:
  \url{https://www.sciencedirect.com/science/article/pii/0020019095001442}
\BIBentrySTDinterwordspacing

\bibitem{alfardanLuckyThirteenBreaking2013}
N.~J. Al~Fardan and K.~G. Paterson, ``Lucky {{Thirteen}}: {{Breaking}} the
  {{TLS}} and {{DTLS Record Protocols}},'' in \emph{2013 {{IEEE Symposium}} on
  {{Security}} and {{Privacy}}}, May 2013, pp. 526--540.

\bibitem{malavoltaAnonymousMultiHopLocks2019}
\BIBentryALTinterwordspacing
G.~Malavolta, P.~{Moreno-Sanchez}, C.~Schneidewind, A.~Kate, and M.~Maffei,
  ``Anonymous {{Multi-Hop Locks}} for {{Blockchain Scalability}} and
  {{Interoperability}},'' in \emph{Proceedings 2019 {{Network}} and
  {{Distributed System Security Symposium}}}.\hskip 1em plus 0.5em minus
  0.4em\relax {San Diego, CA}: {Internet Society}, 2019. [Online]. Available:
  \url{https://www.ndss-symposium.org/wp-content/uploads/2019/02/ndss2019_09-4_Malavolta_paper.pdf}
\BIBentrySTDinterwordspacing

\bibitem{bhargavanProvingTLSHandshake2014}
\BIBentryALTinterwordspacing
K.~Bhargavan, C.~Fournet, M.~Kohlweiss, A.~Pironti, P.-Y. Strub, and
  S.~{Zanella-B{\'e}guelin}, ``Proving the {{TLS Handshake Secure}} ({{As It
  Is}}),'' in \emph{Advances in {{Cryptology}} {\textendash} {{CRYPTO}}
  2014}.\hskip 1em plus 0.5em minus 0.4em\relax {Springer, Berlin, Heidelberg},
  2014, pp. 235--255. [Online]. Available:
  \url{https://link.springer.com/chapter/10.1007/978-3-662-44381-1_14}
\BIBentrySTDinterwordspacing

\bibitem{lippMechanisedCryptographicProof2019}
B.~Lipp, B.~Blanchet, and K.~Bhargavan, ``A {{Mechanised Cryptographic Proof}}
  of the {{WireGuard Virtual Private Network Protocol}},'' in \emph{2019 {{IEEE
  European Symposium}} on {{Security}} and {{Privacy}} ({{EuroS}}\&{{P}})},
  Jun. 2019, pp. 231--246.

\bibitem{koutsos5GAKAAuthenticationProtocol2019a}
A.~Koutsos, ``The {{5G-AKA Authentication Protocol Privacy}},'' in \emph{2019
  {{IEEE European Symposium}} on {{Security}} and {{Privacy}}
  ({{EuroS}}\&{{P}})}, Jun. 2019, pp. 464--479.

\bibitem{loweHierarchyAuthenticationSpecifications1997}
G.~Lowe, ``A hierarchy of authentication specifications,'' in \emph{Proceedings
  10th {{Computer Security Foundations Workshop}}}, Jun. 1997, pp. 31--43.

\bibitem{DBLP:conf/csfw/ClarksonS08}
\BIBentryALTinterwordspacing
M.~R. Clarkson and F.~B. Schneider, ``Hyperproperties,'' in \emph{Proceedings
  of the 21st {{IEEE}} Computer Security Foundations Symposium, {{CSF}} 2008,
  Pittsburgh, Pennsylvania, {{USA}}, 23-25 June 2008}.\hskip 1em plus 0.5em
  minus 0.4em\relax {IEEE Computer Society}, 2008, pp. 51--65. [Online].
  Available: \url{https://doi.org/10.1109/CSF.2008.7}
\BIBentrySTDinterwordspacing

\bibitem{dolev-yao}
D.~Dolev and A.~Yao, ``On the security of public key protocols,'' \emph{IEEE
  Transactions on Information Theory}, vol.~29, no.~2, pp. 198--208, Mar. 1983.

\bibitem{proverif}
B.~Blanchet, ``An {{Efficient Cryptographic Protocol Verifier Based}} on
  {{Prolog Rules}},'' in \emph{Proceedings of the 14th {{IEEE}} Workshop on
  {{Computer Security Foundations}}}, ser. {{CSFW}} '01.\hskip 1em plus 0.5em
  minus 0.4em\relax {USA}: {IEEE Computer Society}, Jun. 2001, p.~82.

\bibitem{tamarin}
S.~Meier, B.~Schmidt, C.~Cremers, and D.~Basin, ``The {{TAMARIN Prover}} for
  the {{Symbolic Analysis}} of {{Security Protocols}},'' in \emph{Computer
  {{Aided Verification}}}, ser. Lecture {{Notes}} in {{Computer Science}},
  N.~Sharygina and H.~Veith, Eds.\hskip 1em plus 0.5em minus 0.4em\relax
  {Berlin, Heidelberg}: {Springer}, 2013, pp. 696--701.

\bibitem{bhargavanTranscriptCollisionAttacks2016}
\BIBentryALTinterwordspacing
K.~Bhargavan and G.~Leurent, ``Transcript {{Collision Attacks}}: {{Breaking
  Authentication}} in {{TLS}}, {{IKE}}, and {{SSH}},'' in \emph{Proceedings
  2016 {{Network}} and {{Distributed System Security Symposium}}}.\hskip 1em
  plus 0.5em minus 0.4em\relax {San Diego, CA}: {Internet Society}, 2016.
  [Online]. Available:
  \url{https://www.ndss-symposium.org/wp-content/uploads/2017/09/transcript-collision-attacks-breaking-authentication-tls-ike-ssh.pdf}
\BIBentrySTDinterwordspacing

\bibitem{easycrypt}
G.~Barthe, B.~Gr{\'e}goire, S.~Heraud, and S.~Z. B{\'e}guelin,
  ``Computer-{{Aided Security Proofs}} for the {{Working Cryptographer}},'' in
  \emph{Advances in {{Cryptology}} {\textendash} {{CRYPTO}} 2011}, ser. Lecture
  {{Notes}} in {{Computer Science}}, P.~Rogaway, Ed.\hskip 1em plus 0.5em minus
  0.4em\relax {Berlin, Heidelberg}: {Springer}, 2011, pp. 71--90.

\bibitem{cryptoverif}
B.~Blanchet, ``A computationally sound mechanized prover for security
  protocols,'' in \emph{2006 {{IEEE Symposium}} on {{Security}} and {{Privacy}}
  ({{S}}\&{{P}}'06)}, May 2006, pp. 15 pp.--154.

\bibitem{long_bana_towards_2012}
\BIBentryALTinterwordspacing
G.~Bana and H.~{Comon-Lundh}, ``Towards {{Unconditional Soundness}}:
  {{Computationally Complete Symbolic Attacker}},'' in \emph{Principles of
  {{Security}} and {{Trust}}}, D.~Hutchison, T.~Kanade, J.~Kittler, J.~M.
  Kleinberg, F.~Mattern, J.~C. Mitchell, M.~Naor, O.~Nierstrasz,
  C.~Pandu~Rangan, B.~Steffen, M.~Sudan, D.~Terzopoulos, D.~Tygar, M.~Y. Vardi,
  G.~Weikum, P.~Degano, and J.~D. Guttman, Eds.\hskip 1em plus 0.5em minus
  0.4em\relax {Berlin, Heidelberg}: {Springer Berlin Heidelberg}, 2012, vol.
  7215, pp. 189--208. [Online]. Available:
  \url{http://link.springer.com/10.1007/978-3-642-28641-4_11}
\BIBentrySTDinterwordspacing

\bibitem{bana_computationally_2014}
------, ``A computationally complete symbolic attacker for equivalence
  properties,'' in \emph{Proceedings of the 2014 {{ACM SIGSAC Conference}} on
  {{Computer}} and {{Communications Security}}}, 2014, pp. 609--620.

\bibitem{banaComputationallyCompleteSymbolic2013}
G.~Bana, K.~Hasebe, and M.~Okada, ``Computationally complete symbolic attacker
  and key exchange,'' in \emph{Proceedings of the 2013 {{ACM SIGSAC}}
  Conference on {{Computer}} \& Communications Security}, 2013, pp. 1231--1246.

\bibitem{BaeldeKoutsosLallemand2023}
D.~Baelde, A.~Koutsos, and J.~Lallemand, ``A {{Higher-Order
  Indistinguishability Logic}} for {{Cryptographic Reasoning}},'' in \emph{2023
  38th {{Annual ACM}}/{{IEEE Symposium}} on {{Logic}} in {{Computer Science}}
  ({{LICS}})}, Jun. 2023, pp. 1--13.

\bibitem{squirrel}
\BIBentryALTinterwordspacing
D.~Baelde, S.~Delaune, C.~Jacomme, A.~Koutsos, and S.~Moreau, ``An
  {{Interactive Prover}} for {{Protocol Verification}} in the {{Computational
  Model}},'' in \emph{{{SP}} 2021 - 42nd {{IEEE Symposium}} on {{Security}} and
  {{Privacy}}}, {San Fransisco / Virtual, United States}, May 2021. [Online].
  Available: \url{https://hal.archives-ouvertes.fr/hal-03172119}
\BIBentrySTDinterwordspacing

\bibitem{baelde_cracking_2022}
\BIBentryALTinterwordspacing
D.~Baelde, S.~Delaune, A.~Koutsos, and S.~Moreau, ``Cracking the {{Stateful
  Nut}},'' in \emph{{{CSF}} 2022 - 35th {{IEEE Computer Security Foundations
  Symposium}}}.\hskip 1em plus 0.5em minus 0.4em\relax {Haifa, Israel}:
  {IRISA}, Aug. 2022, Research {{Report}}. [Online]. Available:
  \url{https://hal.archives-ouvertes.fr/hal-03500056}
\BIBentrySTDinterwordspacing

\bibitem{cremersLogicInteractiveProver2022}
C.~Cremers, C.~Fontaine, and C.~Jacomme, ``A {{Logic}} and an {{Interactive
  Prover}} for the {{Computational Post-Quantum Security}} of {{Protocols}},''
  in \emph{2022 {{IEEE Symposium}} on {{Security}} and {{Privacy}} ({{SP}})},
  May 2022, pp. 125--141.

\bibitem{composition_framework_2020}
\BIBentryALTinterwordspacing
H.~Comon, C.~Jacomme, and G.~Scerri, ``Oracle {{Simulation}}: {{A Technique}}
  for {{Protocol Composition}} with {{Long Term Shared Secrets}},'' in
  \emph{Proceedings of the 2020 {{ACM SIGSAC Conference}} on {{Computer}} and
  {{Communications Security}}}.\hskip 1em plus 0.5em minus 0.4em\relax {Virtual
  Event USA}: {ACM}, Oct. 2020, pp. 1427--1444. [Online]. Available:
  \url{https://dl.acm.org/doi/10.1145/3372297.3417229}
\BIBentrySTDinterwordspacing

\bibitem{venkataraman1987decidability}
\BIBentryALTinterwordspacing
K.~N. Venkataraman, ``Decidability of the purely existential fragment of the
  theory of term algebras,'' \emph{Journal of the ACM}, vol.~34, no.~2, pp.
  492--510, Apr. 1987. [Online]. Available:
  \url{https://dl.acm.org/doi/10.1145/23005.24037}
\BIBentrySTDinterwordspacing

\bibitem{z3}
L.~{de Moura} and N.~Bj{\o}rner, ``Z3: {{An Efficient SMT Solver}},'' in
  \emph{Tools and {{Algorithms}} for the {{Construction}} and {{Analysis}} of
  {{Systems}}}, ser. Lecture {{Notes}} in {{Computer Science}}, C.~R.
  Ramakrishnan and J.~Rehof, Eds.\hskip 1em plus 0.5em minus 0.4em\relax
  {Berlin, Heidelberg}: {Springer}, 2008, pp. 337--340.

\bibitem{vampire}
L.~Kov{\'a}cs and A.~Voronkov, ``First-{{Order Theorem Proving}} and
  {{Vampire}},'' in \emph{Computer {{Aided Verification}}}, ser. Lecture
  {{Notes}} in {{Computer Science}}, N.~Sharygina and H.~Veith, Eds.\hskip 1em
  plus 0.5em minus 0.4em\relax {Berlin, Heidelberg}: {Springer}, 2013, pp.
  1--35.

\bibitem{cvc5}
\BIBentryALTinterwordspacing
H.~Barbosa, C.~W. Barrett, M.~Brain, G.~Kremer, H.~Lachnitt, M.~Mann,
  A.~Mohamed, M.~Mohamed, A.~Niemetz, A.~N{\"o}tzli, A.~Ozdemir, M.~Preiner,
  A.~Reynolds, Y.~Sheng, C.~Tinelli, and Y.~Zohar, ``Cvc5: {{A Versatile}} and
  {{Industrial-Strength SMT Solver}},'' in \emph{Tools and {{Algorithms}} for
  the {{Construction}} and {{Analysis}} of {{Systems}} - 28th {{International
  Conference}}, {{TACAS}} 2022, {{Held}} as {{Part}} of the {{European Joint
  Conferences}} on {{Theory}} and {{Practice}} of {{Software}}, {{ETAPS}} 2022,
  {{Munich}}, {{Germany}}, {{April}} 2-7, 2022, {{Proceedings}}, {{Part I}}},
  ser. Lecture {{Notes}} in {{Computer Science}}, D.~Fisman and G.~Rosu, Eds.,
  vol. 13243.\hskip 1em plus 0.5em minus 0.4em\relax {Springer}, 2022, pp.
  415--442. [Online]. Available:
  \url{https://doi.org/10.1007/978-3-030-99524-9\_24}
\BIBentrySTDinterwordspacing

\bibitem{brusoFormalVerificationPrivacy2010}
M.~Brus{\'o}, K.~Chatzikokolakis, and J.~{den Hartog}, ``Formal
  {{Verification}} of {{Privacy}} for {{RFID Systems}},'' in \emph{2010 23rd
  {{IEEE Computer Security Foundations Symposium}}}, Jul. 2010, pp. 75--88.

\bibitem{kovacsComingTermsQuantified2017}
\BIBentryALTinterwordspacing
L.~Kov{\'a}cs, S.~Robillard, and A.~Voronkov, ``Coming to terms with quantified
  reasoning,'' in \emph{Proceedings of the 44th {{ACM SIGPLAN Symposium}} on
  {{Principles}} of {{Programming Languages}}}, ser. {{POPL}} '17.\hskip 1em
  plus 0.5em minus 0.4em\relax {New York, NY, USA}: {Association for Computing
  Machinery}, Jan. 2017, pp. 260--270. [Online]. Available:
  \url{https://dl.acm.org/doi/10.1145/3009837.3009887}
\BIBentrySTDinterwordspacing

\bibitem{cremersSubtermbasedProofTechniques2022}
\BIBentryALTinterwordspacing
C.~Cremers, C.~Jacomme, and P.~Lukert, ``Subterm-based proof techniques for
  improving the automation and scope of security protocol analysis,'' in
  \emph{{{CSF}} 2022 - 36th {{IEEE Computer Security Foundations Symposium}}},
  {Dubrovnik, Croatia}, Aug. 2022, pp. 1--14. [Online]. Available:
  \url{https://eprint.iacr.org/2022/1130}
\BIBentrySTDinterwordspacing

\bibitem{shannonCommunicationTheorySecrecy1949}
\BIBentryALTinterwordspacing
C.~E. Shannon, ``Communication theory of secrecy systems,'' \emph{The Bell
  System Technical Journal}, vol.~28, no.~4, pp. 656--715, Oct. 1949. [Online].
  Available: \url{https://ieeexplore.ieee.org/document/6769090}
\BIBentrySTDinterwordspacing

\bibitem{skolem}
T.~Skolem, ``Logisch-kombinatorische untersuchungen {\"u}ber die
  erf{\"u}llbarkeit oder bewiesbarkeit mathematischer s{\"a}tze nebst einem
  theorem {\"u}ber dichte mengen,'' in \emph{Selected Works in Logic}, 1920.

\bibitem{bundyChapter13Automation2001}
\BIBentryALTinterwordspacing
A.~Bundy, ``Chapter 13 - {{The Automation}} of {{Proof}} by {{Mathematical
  Induction}},'' in \emph{Handbook of {{Automated Reasoning}}}, ser. Handbook
  of {{Automated Reasoning}}, A.~Robinson and A.~Voronkov, Eds.\hskip 1em plus
  0.5em minus 0.4em\relax {Amsterdam}: {North-Holland}, Jan. 2001, pp.
  845--911. [Online]. Available:
  \url{https://www.sciencedirect.com/science/article/pii/B9780444508133500151}
\BIBentrySTDinterwordspacing

\bibitem{juelsDefiningStrongPrivacy2009}
\BIBentryALTinterwordspacing
A.~Juels and S.~A. Weis, ``Defining strong privacy for {{RFID}},'' \emph{ACM
  Transactions on Information and System Security}, vol.~13, no.~1, pp.
  7:1--7:23, Nov. 2009. [Online]. Available:
  \url{https://dl.acm.org/doi/10.1145/1609956.1609963}
\BIBentrySTDinterwordspacing

\bibitem{hirschiMethodUnboundedVerification2019b}
\BIBentryALTinterwordspacing
L.~Hirschi, D.~Baelde, and S.~Delaune, ``A method for unbounded verification of
  privacy-type properties,'' \emph{Journal of Computer Security}, vol.~27,
  no.~3, pp. 277--342, Jan. 2019. [Online]. Available:
  \url{https://doi.org/10.3233/JCS-171070}
\BIBentrySTDinterwordspacing

\bibitem{feldhoferStrongAuthenticationRFID2004}
M.~Feldhofer, S.~Dominikus, and J.~Wolkerstorfer, ``Strong {{Authentication}}
  for {{RFID Systems Using}} the {{AES Algorithm}},'' in \emph{Cryptographic
  {{Hardware}} and {{Embedded Systems}} - {{CHES}} 2004}, ser. Lecture
  {{Notes}} in {{Computer Science}}, M.~Joye and J.-J. Quisquater, Eds.\hskip
  1em plus 0.5em minus 0.4em\relax {Berlin, Heidelberg}: {Springer}, 2004, pp.
  357--370.

\bibitem{molnarPrivacySecurityLibrary2004}
\BIBentryALTinterwordspacing
D.~Molnar and D.~Wagner, ``Privacy and security in library {{RFID}}: Issues,
  practices, and architectures,'' in \emph{Proceedings of the 11th {{ACM}}
  Conference on {{Computer}} and Communications Security}, ser. {{CCS}}
  '04.\hskip 1em plus 0.5em minus 0.4em\relax {New York, NY, USA}: {Association
  for Computing Machinery}, Oct. 2004, pp. 210--219. [Online]. Available:
  \url{https://dl.acm.org/doi/10.1145/1030083.1030112}
\BIBentrySTDinterwordspacing

\bibitem{ISO9798-3}
``{{IT Security}} techniques {\textendash} {{Entity}} authentication
  {\textendash} {{Part}} 3: {{Mechanisms}} using digital signature
  techniques,'' {International Organization for Standardization}, {Geneva, CH},
  Standard, Jan. 2019.

\bibitem{kbo}
\BIBentryALTinterwordspacing
D.~E. Knuth and P.~B. Bendix, ``Simple {{Word Problems}} in {{Universal
  Algebras}},'' in \emph{Computational {{Problems}} in {{Abstract Algebra}}},
  {\relax JOHN}.~Leech, Ed.\hskip 1em plus 0.5em minus 0.4em\relax {Pergamon},
  Jan. 1970, pp. 263--297. [Online]. Available:
  \url{https://www.sciencedirect.com/science/article/pii/B978008012975450028X}
\BIBentrySTDinterwordspacing

\bibitem{Herbrand1930}
\BIBentryALTinterwordspacing
J.~Herbrand, \emph{{Recherches sur la th{\'e}orie de la d{\'e}monstration}},
  1930. [Online]. Available: \url{http://eudml.org/doc/192791}
\BIBentrySTDinterwordspacing

\bibitem{smtlib}
C.~Barrett, P.~Fontaine, and C.~Tinelli, ``The {{SMT-LIB Standard}}:
  {{Version}} 2.6,'' {Department of Computer Science, The University of Iowa},
  Tech. Rep., 2017.

\bibitem{bachmairChapterResolutionTheorem2001}
\BIBentryALTinterwordspacing
L.~Bachmair, H.~Ganzinger, D.~McAllester, and C.~Lynch, ``Chapter 2 -
  {{Resolution Theorem Proving}},'' in \emph{Handbook of {{Automated
  Reasoning}}}, ser. Handbook of {{Automated Reasoning}}, A.~Robinson and
  A.~Voronkov, Eds.\hskip 1em plus 0.5em minus 0.4em\relax {Amsterdam}:
  {North-Holland}, Jan. 2001, pp. 19--99. [Online]. Available:
  \url{https://www.sciencedirect.com/science/article/pii/B9780444508133500047}
\BIBentrySTDinterwordspacing

\end{thebibliography}


    \appendices
    \section{Complete semantics}\label{sec:full-semantics}

\begin{figure}

    \begin{align*}
        \mevalidxT{\ib}&\coloneqq \sigmI\left( \ib \right)\\
        \mevaltimeT{\mpred\left( T \right)}&\coloneqq \mepred{\mevaltimeT{T}}\\
        \mevaltimeT{\stpa\left[ \vecI \right]}&\coloneqq
            \begin{cases}
                \stpa\left[ \mevalT{\vecI} \right]&\text{if it is in $\domE$}\\
                \mundef&\text{otherwise}
            \end{cases}\\
        \mevalT{T<T'}&\coloneqq \dirac\left( \mevaltimeT{T}\precE \mevaltimeT{T'} \right)\\
        \mevalT{T=T'}&\coloneqq \dirac\left( \mevaltimeT{T} = \mevaltimeT{T'} \right)\\
        \mevalT{I=I'}&\coloneqq \dirac\left( \mevalidxT{I} = \mevalidxT{I'} \right)\\
        \mevalT{\mhappens\left( T \right)}&\coloneqq \dirac\left( \mevaltimeT{T}\neq \mundef \right)\\
        \mevalT{\n\left[ \vecI \right]}&\coloneqq \n_{\mevalidxT{ \vecI }}\\
        \mevalT{f\left[ \vecI \right]\left( \vec t \right)}&\coloneqq f_{\mevalidxT{\vecI}}\left( \mevalT{\vec t} \right)\\
        \mevalT{\minput\left( T \right)}&\coloneqq \mminput{\mevaltimeT{T}}\\
        \mevalT{\cc\left[ \vecI \right]!\left( T \right)}&\coloneqq
            \begin{cases}
                \emptyset&\text{if }\mevaltimeT{T}\text{ is $\mundef$}\\
                \mevalT{u_{\mevaltimeT{T}}\left( \cc\left[ \mevalidxT{\vecI} \right] \right)}&\text{otherwise}
            \end{cases}
    \end{align*}

    \begin{mathpar}
        \mminput{\minit},\mminput{\mundef},\mmframe{\mundef}\coloneqq\emptyset

        \mmexec{\mundef}\coloneqq\bbot

        \mmexec{\minit}\coloneqq\btop

        \mmframe{\minit}\coloneqq\mmsg\left( \minit \right)

        \mminput{a}\coloneqq\matt\left( \mmframe{\mepred{a}} \right)

        \mmexec{a}\coloneqq\mcond\left( a \right)\bwedge \mmexec{\mepred{a}}

        \mmframe{a}\coloneqq\ll \mmexec{a}; \ll\mifthenelse{\mmexec{a}}{\mmsg\left( a \right)}{\mfail}; \mmframe{\mepred{a}}\gg\gg
    \end{mathpar}

    \begin{multline*}
        \mfindst{\veci}{t_1\left[ \veci \right]}{t_2\left[ \veci \right]}{t_3}\coloneqq\\
        \mifthenelse{t_1\left[ \vecib_1 \right]}{t_2\left[ \vecib_1 \right]}{}
            \dots\\\mifthenelse{t_1\left[ \vecib_n \right]}{t_2\left[ \vecib_n \right]}{t_3}
    \end{multline*}
    \begin{multline*}
        \mfindst{\vect}{t_1\left[ \vect \right]}{t_2\left[ \vect \right]}{t_3}\coloneqq\\
        \mifthenelse{t_1\left[ \vectb_1 \right]}{t_2\left[ \vectb_1 \right]}{}
            \dots\\\mifthenelse{t_1\left[ \vectb_k \right]}{t_2\left[ \vectb_k \right]}{t_3}
    \end{multline*}
    Where $\left\{ \vecib_1,\dots,\vecib_n \right\}=\domI^{\marity{\veci}}$ and $\left\{ \vectb_1,\dots,\vectb_k \right\}={\domSE}^{\marity{\vect}}$.

    \caption{Unfolding of \msymbl}\label{fig:unfolding}
\end{figure}

We assume that the following functions are part of $\mfunctions$: $\mifthenelse{\_\allowbreak}{\_\allowbreak}{\_}$, $\ll_;\_\gg$, $\emptyset$, $\btop$, $\bbot$ and $\mfail$.

We use a pseudo-Dirac function in \autoref{fig:unfolding} for conciseness:
\[
    \delta\left( P \right)\coloneqq\begin{cases}
        \btop&\text{ if }P\\
        \bbot&\text{otherwise}
    \end{cases}
\]

\begin{mdefenv}{Previous Step}{previous-step}
    Given $\mathbb{T}$ and a step $a\in \domE\cup\left\{ \mundef \right\}$, we define the \emph{previous step} $\mepred{a}$ as such:
    \begin{equation}
        \mepred{a}\coloneqq \begin{cases}
            \minit & \text{if $a$ is $\minit$}\\
            \mundef & \text{if $a$ is $\mundef$}\\
            \max_{\preceqE}\left\{ b \middle| b\in \domE, b\precE a \right\}&\text{otherwise}
        \end{cases}
    \end{equation}
\end{mdefenv}

\begin{mdefenv}{Unfolding}{Unfolding}
    Given a trace $\mathbb{T}$, we unfold terms of \msymbl into \mbcl according to \autoref{fig:unfolding}, with
    \begin{align}
        \mnoncesbc&\coloneqq\left\{ \n_{\vecib} \middle| \n\left[ \vec\_ \right]\in\mnonces, \vecib\in\domI^{\marity{\vec\_}} \right\}\\
        \mfunctionsbc&\coloneqq\left\{ f_{\vecib } \middle| f\left[ \vec\_ \right]\in\mfunctions, \vecib\in\domI^{\marity{\vec\_}} \right\}\\
        \mattsbc&\coloneqq\left\{ \matt \right\}
    \end{align}
\end{mdefenv}
    \section{Base Axioms}\label{sec:base-axioms}
\begin{figure}
    \begin{multicols}{2}
        \allowdisplaybreaks
        \belowdisplayskip=0.67\belowdisplayskip
        \begin{gather}
            \meval{A \bwedge B} \approx \meval{A} \wedge \meval{B}\label{eq:base-axioms:1}\\
            \meval{A \bvee B} \approx \meval{A} \vee \meval{B}\label{eq:base-axioms:2}\\
            \meval{\bneg A}  \approx \neg \meval{A}\label{eq:base-axioms:3}\\
            \meval{\btop} \approx \top\label{eq:base-axioms:4}\\
            \meval{\bbot} \approx \bot\label{eq:base-axioms:5}\\
            \meval{ \bexists \vecalpha.\ t} \approx \exists \vecalpha.\ \meval{t}\label{eq:base-axioms:6}\\
            \meval{\bforall \vecalpha.\ t} \approx \forall \vecalpha.\ \meval{t}\label{eq:base-axioms:7}
        \end{gather}
    \end{multicols}%
    \vspace{-0.7cm}\begin{equation}
        \meval{A \bimplies B} \approx \meval{A} \Rightarrow \meval{B}\label{eq:base-axioms:8}
    \end{equation}
    Where $A\approx B$ stands for $\modelc^\eta\left( A \right)\left( \rho \right)=\modelc^\eta\left( B \right)\left( \rho \right)$ for all $\eta$ and $\rho$.
    \caption{Collection of valid boolean axioms}\label{fig:sound-b-axioms}
\end{figure}

\begin{mnotationenv}{}{}
    In this section, we use $\alpha$ and $\beta$ to refer to indices and/or timepoints when the distinction between the two is irrelevant.
\end{mnotationenv}

\begin{mprop}\label{prop:base-axioms}
    In any model $\mifthenelse{\_}{\_}{\_}$, $\btop$ and $\bbot$ have the function evaluation that we would usually expect, the equations of \autoref{fig:sound-b-axioms} hold.
\end{mprop}

\begin{mproof}
    Let $\modelc$ be any model in which $\mifthenelse{\_}{\_}{\_}$, $\btop$ and $\bbot$ have the function evaluation that we would usually expect. Then equations \ref{eq:base-axioms:1}-\ref{eq:base-axioms:5} and \ref{eq:base-axioms:8} hold using a truth table on $\mevalbc{\mevalT{A}}$ and $\mevalbc{\mevalT{B}}$.
    \begin{enumerate}
        \item[\eqref{eq:base-axioms:6}] If $\modelc^\eta\left( \exists \alpha.\ \meval{t} \right)\left( \rho \right)=1$ with $\alpha=\mfv(t)$, let $\alphab_0$ be an element such that $\modelc^\eta\left( \meval{\msubstit{t}{\alpha}{\alphab_0}} \right)\left( \rho \right)=1$ and $\betab$ an element from $\mathbb{T}$ (i.e., an element of $\domI$ or $\domSE$) that unfolds like $\alphab_0$ (there has to be one, as unfolding does not change member of $\mathbb{T}$).
        Then $\modelc^\eta\left( \meval{\msubstit{t}{\alpha}{\betab}} \right)\left( \rho \right)=1$. We conclude by the construction of the symbolic existential quantifier (see \eqref{eq:def-exitsts}) and the unfolding of lookups.

        Else $\modelc^\eta\left( \exists \alpha.\ \meval{t} \right)\left( \rho \right)=0$ and the result is trivial.
        \item[\eqref{eq:base-axioms:7}] Consequence of \eqref{eq:base-axioms:6}.
    \end{enumerate}
\end{mproof}
This serves as a proof of \autoref{prop:conn-commutes}.

    \section{Soundness Theorems}\label{sec:soundness-th}
In this section, we formally define the notion of formulas with bounded Skolem normal form and proceed to prove \autoref{th:main}.

\begin{mnotationenv}{}{}
    In this section, we use $\alpha$ and $\beta$ to refer to indices and/or timepoints when the distinction between the two is irrelevant.

    We write $\valid \phi$ to say that $\phi$ is valid in FOL, i.e., $\phi$ is satisfied by any model.
\end{mnotationenv}

\subsection{Some definitions}
Let us begin by introducing some definitions. For conciseness, let $\Ez\coloneqq\mevall$. We define $\mSk$ as a set of Skolem symbols, and \mevallskinner{\Ez}{\mskolems} as the logic extending \Ez with those Skolem symbols.

\begin{mdefenv}{Sequences of Sets of Random Tapes}{sequence-of-sets-of-random-tapes}
    We write $\Omega^\mathbb{N}$ as the set of sequences of $\Omega$. We also define the ring $\left( \Omega, \symdiff, \cap \right)$ as follows:
    \begin{enumerate}[ref=(\arabic{enumi})]
        \item $\left( A_\eta \right)_{\eta\in \mathbb{N}}\symdiff\left( B_\eta \right)_{\eta\in \mathbb{N}}\coloneqq \left( A_\eta\symdiff B_\eta \right)_{\eta\in \mathbb{N}}$;
        \item $\left( A_\eta \right)_{\eta\in \mathbb{N}}\cap\left( B_\eta \right)_{\eta\in \mathbb{N}}\coloneqq \left( A_\eta\cap B_\eta \right)_{\eta\in \mathbb{N}}$;
    \end{enumerate}
    where $\symdiff$ is the symmetric difference.
    We also include the union $\cup$ as $A\cup B\coloneqq A\symdiff B \symdiff \left( A \cap B \right)$.
\end{mdefenv}

\begin{mdefenv}{Negligible Family}{negligible-family}
    A family (or sequence) $\left( \mathfrak{S}_{\eta} \right)_{\eta\in\mathbb{N}}$ of subsets of $\Omega$ is \emph{negligible} if 
    $
        \mprob\left( \mathfrak{S}_{\eta} \right)=\mnegl\left( \eta \right)
    $. We write $\Omega_{\mnegl}$ the set of such sequences.
\end{mdefenv}

\begin{mdefenv}{Skolem Model}{skolem-model}
    A Skolem model \modelsk maps terms of \mevallskinner{\Ez}{\mskolems} to \Ez such that it is the identity on every symbol except those of \mskolems.
\end{mdefenv}

\begin{mdefenv}{Finite Term}{finite-term}
    A \emph{domain function} $\domterm$ over \mskolems is a function from \mevallskinner{\Ez}{\mskolems} to the finite sets of \Ez such that for all $\ub=\genf\left( \vecvb \right)$ that is not of the form $\msk\left( \_ \right)$ (with $\msk\in\mSk$), we have $\domterm\left( \genf\left( \vecvb \right) \right)=\left\{ \genf\left( \vecub \right) \middle| \vecub \in \domterm\left( \vecvb \right) \right\}$ where $\genf$ stands for any application-like construction (i.e., anything but constant symbols).
\end{mdefenv}

\begin{mdefenv}{Extended Cryptographic Model}{extended-cryptographic-model}
    An \emph{extended cryptographic model} $\modelec$ is composed of
    \begin{itemize}
        \item a cryptographic model $\modelc$;
        \item a family of Skolem models $\left( \modelsk^{\eta,\rho} \right)_{\eta,\rho}$;
        \item a domain function $\domterm$ over \mskolems.
    \end{itemize}
    such that
    \begin{enumerate}[ref=(\arabic{enumi})]
        \item\label{item:def:extended-cryptographic-model:1} $\modelec^\eta\left( \meval{t} \right)\left( \rho \right)=\modelc^\eta\left( \meval{\modelsk^{\eta,\rho}\left( t \right)} \right)\left( \rho \right)$
        \item\label{item:def:extended-cryptographic-model:2} $\modelec^\eta\left( S \right)\left( \rho \right)=\modelc^\eta\left( \modelsk^{\eta, \rho}\left( S \right) \right)$
        \item\label{item:def:extended-cryptographic-model:3} for all term $u$, we have 
        \begin{equation}
            \modelec^\eta\left( \bigvee\nolimits_{\substack{v\in \domterm\left( u \right),\, \vecbeta = \mfv\left( v \right)}}\exists \vecbeta.\  v = u\right)\left( \rho \right)=1
        \end{equation}
        where $\vecbeta$ is a list of variables over timepoints and/or indices.
    \end{enumerate}
\end{mdefenv}

\begin{mdefenv}{Model Extension}{model-extension}
    Let $L$ be a model over a logic $\mathcal{E}$. A model $K$ over $\mathcal{E}'$ \emph{extends} $L$ when $\mathcal{E}\subseteq \mathcal{E}'$ and for all $\eta$, $\rho$, and $\varphi\in \mathcal{E}$, we have $L^\eta\left( \varphi \right)\left( \rho \right) = K^\eta\left( \varphi \right)\left( \rho \right)$. We write it $L \submodel K$. We also say that $K$ \emph{covers} $L$.
\end{mdefenv}



\begin{mdefenv}{Formula with Bounded Skolem Normal Form}{formula-bounded-snf}
    A formula $\varphi\in \Ez$ with $\vecu=\mfv\left(\varphi\right)$ has a \emph{bounded Skolem normal form} if we can find a finite \mskolems and a domain function $\domterm$ over \mskolems such that:
    \begin{enumerate}[ref=(\arabic{enumi})]
        \item\label{item:def:acceptable-set-of-formulas:1}
            the Skolem normal form $\forall \vecu,\vecv.\ \varphi_{\msk}$ of $\forall \vecu.\ \varphi$ uses only symbols of $\mevallskinner{\Ez}{\mSk}$;
        \item\label{item:def:acceptable-set-of-formulas:2}
            for all cryptographic model \modelc, we can find an extended cryptographic model $\modelec^\modelc$ using $\domterm$ such that:
            \begin{enumerate}
                \item\label{item:def:acceptable-set-of-formulas:2:1}
                    $\modelc \submodel \modelec^\modelc$;
                \item\label{item:def:acceptable-set-of-formulas:2:2}
                    for all $\vecub\in \Ez$,
                    \begin{multline}
                        \text{if }\modelc^\eta\left( \msubstit{\varphi}{\vecu}{\vecub} \right)\left( \rho \right)=1\\\text{ then }\modelec^\eta\left(  \forall\vecv.\ \msubstit{\varphi_{\msk}}{\vecu}{\vecub} \right)\left( \rho \right)=1
                    \end{multline}
            \end{enumerate}
    \end{enumerate}
\end{mdefenv}

\subsection{Some properties}
\begin{mpropenv}{Model Extension as Order}{model-extension-as-order}
    $\submodel$ is a partial order over models and a preorder over sets of models.
\end{mpropenv}

\begin{mpropenv}{}{the-ideal-of-negligible-sequences}
    $\Omega_{\mnegl}$ is an ideal of $\left( \Omega^{\mathbb{N}}, \symdiff, \cap \right)$.
\end{mpropenv}
\begin{mdemoenv}{}
    \begin{enumerate}[ref=(\arabic{enumi})]
        \item Let $\left( A_\eta \right)_{\eta\in \mathbb{N}}, \left( B_\eta \right)_{\eta\in\mathbb{N}}\in \Omega_{\mnegl}$:
        \begin{multline*}
            \mprob\left( A_\eta\symdiff B_\eta \right)\\
                =\mprob\left( A_\eta \right)+\mprob\left( B_\eta \right)-2\mprob\left( A_\eta\cap B_\eta \right)\\
                \leq \mprob\left( A_\eta \right)+\mprob\left( B_\eta \right) = \mnegl\left( \eta \right)
        \end{multline*}
        \item Let $\left( A_\eta \right)_{\eta\in \mathbb{N}}\in\Omega^\mathbb{N}$ and $\left( B_\eta \right)_{\eta\in\mathbb{N}}\in \Omega_{\mnegl}$:
        \[
            \mprob\left( A_\eta \cap B_\eta \right)\leq \mprob\left( B_\eta \right)=\mnegl\left( \eta \right)
        \]
    \end{enumerate}
\end{mdemoenv}

\begin{figure}
    \input{101-figures/010-sequent-rules.tex}
    \caption{Some Sequent Rules}\label{fig:sequent-rules}
\end{figure}

\begin{mpropenv}{}{sequent-rules}
    The rules in \autoref{fig:sequent-rules} are sound.
\end{mpropenv}
\begin{mproof}
    Consequences of \autoref{prop:the-ideal-of-negligible-sequences}.
\end{mproof}

\ifreport
    \begin{mlemmeenv}{}{union-acceptable}
    If $\Delta$ and $\Gamma$ have a bounded Skolem normal form, then we can find a $\mSk$ and $\domterm$ such that both formulas' Skolem normal forms are in \mevallskinner{\Ez}{\mSk} with $\domterm$, and we can make use of the same $\modelec$.
\end{mlemmeenv}
\begin{mdemoenv}{}
    $\Delta$ has a bounded Skolem normal form with $\mSk_\Delta$ and $\domterm_\Delta$, and $\Gamma$ with $\mSk_\Gamma$ and $\domterm_\Gamma$. W.l.o.g. we can assume $\mSk_\Delta\cap \mSk_\Gamma=\varnothing$. Let $\mSk=\mSk_\Delta\cup\mSk_\Gamma$, $\vecu=\mfv\left( \Delta\right)$, and  $\vecv=\mfv\left( \Gamma\right)$.

    We define:
    \begin{equation}
        \domterm\left( \ub \right)=\begin{cases}
            \begin{multlined}[6cm]
                \bigcup\nolimits_{\vecvb\in\domterm\left( \vecub \right)}\domterm_\Delta\left( \msk[\Delta]\left( \vecvb \right)\right)\vspace*{-10pt}\\
                \text{if }\ub=\msk[\Delta]\left( \vecub \right)\text{ with }\msk\in\mSk_\Delta\vspace*{3pt}
            \end{multlined}\\
            \begin{multlined}[6cm]
                \bigcup\nolimits_{\vecvb\in\domterm\left( \vecub \right)}\domterm_\Gamma\left( \msk[\Gamma]\left( \vecvb \right)\right)\vspace*{-10pt}\\
                \text{if }\ub=\msk[\Gamma]\left( \vecub \right)\text{ with }\msk\in\mSk_\Gamma\vspace*{3pt}
            \end{multlined}\\
            \begin{multlined}
                \left\{ \genf\left( \vecub \right) \middle| \vecub\in\domterm\left( \vecvb \right) \right\}\text{ if }\ub=\genf\left( \vecvb \right)
            \end{multlined}
        \end{cases}
    \end{equation}

    Let us show that $\Delta$ has a bounded Skolem normal form with $\mSk_\Delta\cup\mSk_\Gamma$ and $\domterm$:
    \begin{enumerate}
        \item because $\mevallskinner{\Ez}{\mSk_\Delta}\subseteq \mevallskinner{\Ez}{\mSk_\Delta\cup\mSk_\Gamma}$
        \item Let \modelc be a cryptographic model and $\modelec_\Delta$ and $\modelec_\Gamma$ its extension according to \autoref{def:formula-bounded-snf}'s~\ref{item:def:acceptable-set-of-formulas:2}.
        Let us build $\modelec$ such that $\modelec_\Delta\submodel \modelec$ and $\modelec_\Gamma\submodel \modelec$. We choose $\mrestriction{\modelsk^\modelec}{\mSk_\Delta}=\mrestriction{\modelsk^{\modelec_\Delta}}{\mSk_\Delta}$ and $\mrestriction{\modelsk^\modelec}{\mSk_\Gamma}=\mrestriction{\modelsk^{\modelec_\Gamma}}{\mSk_\Gamma}$. \modelec is fully defined.
        
        Moreover, by induction, we show that $\modelec$ follows \autoref{def:extended-cryptographic-model}'s~\ref{item:def:extended-cryptographic-model:3} and \autoref{def:model-extension}. Thus,
        \begin{equation}\label{eq:proof:prop:union-acceptable}
            \modelec_\Delta\submodel \modelec\text{ and }\modelec_\Gamma\submodel \modelec
        \end{equation}
        
        By \autoref{prop:model-extension-as-order} we get \autoref{def:formula-bounded-snf}'s~\ref{item:def:acceptable-set-of-formulas:2:1}, and \eqref{eq:proof:prop:union-acceptable} gives us \autoref{def:formula-bounded-snf}'s~\ref{item:def:acceptable-set-of-formulas:2:2}.
    \end{enumerate}

    By symmetry, we also get the result for $\Gamma$.
\end{mdemoenv}
\smallskip

\begin{mprop*}[\ref{prop:bsnf-and-or}]
    The notion of bounded Skolem normal form is stable by conjunction and disjunction.
\end{mprop*}
\begin{mdemoenv}{}
    Let $\Delta$ and $\Gamma$ be two formulas with a bounded Skolem normal form using $\mSk$ and $\domterm$ (as per \autoref{lemme:union-acceptable}) and $\vecu$ and $\vecv$ be their respective free variables. Let $\star$ stand for $\vee$ or $\wedge$.

    Let $\forall\vecu,\vecu'.\ \Delta_{\msk}$ and $\forall\vecv,\vecv'.\ \Gamma_{\msk}$ be the Skolem normal form of $\Delta$ and $\Gamma$ respectively as defined in \autoref{def:formula-bounded-snf}'s~\ref{item:def:acceptable-set-of-formulas:1}.

    \begin{enumerate}
        \item $\forall\vecu,\vecu',\vecv,\vecv'. \Delta_{\msk}\star\Gamma_{\msk}$ is a Skolem normal form of  $\forall\vecu,\vecv. \Delta\star\Gamma$ and is in \mevallskinner{\Ez}{\mSk}.
        \item Let \modelc be a cryptographic model and \modelec its extension as defined in \autoref{def:formula-bounded-snf}.
        \begin{enumerate}
            \item $\modelc \submodel \modelec$ by definition.
            \item For all $\vecub$ and $\vecvb$, \eqref{eq:proof:prop:bsnf-and-or} is valid in FOL.
            \begin{multline}\label{eq:proof:prop:bsnf-and-or}
                \left( \begin{multlined}
                    \left( \msubstit{\Delta}{\vecu}{\vecub}\implies \forall\vecu'.\ \msubstit{\Delta_{\msk}}{\vecu}{\vecub} \right)\\
                    \wedge\left( \msubstit{\Gamma}{\vecv}{\vecvb} \implies \forall\vecv'.\ \msubstit{\Gamma_{\msk}}{\vecv}{\vecvb} \right)
                \end{multlined} \right)\\
                \implies \left(
                    \begin{multlined}
                        \msubstit{\left( \Delta \star \Gamma \right)}{\vecu,\vecv}{\vecub,\vecvb}\\
                        \implies \forall\vecu',\vecv'.\ \msubstit{\left(\Delta_{\msk}\star\Gamma_{\msk}\right)}{\vecu,\vecv}{\vecub,\vecvb}
                    \end{multlined}
                \right)
            \end{multline}
            We conclude using \nameref{rule:implies} and the definition of $\submodel$.
        \end{enumerate}
    \end{enumerate}
\end{mdemoenv}

Before proving \autoref{prop:existential-quantifiers-and-acceptability}, we first prove this convenient property:

\begin{mpropenv}{Model Extension}{model-extension}
    If $\modelc$ and $\modelec$ agree on all quantifier-free terms of \Ez, then $\modelc\submodel \modelec$.
\end{mpropenv}
\begin{mdemoenv}{}
    By induction over $\varphi$.
    \begin{enumerate}[wide, label=\textbf{\arabic{enumi})},ref=\arabic{enumi}]
        \item The base cases are trivial (they are quantifier-free formulas).
        \item\label{demo:prop:model-extension:easy} $\varphi\vee\varphi'$ and $\neg\varphi$ are also trivial due to the semantics of $\modelc$ and $\modelec$.
        \item\label{demo:prop:model-extension:exists} If $\varphi$ is of the form $\exists u.\ \varphi'$, we then suppose that the property holds for $\msubstit{\varphi'}{u}{\ub}$ when $\ub\in \Ez$ ($\ub$ is quantifier free).
        
        If $\modelc^\eta\left( \varphi \right)\left( \rho \right)=1$ then $\modelec^\eta\left( \varphi \right)\left( \rho \right)=1$ as a term of \Ez is also a term of \mevallskinner{\Ez}{\mskolems}. If $\modelc^\eta\left( \varphi \right)\left( \rho \right)=0$ then let us assume that $\modelec^\eta\left( \varphi \right)\left( \rho \right)=1$. It means that there is a term $\ub$ of \mevallskinner{\Ez}{\mskolems} such that $\modelec^\eta\left( \varphi'\left[ \ub \right] \right)\left( \rho \right)=1$.
        
        By \autoref{def:extended-cryptographic-model}'s~\ref{item:def:extended-cryptographic-model:3} and superposition we can find a $\vb\in\domterm\left( \ub \right)$ such that $\modelec^\eta\left( \exists\vecalpha.\ \msubstit{\varphi'}{u}{\vb} \right)\left( \rho \right)=1$ and $\vecalpha=\mfv\left(\vb\right)$. Thus, we can find a $\vecalphab$ such that $\modelec^\eta\left( \msubstit{\varphi'}{u}{\vb_{\vecalphab}} \right)\left( \rho \right)=1$ where $\vb_{\vecalphab}\coloneqq \msubstit{\vb}{\vecalpha}{\vecalphab}$.
        
        Then let $\vecbetab=\modelsk\left( \vecalphab \right)$, we still have $\modelec^\eta\left( \msubstit{\varphi'}{u}{\vb_{\vecbetab}} \right)\left( \rho \right)=1$ with $\msubstit{\varphi'}{u}{\vb_{\vecbetab}}\in\Ez$ and $\vb_{\vecbetab}\coloneqq \msubstit{\vb}{\vecalpha}{\vecbetab}$. This contradicts the induction hypothesis.
        \item $\varphi$ is a universal quantifier. The result is a consequence of \ref{demo:prop:model-extension:easy} and \ref{demo:prop:model-extension:exists}.
    \end{enumerate}
\end{mdemoenv}

\begin{mprop*}[\ref{prop:existential-quantifiers-and-acceptability}]
    If $\varphi$ is bounding and has a bounded Skolem normal form, then $\exists x.\ \varphi$ has a bounded Skolem normal form.
\end{mprop*}
\begin{mdemoenv}{}
    Let $x\uplus\vecu=\mfv(\varphi)$ and $\domterm_0$ and $\mSk_0$ be what $\forall x, \vecu.\ \varphi$ has a bounded Skolem normal form with. Let $\modelc$ be a cryptographic model and $\modelec_0=\left( {\modelsk}^{(0)},\dots \right)$ its extension as defined in \autoref{def:formula-bounded-snf} for $\varphi$.

    \begin{enumerate}[wide, label=\textbf{\arabic{enumi})},ref=\arabic{enumi}]
        \item We define $\mSk\coloneqq\mSk_0\cup\left\{ \msk\left( \__1,\dots,\__n \right) \right\}$ such that $\msk\not\in\mSk$ and $\marity{\vecu}=n$.
        Then $\forall \vecu.\ \exists x.\ \varphi$ can be skolemed in $\mevallskinner{\Ez}{\mSk}$.
        Indeed, let $\forall\vecv.\ \varphi_{\msk}$ be a Skolem normal form of $\forall x,\vecu.\ \varphi$, then $\forall \vecu. \msubstit{\varphi}{x}{\msk\left(\vecu\right)}$ is a Skolem normal form of $\forall \vecu.\ \exists x.\ \varphi$.
        
        \item Let us build $\modelec$, an extension of $\modelc$ as in \autoref{def:formula-bounded-snf}'s~\ref{item:def:acceptable-set-of-formulas:2}.
        We define:
        \begin{equation}
            \domterm\left( \msk'\left( \vecu \right) \right)=\begin{cases}
                D^{\vecub}_{\varphi}\cup\left\{ \mfail \right\}&\text{if }\msk'=\msk\\
                \bigcup_{\vecvb\in\domterm\left( \vecub \right)}\domterm\left( \msk'\left( \vecvb \right) \right)&\text{otherwise}
            \end{cases}
        \end{equation}
        the rest of $\domterm$ is defined recursively by \autoref{def:finite-term}. Notice that $\domterm$ is $\domterm_0$ over $\mevallskinner{\Ez}{\mSk_0}$.

        Since $\modelc^\eta\left( \_ \right)\left( \rho \right)$ is a first-order model we can construct the set:
        \begin{equation}
            SK^{\eta, \rho}_{\vecub}\coloneqq \begin{cases}
                \begin{multlined}
                    \left\{ t' \middle| \modelc^\eta\left( \msubstit{\varphi}{x,\vecu}{t',\vecub} \right)\left( \rho \right)=1 \right\}
                    \vspace*{-10pt}\\
                    \text{if }\modelc^\eta\left( \exists x.\ \msubstit{\varphi}{\vecu}{\vecub} \right)\left( \rho \right)=1\vspace*{3pt}
                \end{multlined}\\
                \left\{ \mfail \right\}\text{ otherwise}
            \end{cases}
        \end{equation}
        By noticing that we can find an ordering $\prec_{\Evalname}$ over the terms such that it has a smallest element (e.g., a KBO~\cite{kbo}), we build
        \[
            \modelsk^{\eta, \rho}\left( \phi \right)\coloneqq\begin{cases}
                \min\nolimits_{\prec_{\Evalname}}\left( SK^{\eta, \rho}_{\modelsk^{\eta, \rho}\left( \vecub\right)} \right)&\text{if }\phi=\msk\left( \vecub \right)\\
                {\modelsk^{\eta, \rho}}^{(0)}\left( \phi \right)&\text{otherwise}
            \end{cases}
        \]
        notice that $\modelsk$ is well defined as the number of $\msk$ strictly decreases at each recursive call.
        $\modelsk$ is a Skolem model that matches ${\modelsk}^{(0)}$ over $\mevallskinner{\Ez}{\mSk_0}$.

        Let $\modelec\coloneqq\left( \domterm, \modelsk,\dots\modelec_0 \right)$. We note that $\modelec$ is indeed an extended model: it verifies \autoref{def:extended-cryptographic-model}'s~\ref{item:def:extended-cryptographic-model:3} because $\varphi$ is bounding. Indeed, reusing \autoref{def:bounding-formula}'s notations, we have:
        \begin{equation}
            SK^{\eta, \rho}_{\vecub}\subseteq D^{\vecub}_{\varphi}\cup\left\{ \mfail \right\}\text{ (up to $\alpha$-renaming)}
        \end{equation}
        
        Then
        \begin{enumerate}
            \item It agrees with $\modelec_0$ all ground terms on $\mevallskinner{\Ez}{\mSk_0}$ (all its components are the same on over that set), thus it agrees with $\modelc$ on all ground terms of \Ez.
            Then by \autoref{prop:model-extension} we have $\modelc\submodel\modelec$;
            \item let $\vecub\in\Ez$, let us assume we have $\modelc^\eta\left( \exists x.\ \msubstit{\varphi}{\vecu}{\vecub} \right)\left( \rho \right)=1$.
            By construction, we can find $t\in SK^{\eta, \rho}_{\vecub}$ such that $\modelc^\eta\left( \msubstit{\varphi}{x,\vecu}{t,\vecub} \right)\left( \rho \right)=1$.
            Also, by construction, we can suppose $\modelsk^{\eta, \rho}\left( \msk\left( \vecub \right) \right)=\modelsk^{\eta, \rho}\left( t \right)$. Which lets us conclude.
        \end{enumerate}
    \end{enumerate}
\end{mdemoenv}

\else
    \autoref{prop:bsnf-in-base-logic}, \autoref{prop:bsnf-and-or} and \autoref{prop:existential-quantifiers-and-acceptability} are proven in the technical report~\cite{CRYPTOVAMPIREAutomatedReasoningTR}.
\fi
\subsection{The Theorems}
This section provides a proof of \autoref{th:main}.

\begin{mth*}[\ref{th:main}]
    Let $\varphi$ be a formula s.t. $\mfv\left(\varphi\right)=\vecu$, $t$ a term, and $\modelclass$ a set of cryptographic models such that
    \begin{enumerate}[ref=(\arabic{enumi})]
        \item $\varphi$ has a bounded Skolem normal form;
        \item for all $\vecvb$, $\modelclass\validp \msubstit{\varphi}{\vecu}{\vecvb}$;
        \item $\left( \forall\vecu.\ \varphi \right)\implies \meval{t}$ is valid in \emph{FOL};
    \end{enumerate}
    then $\modelclass\validp \meval{t}$.
\end{mth*}

\begin{mex*}[\ref{ex:no-guessing-rejected}]
    Recall \autoref{ex:no-guessing-rejected}. \Ez rejects \eqref{eq:ex:no-guessing-rejected}.
    \begin{equation}\tag{\ref{eq:ex:no-guessing-rejected}}
        \forall \veci, x.\ \meval{\n\left[ \veci \right]} = \meval{x} \Rightarrow \n\left[ \veci \right]\sqsubseteq x
    \end{equation}
\end{mex*}

At first glance, it seems we can recover \autoref{ex:no-guessing-rejected} using Herbrand's Theorem~\cite{Herbrand1930}. We indeed can show the following lemma:

\begin{mtheoremenv}{}{relation-to-fol}
    Let $\varphi$ be a formula of \Ez with $\vecu\coloneqq \mfv\left(\varphi\right)$, $t$ be a term of \Ez and $\modelclass$ a class of cryptographic models, if we can find $\modelclass'$ and \mskolems such that
    \begin{enumerate}[ref=(\arabic{enumi})]
        \item\label{th:relation-to-fol:1} $\forall \vecu,\vecv.\ \varphi_{\msk}$ is a Skolem normal form of $\forall \vecu.\ \varphi$ and is part of \mevallskinner{\Ez}{\mSk};
        \item\label{th:relation-to-fol:2} $\modelclass\submodel\modelclass'$;
        \item\label{th:relation-to-fol:3} for all $\vecub\in \mevallskinner{\Ez}{\mSk}$, $\modelclass'\validp \forall \vecv.\ \msubstit{\varphi_{\msk}}{\vecu}{\vecub}$;
        \item\label{th:relation-to-fol:4} $\valid \left( \forall\vecu.\ \varphi\right) \Rightarrow \meval{t}$ (in FOL);
    \end{enumerate}
    then $\modelclass \validp \meval{t}$.
\end{mtheoremenv}
\begin{mproof}
    \ref{th:relation-to-fol:1} and \ref{th:relation-to-fol:4} give us that $\left( \forall\vecu,\vecv.\ \varphi_{\msk} \right) \wedge \neg\meval{t}$ is unsatisfiable.

    Then Herbrand's theorem~\cite{Herbrand1930} gives us $\vecub_1,\dots,\vecub_n,\allowbreak\vecvb_1,\dots,\vecvb_n$ such that $\left(\bigwedge_{k=1}^n\msubstit{\varphi_{\msk}}{\vecu, \vecv}{\vecub_k,\vecvb_k} \right) \wedge\neg\meval{t}$ is unsatisfiable. Thus
    \[
        \valid\left(\bigwedge\nolimits_{k=1}^n\forall \vecv.\ \msubstit{\varphi_{\msk}}{\vecu}{\vecub_k} \right) \implies\meval{t}
    \]
    We conclude using \ref{th:relation-to-fol:2}, \ref{th:relation-to-fol:3}, and \autoref{prop:sequent-rules}.
\end{mproof}

Unfortunately, \autoref{th:relation-to-fol:3} is tricky to show as we have very little control over the interpretation of the $\ub$ if it contains Skolems. We use the class of models from \autoref{def:extended-cryptographic-model} and formula with bounded Skolem normal to regain control:

\begin{mtheoremenv}{}{extending-sat}
    Reusing the notation of \autoref{def:formula-bounded-snf}.
    Let $\varphi\in \Ez$ be a formula with a bound Skolem normal form and \modelc a cryptographic models with $\vecu\coloneqq \mfv(\varphi)$ such that for all ground $\vecub\in\Ez$ we have $\modelc\validp \msubstit{\varphi}{\vecu}{\vecub}$.
    Then for all $\vecub'\in\mevallskinner{\Ez}{\mSk}$ we have $\modelec^\modelc\validp \forall\vecv.\, \msubstit{\varphi_{\msk}}{\vecu}{\vecub'}$.
\end{mtheoremenv}

\begin{mlemmeenv}{}{quantifiers}
    Let $\modelec$ be as in \autoref{def:extended-cryptographic-model},
    \begin{gather}
        \modelec^\eta\left( \bigwedge\nolimits_{\alpha\in\mathbb{T}} \varphi \right)\left( \rho \right) = \modelec^\eta\left( \forall\alpha.\ \varphi \right)\left( \rho \right)\label{eq:lemma:quantifiers:forall}\\
        \modelec^\eta\left( \bigvee\nolimits_{\alpha\in\mathbb{T}} \varphi \right)\left( \rho \right) = \modelec^\eta\left( \exists\alpha.\ \varphi\right)\left( \rho \right)\label{eq:lemma:quantifiers:exists}
    \end{gather}
    where $\alpha$ is an index or a timepoint. This naturally extends to quantification over multiple variables.
\end{mlemmeenv}
\begin{mdemoenv}{}
    Let $\varphi$ be a formula and $\alpha$ its free variable.

    \begin{itemize}
        \item[\textbf{\eqref{eq:lemma:quantifiers:forall}}]
        If $\modelec^\eta\left( \forall\alpha.\ \varphi\left[ \alpha \right]\right)\left( \rho \right)=0$, then we can find $\alphab$ such that $\modelec^\eta\left(\msubstit{\varphi}{\alpha}{\alphab}\right)\left( \rho \right)=0$.
        However, the unfolding $\betab$ of $\modelsk^{\eta,\rho}\left( \alphab \right)$ is in $\mathbb{T}$. Thus, by construction of the unfolding and $\modelec$, we get $\modelec^\eta\left(\msubstit{\varphi}{\alpha}{\betab} \right)\left( \rho \right)=0$. Hence, the equality.
        
        Else $\modelec^\eta\left( \forall\alpha.\ \varphi\right)\left( \rho \right)=1$ and the result is trivial.
        \item[\eqref{eq:lemma:quantifiers:exists}] Using \eqref{eq:lemma:quantifiers:forall}.
    \end{itemize}
\end{mdemoenv}

\begin{mdemoenv}{th:extending-sat}
    Let $\vecub'\in\mevallskinner{\Ez}{\mSk}$ and $\vecub\in\domterm\left( \vecub' \right)$ with $\vecalpha\coloneqq\mfv\left(\vecub\right)$. We also use notations
    \begin{align}
        \vecub_{\vecbetab}&\coloneqq \msubstit{\vecub}{\vecalpha}{\vecbetab}\\
        \Pi_{\vecub}&\coloneqq\bigwedge\nolimits_{\vecalphab\in\mathbb{T}} \forall \vecv.\ \msubstit{\varphi_{\msk}}{\vecu}{\vecub_{\vecalphab}}\\
        \Gamma&\coloneqq\bigwedge\nolimits_{\substack{\vecub\in\domterm\left( \vecub' \right)\\\vecalpha=\mfv\left(\vecub\right)}} \forall \vecalpha,\vecv.\ \msubstit{\varphi_{\msk}}{\vecu}{\vecub}\\
        \Delta&\coloneqq\bigvee\nolimits_{\substack{\vecvb\in\domterm\left( \vecub' \right)\\\vecalpha=\mfv\left(\vecvb\right)}} \exists \vecalpha.\ \vecvb = \vecub'
    \end{align}
    
    By assumption, for all $\vecalphab\in\Ez$ we have $\modelc\validp \msubstit{\varphi}{\vecu}{\vecub_{\vecalphab}}$. Thus, by \autoref{def:formula-bounded-snf}'s~\ref{item:def:acceptable-set-of-formulas:2:2}, $\modelec^\modelc\validp \forall \vecv.\ \allowbreak\msubstit{\varphi_{\msk}}{\vecu}{\vecub_{\vecalphab}}$.

    Then, rule \nameref{rule:and}, gives us $\modelec^\modelc\validp \Pi_{\vecub}$.
    Finally, using \autoref{lemme:quantifiers}, we get $\modelec^\modelc\validp \forall \vecalpha,\vecv.\ \msubstit{\varphi_{\msk}}{\vecu}{\vecub_{\vecalphab}}$.

    Then, rule \nameref{rule:and}, gives us $\modelec^\modelc\validp \Gamma$.
    And we also have by construction of $\modelec^\modelc$ (\autoref{def:extended-cryptographic-model}'s~\ref{item:def:extended-cryptographic-model:3}) that $\modelec^\modelc\validp \Delta$.

    Moreover $\valid \Gamma\wedge\Delta\implies \forall\vecv.\, \msubstit{\varphi_{\msk}}{\vecu}{\vecub'}$. Thus we can conclude using \nameref{rule:implies} and \nameref{rule:and}.
\end{mdemoenv}

\smallskip
Then \autoref{th:main} is the result of chaining \autoref{th:extending-sat} and \autoref{th:relation-to-fol}.

    \section{Subterm definition}\label{sec:subterm-appendix}

\begin{figure}
    \input{101-figures/102-subterm.tex}
    \caption{Subterm sets}\label{fig:subterm-sets}
\end{figure}

We define the overapproximated sets of subterms $\mmst\left( t \right)$ for a protocol $\mathcal{P}$ used in \autoref{def:subterm} in \autoref{fig:subterm-sets}.

Let $\mathfrak{F}$ be a set of functions \enquote{ignored} by $\mmst\left( t \right)$ and $\mathfrak{H}$ the set of terms using function as their head.
\begin{align}
    \mathfrak{F}&\coloneqq\left\{ 
        \mifthenelse{\_}{\_}{\_}
        \ll\_;\_\gg, \emptyset
    \right\}\\
    \mathfrak{H}&\coloneqq \left\{ f\left( \vec t \right)\middle| f\in\mathfrak{F} \right\}
\end{align}

\newcommand{\expsubt}[1]{\boxed{#1}_{\modelc}^{\eta,\rho}}
\begin{mnotationenv}{}{subterm-set-expansion}
    Let $S$ be a set of $\Ez^2$, we write
    \begin{equation}
        \expsubt{S}\coloneqq \left\{ \mevalT{\msubstit{t}{\veci}{\sigma\left(\veci\right)}}\middle| 
            \begin{gathered}
                (\phi, t)\in S,\\
                \veci=\mfv\left(\phi\right)\cup\mfv\left(t\right),\\
                \modelc^\eta\left( \msubstit{\phi}{\veci}{\sigma\left(\veci\right)} \right)\left( \rho \right)=1 
            \end{gathered}\right\}
    \end{equation}

    Where the $\sigma$ are assignments of the variables.
\end{mnotationenv}

For a BC~term $t$, let $\mstbc\left( t \right)$ be the set of its subterms.

In this section, we will focus on proving our claim that $\sqsubseteq$ really produces an overapproximation of $\mstbc\left( \_ \right)$. We achieve this by showing \autoref{th:subterm-soundness}.

\begin{mtheoremenv}{}{subterm-soundness}
    Let $t\in\Ez$, and $\modelc$ a cryptographic model
    \begin{equation}
        \mstbc\left( \mevalT{t} \right)\backslash\mathfrak{H}\subseteq\expsubt{\mmst\left( t \right)}
    \end{equation}
\end{mtheoremenv}

\ifreport
    
\begin{figure}
    \input{101-figures/103-subterm-tmp.tex}
    \caption{Temporary Subterm Set}\label{fig:subterm-sets-tmp}
\end{figure}

Consider the sets $\mstsq\left( \_ \right)$ defined in \autoref{fig:subterm-sets-tmp}.

\begin{mlemmeenv}{}{sbtm-works-with-the-tmp}
    Let $t\in\Ez$, and $\modelc$ a cryptographic model
    \begin{equation}
        \mstbc\left( \mevalT{t} \right)\backslash\mathfrak{H}\subseteq\expsubt{\mstsq\left( t \right)}
    \end{equation}
\end{mlemmeenv}
\begin{mdemoenv}{}
    By induction over the size of $\mevalT{t}$ and then by case analysis over $t$. Consider for instance the case $t=\minput\left( T \right)$. Let $\Delta\coloneqq\mstbc\left( \mevalT{\minput\left( T \right)} \right)\backslash\mathfrak{H}$

    \underline{$\mevaltimeT{T}=\minit,\mundef$:} Then $\Delta=\varnothing$ and we conclude.

    \underline{$\mevaltimeT{T}=\stpa\left[ \vecib \right]$:} A quick induction shows us that (remember that $\bwedge$ is simply sugar over an \textsc{If}):
    \begin{equation}
        \Delta=\left\{ \mstbc\left\{ t' \right\}\backslash\mathfrak{H} \middle| b\precE \stpa\left[ \vecib \right], t'\in\left\{ \mmsg\left( b \right), \mcond\left( b \right) \right\}\right\}
    \end{equation}

    Then let $u\in \Delta$ with $u\neq \minput\left( T \right)$.
    We know that $u\in \mstbc\left( \mevalT{\mmsg\left( \stpb\left[ \vecjb \right] \right)} \right)$ (or $\mcond\left( \stpb\left[ \vecjb \right] \right)$) with $\stpb\left[ \vecjb \right]\in\domE$ such that $\stpb\left[ \vecjb \right]\precE \stpa\left[ \vecib \right]$.
    That is we can find a assignment $\sigma$ such that $u\in \mstbc\left( \mevalT{\mmsg\left( \stpb\left[ \sigma\left( \vecj \right) \right] \right)} \right)$ (or $\mcond\left( \stpb\left[ \sigma\left( \vecj \right) \right] \right)$).
    By induction $u\in \expsubt{\mmsg\left( \stpb\left[ \vecj\right] \right)}$ and since $\modelc^\eta\left( \stpb\left[ \vecjb \right] <  \stpa\left[ \vecib \right] \right)$, we get $u\in \expsubt{\mstbc\left( \minput\left( T \right) \right)}$.

    Thus we conclude that 
    \begin{equation}
        \Delta\in \expsubt{\mstbc\left( \minput\left( T \right) \right)}
    \end{equation}
\end{mdemoenv}

\begin{mlemmeenv}{}{inclusion-sbtm}
    \begin{equation}
        \expsubt{\mstsq\left( t \right)} \subseteq \expsubt{\mmst\left( t \right)}
    \end{equation}
\end{mlemmeenv}
\begin{mdemoenv}{}
    By induction. The second set is less constraining on the side condition.
\end{mdemoenv}

\begin{mdemoenv}{th:subterm-soundness}
    By Lemma \ref{lemme:sbtm-works-with-the-tmp} then \ref{lemme:inclusion-sbtm}.
\end{mdemoenv}

\smallskip
Finally to make $\mmst\left( \_ \right)$ fully usable we must ensure it is effectively computable:
\begin{mpropenv}{}{st-is-finite}
    For all $t$, $\mmst\left( t \right)$ is finite.
\end{mpropenv}
\begin{mdemoenv}{}
    This comes from the fact that $\mmsttmp{1}{1}\left( \_ \right)$ do not have conditions and are subsets of all the terms appearing in the description of the protocol (which is finite).
    They are the only recursively defined sets that do not decrease with the structural ordering.
\end{mdemoenv}

In practice, we compute $\mmst\left( \_ \right)$ by memoizing the calls to the inputs and the memory cells. This in turn is equivalent to looking for connected parts of the graph of calls to inputs and memory cells.

\else
    \begin{mproof}
        The proof is in the technical report~\cite{CRYPTOVAMPIREAutomatedReasoningTR}.
    \end{mproof}
\fi
    \ifreport
        \section{Compatibility with squirrel}\label{sec:comp-w-squirrel}

We remind in \autoref{fig:squirrel-syntax} the grammar used in the tool \msquirrel as defined in~\cite{baelde_cracking_2022}.

\begin{figure}
    \begin{align*}
        T &\coloneqq \tau \gsep \mathsf{a}\left[ \vec\imath \right] \gsep \mpred(T)\\
        t &\coloneqq x \gsep \mathsf{c}\left[ \vec\imath \right]\mat T \gsep \mathsf{n}\left[\vec\imath\right] \gsep f\left[ \vec\imath \right]\left( \vec t \right) \\
        &\qquad \gsep \textsf{input}\mat T {\color{gray!80}\gsep \textsf{output} \mat T \gsep \textsf{frame}\mat T}\\
        &\qquad \gsep \textsf{if }\phi \textsf{ then } t \textsf{ else }t'\\
        &\qquad \gsep \textsf{find }\vec\imath \textsf{ such that }\phi \textsf{ then } t \textsf{ else }t'\\
        A &\coloneqq t =t' \gsep  i = i' \gsep T=T'\gsep T< T' \gsep T\leq T' \\
        &\qquad \gsep \textsf{happens}(T) {\color{gray!80}\gsep \textsf{cond}\mat T \gsep \textsf{exec} \mat T}\\
        \phi &\coloneqq A \gsep \top \gsep \bot \gsep \phi \wedge \phi' \gsep \phi \vee \phi' \gsep \phi \Rightarrow \phi \gsep \neg\phi\\
        &\qquad \gsep \forall i.\phi \gsep \exists i.\phi \gsep \exists \tau. \phi \gsep \forall \tau. \phi
    \end{align*}
    \caption{\msquirrel~\cite{baelde_cracking_2022}'s syntax}\label{fig:squirrel-syntax}
\end{figure}

\newcommand{\mmap}{\mathscr{T}}

\begin{figure*}
    \begin{mathpar}
        \mmmap{\tau}\coloneqq\tau

        \mmmap{i}\coloneqq i

        \mmmap{\mathsf{a}\left[ \vec\imath \right]}\coloneqq\mathsf{a}\left[ \vec\imath \right]

        \mmmap{\mpred(T)}\coloneqq\mpred\left( \mmmap{T}\right)

        \mmmap{\mathsf{happens}(T)}\coloneqq\mhappens\left( \mmmap{T}\right)

        \mmmap{\mathsf{c}\left[ \vec\imath \right]\mat T}\coloneqq\mathsf{c}\left[ \vec\imath \right]\left( \mmmap{T}\right)

        \mmmap{\mathsf{n}\left[\vec\imath\right]} \coloneqq \mathsf{n}\left[\vec\imath\right]

        \mmmap{f\left[ \vec\imath \right]\left( t_1, \dots, t_n \right)}\coloneqq f\left[ \vec\imath \right]\left( \mmmap{t_1}, \dots, \mmmap{t_n} \right)

        \mmmap{\textsf{input}\mat T} \coloneqq \minput\left( \mmmap{T} \right)
        
        \mmmap{\textsf{output}\mat \mathsf{a}\left[ \veci \right]} \coloneqq \mmsg\left( \mathsf{a}\left[ \veci \right] \right)

        \mmmap{\textsf{cond}\mat \mathsf{a}\left[ \veci \right]} \coloneqq \mcond\left( \mathsf{a}\left[ \veci \right] \right)

        \mmmap{t = t'} \coloneqq \mmmap{t} \equiv \mmmap{t'}

        \mmmap{i = i'} \coloneqq \mmmap{i} = \mmmap{i'}

        \mmmap{T = T'} \coloneqq \mmmap{T} = \mmmap{T'}

        \mmmap{T < T'} \coloneqq \mmmap{T} < \mmmap{T'}

        \mmmap{\top} \coloneqq \btop

        \mmmap{\bot} \coloneqq \bbot

        \mmmap{ \phi \square \phi} \coloneqq \mmmap{\phi} \mathbin{\bar\square} \mmmap{\phi}




        \mmmap{\neg\phi} \coloneqq \bneg\mmmap{\phi}

        \mmmap{\mathfrak{Q} \vec\imath, \vec\tau. \phi} \coloneqq \bar{\mathfrak{Q}} \vec\imath, \vec\tau . \mmmap{\phi}



        \mmmap{\textsf{if }\phi \textsf{ then } t \textsf{ else }t'} \coloneqq \mifthenelsenb{\mmmap{\phi}}{\mmmap{t}}{\mmmap{t'}}

        \mmmap{\textsf{find }\vec\imath \textsf{ such that }\phi \textsf{ then } t \textsf{ else }t'} \coloneqq
        \mfindst{\vec\imath}{\mmmap{\phi}}{\mmmap{t}}{\mmmap{t'}}

    \end{mathpar}
    where $\square\in \left\{\wedge, \vee, \Rightarrow \right\}$ and $\mathfrak{Q}\in \left\{ \forall, \exists \right\}$.

    \caption{Mapping between \msquirrel's terms and \tool's}\label{fig:sq-cv-mapping}
\end{figure*}

We write $\mathcal{T}^S$ for the set of \msquirrel terms that can be built out of the non-grayed out grammar from \autoref{fig:squirrel-syntax}, and $\mathcal{T}^C=\msymbl$ for \tool's symbolic terms (from \autoref{fig:symbolic-logic}). We define $\mmap:\mathcal{T}^S\mapsto \mathcal{T}^C$ to map between the two. Intuitively, $\mmap$ commutes with all non-macro operators (including memory cells). Then
\begin{gather}
    \mmmap{\textsf{input}\mat T} \coloneqq \minput\left( \mmmap{T} \right)
\end{gather}
Any other term is not supported. This corresponds to at least all non-grayed out terms in \autoref{fig:squirrel-syntax}. Most of the remaining is recovered using \autoref{prop:exec-is-forall}. Effectively, the only non-supported term is $\mathsf{frame}\mat T$, which is rarely directly used.

The full description is in \autoref{fig:sq-cv-mapping}.

\begin{mlemma}[Steps]
    Let
    \[
        \underline{\mathsf{a}\left[ \vec\imath \right]}.
            \left(
                \phi_{\mathsf{a}\left[ \vec\imath \right]},
                o_{\mathsf{a}\left[ \vec\imath \right]},
                \left\{ \mathsf{s}\left[ \vec\jmath \right]\leftarrow u_{\mathsf{a}\left[ \vec\imath \right],\mathsf{s}\left[ \vec\jmath \right]} \middle| s\in \mathcal{C} \right\}
            \right)
    \]
    be a \msquirrel action, then, when applicable,
    \[
        \mathsf{a}\left[ \vec\imath \right] \coloneqq \left( \mmap\left( \phi_{\mathsf{a}\left[ \vec\imath \right]} \right), \mmap\left( o_{\mathsf{a}\left[ \vec\imath \right]} \right), \lambda \mathsf{c}\left[ \vec j \right]. \mmap\left( u_{\mathsf{a}\left[ \vec\imath \right],\mathsf{c}\left[ \vec\jmath \right]} \right) \right)
    \]
    is a \tool step.

    Using the $\lambda$ notation to define functions.
\end{mlemma}

We will then assume \tool and \msquirrel share their set of step/action names $\mathcal{S}$.

\begin{mlemma}[Protocol]
    Let $\underline{P}=\left( P_{\mathrm{act}}, \mathcal{U}_0, < \right)$ be a \msquirrel protocol, then $\mathcal{P}=\left( \left\{ \mathsf{a}\left[ \_ \right] \middle| \mathsf{a}\left[ \vec\imath \right] \in P_{\mathrm{act}} \right\}, < \right)$ is a \tool abstract protocol.
\end{mlemma}
\begin{mproof}
    $<$ is a partial order over $\left\{ \mathsf{a}\left[ \vec\imath \right] \middle| \mathsf{a}\left[ \_ \right] \in S \right\}$, thus it is a preorder. Moreover, $<$ is insensitive to the indices. Finally, all steps may only refer to previous steps, and memory cells do not have cyclic calls (as they only refer to previous steps).
\end{mproof}
\begin{mlemma}[Trace]
    Let $\mathbb{T}_{\mathrm{sq}}=\left( \domI, \mathcal{D}_\mathcal{T}^{\mathrm{sq}}, <_T, \sigmI, \sigma_T \right)$ be a \msquirrel trace. Then $\mathbb{T}=\left(  \domI, \mathcal{D}_\mathcal{T}^{\mathrm{sq}},\sigmI, <_T\right)$ is a \tool trace.
\end{mlemma}

\bigskip
We write $\mevalTsq{\_}$ for the \msquirrel expansion.

\begin{mth*}[Interoperability~(\ref{th:sq-is-cv})]
    For any computational model\footnote{The notion is the same for \tool and \msquirrel as it comes from the \BCLogic and they share their function symbols}, \msquirrel protocol (and its \tool variant), trace $\mathbb{T}$ over it, security parameter $\eta$ and random tapes $\rho$, we have
    \begin{equation}
        \modelc^\eta\left(\meval{\mmmap{t}} \right)\left( \rho \right)=\mevalsq{t}\left( 1^\eta, \rho \right)
        \tag{\ref{eq:sq-cv-th}}
    \end{equation}
\end{mth*}

\begin{mdemoenv}{}
    By induction on the size of $\mevalTsq{t}$ where \autoref{prop:base-axioms} along with \autoref{lemme:quantifiers} and \autoref{prop:conn-commutes} tells us that the unfolding of the quantifiers in \tool and \msquirrel can be assumed to be the same.
\end{mdemoenv}

\begin{mpropenv}{Translation of marcos}{exec-is-forall}
    In \msquirrel
    \begin{gather}
        \proves \mathsf{cond}\mat \tau \Leftrightarrow \bigvee\limits_{\mathsf{a}\left[ \veci \right]\in P_{\mathrm{act}}} \exists \veci. \tau = \mathsf{a}\left[ \veci \right] \wedge \mathsf{cond}\mat \mathsf{a}\left[ \veci \right]\label{eq:cond-is-exhaustiveness}\\
        \proves \mathsf{exec} \mat T \Leftrightarrow \left( \forall \tau. \tau \leq T \Rightarrow \mathsf{cond} \mat T \right)\label{eq:exec-is-forall}
    \end{gather}
    This lets us effectively use $\mathsf{exec} \mat T$ in \tool.
\end{mpropenv}
\begin{mdemoenv}{}
    \eqref{eq:cond-is-exhaustiveness} by exhaustiveness of the steps and \eqref{eq:exec-is-forall} by induction on $T$.
\end{mdemoenv}

        \section{Cryptographic Axioms}\label{sec:crypto-axioms}

In this section, we present the other cryptographic notions we support, along with a proof sketch of their axiom.

\subsection{No Guessing Theorem}\label{sec:app:no-guessing}
In this section, we produce a proof for \autoref{th:no-guessing}.

\begin{mth*}[\ref{th:no-guessing} No Guessing]
    It is not possible to guess honest nonces. Formally, for all nonce $\n\left[ \_ \right]\in \mathcal{N}$, ground indices $\vecib\in \mindices$, and message $m$, we have
    \begin{equation*}\tag{\ref{eq:no-guessing}}
        \meval{\n\left[ \vecib \right]} = \meval{m} \Rightarrow \n\left[ \vecib \right]\sqsubseteq m
    \end{equation*}
\end{mth*}
\begin{mdemoenv}{}
    Let $\n\in\mnonces$ and nonce name $\vecib$ some indices and $m$ a message.

    Suppose $\modelc^\eta\left( \n\left[ \vecib \right]\sqsubseteq m \right)\left( \rho \right)=0$ (notice that this then holds for all $\rho$). 
    By \autoref{th:subterm-soundness} we get that the random variables $\modelc^\eta\left( \n\left[ \vecib \right] \right)$ and $\modelc^\eta\left( m \right)$ are independants as $m$ never gets access to the relevant part of $\rho_h$.

    Thus
    \begin{equation}
        \mprob\left( \modelc^\eta\left( \n\left[ \vecib \right]= m \right) \right)=\frac{1}{2^\eta}=\mnegl\left( \eta \right)
    \end{equation}
\end{mdemoenv}

We continue with a proof of \autoref{sec:final-optimization}'s version of \autoref{th:no-guessing}.
\begin{mth*}[\ref{th:no-guessing-2}]
    For all nonces $\n\in\mnonces$, \eqref{eq:super-no-guessing} can be added to the set of axioms
    \begin{equation}\tag{\ref{eq:super-no-guessing}}
        \forall \veci, m.\ \meval{\n\left[ \veci \right]} = \meval{m} \Rightarrow \n\left[ \veci \right]\bsqsubseteq m
    \end{equation}
    where $\bsqsubseteq$ is defined by \eqref{eq:sq-nonce-def} and is compatible with $\beq$.
    \begin{equation}\tag{\ref{eq:sq-nonce-def}}
        \m \bsqsubseteq \meval{t} \coloneqq \forall t'. \left(\meval{t}=\meval{t'} \Rightarrow \m \sqsubseteq t' \right).
    \end{equation}
\end{mth*}
\begin{mproof}
    Let $\forall \vecu. \Gamma$ be an axiom containing \autoref{th:no-guessing}, that is such that
    \begin{equation}
        \valid \left( \forall \vecu. \Gamma \right)\implies \forall \veci, m.\  \meval{\n\left[ \veci \right]} = \meval{m} \Rightarrow \n\left[ \veci \right]\sqsubseteq m
    \end{equation}
    Then by transitivity of $=$, we have
    \begin{equation}
        \valid \left( \forall \vecu. \Gamma \right)\implies \eqref{eq:super-no-guessing}
    \end{equation}
    Thus, for all $t$, a proof of $\valid \left( \forall \vecu. \Gamma \right)\implies \meval{t}$ implies a proof of $\valid \left( \forall \vecu. \Gamma \right)\wedge \eqref{eq:super-no-guessing} \implies \meval{t}$. Therefore, \autoref{th:main} lets us add \eqref{eq:super-no-guessing} to the axiom pool while retaining the cryptographic semantics of the proof (as long as $\Gamma$ is a described in \autoref{th:main}).

    Then the claimed compatibility with $\beq$ comes from \eqref{eq:th:no-guessing-2-compat}.
    \begin{equation}\label{eq:th:no-guessing-2-compat}
        \valid\forall t, t'.\ \m\bsqsubseteq t \wedge \meval{t} =\meval{t'} \implies \m \bsqsubseteq t'
    \end{equation}

    The nonce itself can be extracted from the reasoning modulo $\beq$ using a type system as described in \appref{sec:general-implementation}.
\end{mproof}

\subsection{Euf-Cma}
We already presented the axiom of \textsc{MAC}s in \autoref{prop:euf-cma}.
We start from the BC~rule from \autoref{ex:challenge}.

\begin{mdemoenv}{ex:challenge}
    We refer to previous work~\cite{squirrel} for a more in-depth proof.

    The intuition is that assuming \eqref{eq:euf-cma-bc} does not hold, then $\left( m, \sigma \right)$ is a winning attacker to the \eufcma security game, which contradicts the assumptions.
\end{mdemoenv}

We can then lift to \autoref{prop:euf-cma}.

\begin{mdemoenv}{prop:euf-cma}
    We start from the BC~formula:
    \[
        \modelclass{}_{\eufcma}\validbc \mverify\left(\sigma, m, \mk \right) \bimplies
            \bar{\bigvee}_{\mhash\left( u, \mk \right)\in \mstbc\left( m, \sigma \right) } u \beq m
        \tag{\ref{eq:euf-cma-bc}}
    \]
    By \autoref{th:subterm-soundness} and \autoref{prop:validity} we get that
    \begin{equation}\label{eq:demo:euf-cma}
        \modelclass{}_{\eufcma}\validbc \mverify\left(\sigma, m, \mk \right) \bimplies
            \bar{\bigvee}_{\mhash\left( u, \mk \right)\in \mmst\left( m, \sigma \right) } u \beq m
    \end{equation}

    Turing \eqref{eq:demo:euf-cma} into a formula of \mevall and using \autoref{prop:conn-commutes}, we get
    \begin{equation}
        \modelclass{}_{\eufcma}\validp \meval{\mverify\left(\sigma, m, \mk \right)} \implies
            \bigvee\nolimits_{\mhash\left( u, \mk \right)\in \mmst\left( m, \sigma \right) } \meval{u} = \meval{m}
    \end{equation}
    Which in turn (including the side condition in the formula) gives us \eqref{eq:eufcma1} for \autoref{prop:euf-cma}.
\end{mdemoenv}

Very much the same way we show the public key version of \eufcma:
\begin{mprop}[Euf-Cma]\label{prop:pub-euf-cma}
    When $\mathsf{sign}$, $\mathsf{verify}$ and $\mathsf{vk}$ form a signature scheme that is existentially unforgeable under chosen message attacks (\eufcma),
    then the protocol $\mathcal{P}$ satisfies the following:
    \begin{multline}
        \meval{\mathsf{verify}\left( \sigma, m, \mathsf{vk}\left( \mnonce{\mathsf{k}} \right) \right)} \Rightarrow \\
            \mleft(\begin{multlined} \mathsf{k}\sqsubseteq_{\mathsf{sign}\left( \_,\bullet \right), \mathsf{vk}\left( \bullet \right)} m, \sigma, \mathcal{P} \\
        \vee \exists u. \mleft( \mathsf{sign}\left( u, \mnonce{\mathsf{k}} \right) \sqsubseteq m, \sigma \wedge \meval{u} = \meval{m} \mright)\end{multlined} \mright)
    \end{multline}
    where $\mathsf{k}$ is a \randomt{}.
\end{mprop}

Since the subterm relation is a bounding formula (\autoref{def:bounding-formula}), Properties \ref{prop:bsnf-in-base-logic} and \ref{prop:bsnf-and-or} tell us that we can expand \autoref{prop:euf-cma} and \autoref{prop:pub-euf-cma} from axioms schemas into formula with bounded Skolem normal form.

\subsection{Int-Ctxt}

\begin{mprop}[Int-Ctxt]
    When $\mathsf{senc}$ and $\mathsf{dec}$ form a symmetric encryption scheme that conserves the integrity of the ciphertext (\textsc{IntCtxt}), any protocol $\mathcal{P}$ verifies:
    \begin{multline}\label{eq:intctxt}
        \meval{\mathsf{verify}\left( c, \bar{\mathsf{k}}\right)} \Rightarrow \\
        \exists m, r. \left( \meval{\mathsf{senc}\left( m, r, \bar{\mathsf{k}} \right)} = \meval{c} \wedge \mathsf{senc}\!\left( m, r, \bar{\mathsf{k}} \right)\sqsubseteq c\right) \\
        \vee \mathsf{k}\sqsubseteq'_{\mathsf{senc}\left( \_, \_, \bullet \right), \mathsf{sdec}\left( \_, \bullet \right), \mathsf{verify}\left( \_, \bullet \right)} c, \mathcal{P} \\
        \vee \neg{\normalfont\textsf{senc-rand}}_\mathsf{k}^\mathcal{P}\left( u \right)
    \end{multline}
    where $\mathsf{k}$ is a \randomt{} and $\mathsf{verify}(c, k)\coloneqq \mathsf{dec}\not\equiv \mathsf{fail}$.

    ${\normalfont\textsf{senc-rand}_\mathsf{k}^\mathcal{P}(u)}$ holds when in all instances of $c_r=\mathsf{senc}\left( m, r, \mnonce{\mathsf{k}} \right)$ appearing in the extension of $u$ in any trace, $r$ is of the form $\mnonce{\mathsf{r}}$ and only appear in $u$ in $c_r$.
\end{mprop}
\begin{mproof}
    This is a similar lifting of a rule from~\cite{squirrel}.
\end{mproof}


        \section{Engineering Details}\label{sec:engineering}

\subsection{Effective First-Order Representation}\label{sec:general-implementation}
As presented in \autoref{sec:form:challenges}, the main first-order formalization challenges revolve around the interactions (or lack thereof) between $\beq$ and $\sqsubseteq$.

Therefore, efficient first-order reasoning within \tool crucially depends on built-in, native support for
\begin{enumerate*}[label=\textbf{(\roman{enumi})}, ref=(\roman{enumi})]
    \item\label{item:sec:general-implementation:1} efficiently distinguishing when to use $=$ and $\beq$ in order to quickly evaluate the subterm relation (e.g., inferring that $x$ is a subterm of $x \bwedge y$).
    Additionally, \item\label{item:sec:general-implementation:2} base formulas should also be handled efficiently (e.g., $\bwedge$ should be treated as a logical conjunction).
\end{enumerate*}

We resolve these issues in \tool, by \ref{item:sec:general-implementation:1} considering multi-sorted first-order representations of \autoref{sec:contributions}, for which we propose extensions to the \smtlib{} type system~\cite{smtlib}.
To this end, we declare symbolic terms ($t$ in \autoref{fig:symbolic-logic}) as \emph{datatypes}, allowing us to distinguish between three main types: \msgt{} (symbolic bitstring computations that can be sent over the network), \condt{} (symbolic boolean computations), and \noncet{}.
We introduce the function $\mnonce{\n}$ to map \noncet{} to \msgt{}, enabling us to \enquote{distinguish} honest nonces in cryptographic axioms and speed up reasoning with $=$ using types.
Similarly, we enumerate possible \tool steps via datatypes, introducing the type \tpt{}.

Further, symbolic quantifiers and lookups (e.g., $\bexists i.\phi$) are named in the following sense: for each term $t$ denoting a symbolic quantifier $\mathfrak{Q}\vec x. \vec y$, where $\mathfrak{Q}$ is $\bforall$, $\bexists$, or $\mfindst{\_\allowbreak}{\_\allowbreak}{\_\allowbreak}{\_}$, we introduce a new honest function  $f_\mathfrak{Q}$.
Here, the function $f_\mathfrak{Q}$ takes the free variables of $t$ as its arguments. We additionally consider the respective axioms for $\meval{f_\mathfrak{Q}\left(\dots\right)}$ and subterm relations, such as instances of axiom~\eqref{eq:subterm-def-quantifiers}.

For efficient reasoning modulo $\beq$, \ref{item:sec:general-implementation:2} we introduce the new sort \bitstrt{} and, for each $f\left[ \vec\_ \right]\in \mathcal{F}$, we consider a free function $\meval{f}$ such that
\begin{equation}
    \meval{f\left[ \vec\imath \right]\left( t_1,\dots, t_n \right)}=\meval{f}\left(\vec\imath, \meval{t_1},\dots, \meval{t_n}\right)\label{eq:rewrite-bitstring}
\end{equation}
As such, \bitstrt{} is used to denote the sort of evaluated \msgt{}, whereas the built-in \boolt{} represents the sort of evaluated \condt{}. \tool's \EvalLogic (\autoref{sec:evaluated}) is represented as is using standard Boolean connectives.

The above considerations \ref{item:sec:general-implementation:1}--\ref{item:sec:general-implementation:2} provide first-order encodings for \tool formalization.
To turn reasoning over such encodings efficient, we introduce various modifications to the saturation-based theorem proving over \tool encodings, complemented with preprocessing heuristics.

\subsection{Customized Subterm Relations}\label{sec:custom-subterm}
Unfortunately, this general subterm relation $\sqsubseteq$ is often not expressive enough for our class of problems.
Already in the \eufcma axiom (\autoref{prop:euf-cma} of \autoref{sec:overview}), we use $\sqsubseteq_{\mverify\left(\_,\_,\bullet\right),\mhash\left(\_,\bullet\right)}$.
This means we ignore the positions of the key when looking for subterms through $\mhash\left( \_, \_ \right)$ and $\mverify\left( \_, \_, \_ \right)$.

In the general case, we write $\sqsubseteq_{ f\left[ \vec\_ \right]\left(\__1,\dots,\bullet_j,\dots,\__n \right)}$ for some $f\left[ \vec\_ \right]\in \mfunctions$ to express that one should ignore the $j^\text{th}$ argument when looking through $f$. Such a relation cannot be expressed solely through the base $\sqsubseteq$. Therefore, we introduce it as a brand-new binary relation that closely resembles $\sqsubseteq$.
\tool computes new sets $\mmst^{ f\left[ \vec\_ \right]\left(\__1,\dots,\bullet_j,\dots,\__n \right)}\left(\_\right)$ and we describe the relation to the theorem prover by adapting the axioms of \autoref{fig:subterm-relation} to $\mmst^{ f\left[ \vec\_ \right]\left(\__1,\dots,\bullet_j,\dots,\__n \right)}\left(\_\right)$. In particular, the instance of axiom~\eqref{eq:subterm-def-function} for $f$ is replaced by:
\begin{gather}
    \begin{multlined}
        \forall t, t_1, \dots, t_n, \vec\imath.\\
        t \sqsubset_{f\left[ \vec\_ \right]\left( \__1,\dots,\bullet_j,\dots,\__n \right)} f\left[ \vec\imath \right]\left(t_1,\dots, t_n  \right) \wedge t = t_j \Rightarrow\\
        \bigvee\nolimits_{\substack{k=1\\k\neq j}}^n t \sqsubseteq_{ f\left[ \vec\_ \right]\left(\__1,\dots,\bullet_j,\dots,\__n \right)} t_k \end{multlined}\\
    \begin{multlined}
        \forall t, t_1, \dots, t_n, \vec\imath.\\
        t \sqsubset_{f\left[ \vec\_ \right]\left( \__1,\dots,\bullet_j,\dots,\__n \right)} f\left[ \vec\imath \right]\left(t_1,\dots, t_n  \right) \wedge t \neq t_j \Rightarrow\\
        \bigvee\nolimits_{k=1}^n t \sqsubseteq_{f\left[ \vec\_ \right]\left( \__1,\dots,\bullet_j,\dots,\__n \right)} t_k \end{multlined}
\end{gather}

\autoref{ex:custom-subterm} gives an example of such a customized subterm relation.
\begin{mex}\label{ex:custom-subterm}
    \newcommand{\hvsqsubseteq}{\sqsubseteq_{\mhash\left(\_,\bullet\right)}}
    \newcommand{\hvsqsubset}{\sqsubset_{\mhash\left(\_,\bullet\right)}}
    With the simpler case of $\hvsqsubseteq$, we replace the instance of \autoref{eq:subterm-def-function} where $f$ is $\mhash$ with
    \begin{gather}
        \forall t, m.\ t\hvsqsubset \mhash\left(m, t\right) \implies t \hvsqsubseteq m\\
        \begin{multlined}
            \forall t, m_1,m_2.\vspace{-10pt}\\\left(t\hvsqsubset \mhash\left(m_1, m_2\right) \wedge t \neq m_2\right) \implies \bigvee_{k=1}^{2}t \hvsqsubseteq m_k
        \end{multlined}
    \end{gather}
\end{mex}

Such customization naturally extends to more than one function and more than one argument position.

\subsection{Forcing Rewritings}
Saturation-based provers rely on term orderings to keep their proof search small~\cite{bachmairChapterResolutionTheorem2001}, by ensuring that smaller terms (w.r.t. the ordering) are not rewritten by equal larger terms.
To further reduce the application of equality (and thus rewritings of equal terms), we use equality reasoning based on $\beq$ by using the \bitstrt{} sort from \autoref{sec:general-implementation}.
Doing so, we orient the axiom~\eqref{eq:rewrite-bitstring} as
\newcommand{\rewriteeq}{\mathbin{\stackrel{\mathclap{\resizebox{!}{0.7\height{}}{$\to$}}}{=}}}
\begin{equation}\label{eq:rewrite}
    \meval{f\left[ \vecI \right]\left( t_1,\dots, t_n \right)}%
    \rewriteeq%
    \meval{f}\left(\vecI, \meval{t_1},\dots, \meval{t_n}\right) \tag{\ref{eq:rewrite-bitstring}}
\end{equation}
where $\rewriteeq$
denotes that $\meval{f\left[ \vecI \right]\left( t_1,\dots, t_n \right)}$ is bigger than (and thus should be rewritten by) $\meval{f}\left(\vecI, \meval{t_1},\dots, \meval{t_n}\right)$.
Such an orientation goes, however, against standard orderings, as, for example, constants/unary functions are usually smaller than function symbols of those of higher arity.

To enforce~\eqref{eq:rewrite}, we introduce additional orderings over \tool terms;
while these extensions may not preserve refutational completeness, our experiments show overall good performance (see \autoref{sec:experiments}).

    \fi
\end{document}